\documentclass[a4paper, 11pt]{article}
\usepackage[total={6in, 8in}]{geometry}

\usepackage{amsmath,amssymb,amsfonts}
\usepackage{bm}
\usepackage{csquotes}
\usepackage{subfigure}
\usepackage{xcolor}
\usepackage{graphicx}
\usepackage[ruled]{algorithm2e}
\usepackage{authblk}
\usepackage{natbib}

\newcommand{\mathsfbi}[1]{\mathsf{\mathbf{#1}}}
\newcommand{\bcdot}{\boldsymbol{\cdot}}

\newcommand{\dd}{\mathrm{d}}
\newcommand{\ddt}[2]{\frac{\displaystyle \partial{#1}}{\displaystyle \partial{#2}}}
\newcommand{\dddt}[2]{\frac{\displaystyle \partial^{2}{#1}}{\displaystyle \partial{#2}^{2}}}
\newcommand{\dBt}{\mathrm{d}\bm{B}_t}

\newcommand{\sdBt}{\boldsymbol{\sigma} \mathrm{d}\bm{B}_t}

\begin{document}

\title{Input-output analysis of the stochastic Navier-Stokes equations:\\ application to turbulent channel flow}

\author[1]{Gilles Tissot}
\affil[1]{INRIA Rennes Bretagne Atlantique, IRMAR -- UMR CNRS 6625, av. Général Leclerc, 35042 Rennes, France}

\author[2]{Andr\'e V. G. Cavalieri}
\affil[2]{Department of Aerospace Engineering, Instituto Tecnol\'{o}gico de Aeron\'{a}utica,Pra\c{c}a Mal. Eduardo Gomes 50, Vila das Ac\'{a}cias, 12228-900, S\~{a}o Jos\'{e} dos Campos, Brazil}

\author[1]{\'Etienne M\'emin}

\maketitle

\begin{abstract}
   Stochastic linear modelling proposed in \textsc{Tissot}, \textsc{Mémin} \& \textsc{Cavalieri} (\textit{J. Fluid Mech.}, vol. 912, 2021, A51) is based on classical conservation laws subject to a stochastic transport.
   Once linearised around the mean flow and expressed in the Fourier domain, the model has proven its efficiency to predict the structure of the streaks of streamwise velocity in turbulent channel flows.
   It has been in particular demonstrated that the stochastic transport by unresolved incoherent turbulence allows to better reproduce the streaks through lift-up mechanism.
   In the present paper, we focus on the study of streamwise-elongated structures, energetic in the buffer and logarithmic layers.
   In the buffer layer, elongated streamwise vortices, named rolls, are seen to result from coherent wave-wave non-linear interactions, which have been neglected in the stochastic linear framework.
We propose a way to account for the effect of these interactions in the stochastic model by introducing a stochastic forcing, which replace the missing non-linear terms.
   In addition, we propose an iterative strategy in order to ensure that the stochastic noise is decorrelated from the solution, as prescribed by the modelling hypotheses.
   We explore the prediction abilities of this more complete model in the buffer and logarithmic layers of channel flows at $Re_\tau=180$, $Re_\tau=550$ and $Re_\tau=1000$.
   We show an improvement of predictions compared to resolvent analysis with eddy viscosity, especially in the logarithmic layer.
\end{abstract}

\section{\label{sec:intro}Introduction}
Coherent structures of the near-wall turbulence is an extensively explored topic.
In the buffer layer, very close to the wall, the flow organises into streamwise vortices, or rolls, and elongated patterns of high/low streamwise velocity denoted as \emph{streaks} \citep{kline1967,smith1983,jimenez1991}.
These structures develop, break and are regenerated in a quasi-cyclical process \citep{panton2001}.
A scenario explaining their behaviour \citep{hamilton1995} considers the cycle where the streaks intensify by the lift-up mechanism \citep{ellingsen1975,brandt2014}, destabilise by spanwise meandering leading to non-linear interactions which give finally birth to new streamwise vortices.
This final step allows to start a new cycle.

In the logarithmic layer, the flow organises as well along streaks \citep{flores2010,smits2011}.
These structures are of larger size with a more disorganised motion due to the higher Reynolds number based on the wall distance (the wall distance in viscous units $y^+=\frac{yu_\tau}{\nu}$ is a Reynolds number based on the friction velocity $u_\tau$, the kinematic viscosity $\nu$ and the wall distance in outer units $y$, typical length scale of the largest structure \citep{popebook}).
Understanding their dynamical behaviour is still today an active research area.
Smaller scales appear to be unnecessary to sustain these structures \citep{hwangPRL2010}, suggesting the presence of a self-sustaining mechanism at large scale.
Several evidences indicate that this mechanism is similar to the one active in the buffer layer \citep{cossu2017,lozano2020}.
As noted in \citet{cossu2017}, these large scale coherent structures exist in the sense of (ensemble) averaging or filtering as associated to large eddy simulation (LES). 
As a consequence, it is crucial to include the effect of small scales on these large scales, through Reynolds stress models for instance, or, as we propose here, by stochastic modelling.
As a practical example in \citet{bae2021}, resolvent analysis has been used to extract these large coherent structures in view of performing diagnostics of their action in removing their contributions in a numerical simulation.
This procedure yields a drastic reduction of the turbulence intensity.
The reduction is significant in the buffer layer and slightly less so in the logarithmic-layer, highlighting the requirement of modelling improvements in this region.
Besides, practical control strategies require an accurate prediction of these structures, and providing simplified models predicting coherent structures at a given scale with a high fidelity is still today challenging.

By the knowledge of the time-averaged velocity field and possibly of some higher-order statistics, predicting coherent structures in a turbulent flow without resolving the whole space-time dependent solution has become an important research direction, to which many groups have devoted strong efforts.
Considering a linearisation of the Navier--Stokes operator around a suitably chosen flow -- often taken as the time-averaged flow \citep{barkley2006} -- it is natural to search for wave-solutions in the Fourier domain, which beyond a natural physical meaning gives access to efficient linear-algebra techniques.
Since turbulence interacts with these wavy coherent structures, linearised solutions are often insufficient, and a closure is required.

Resolvent analysis \citep{schmid2001,trefethenbook,jovanovic2005} has become widely used to model coherent structures in turbulent flows since it considers the response of the linearised system to a forcing interpreted as the unknown non-linear term \citep{mckeon2010}.
By singular value decomposition (SVD) of the resolvent operator, optimal harmonic forcing modes and associated responses are found.
Resolvent analysis is used for the modelling of dominant coherent structures in turbulent flows \citep{mckeon2010,bae2021}, data assimilation \citep{gomez2016b,symon2019,martini2020,towne2020,amaral2021,franceschini2021}, as well as flow control \citep{leclercq2019}.

In the context of a triple decomposition, where the velocity field is split into a time-average, a coherent-structure component and an inchoherent turbulent field, an eddy viscosity can be introduced to the generalised Reynolds stresses induced by the incoherent part \citep{reynolds1972}.
For streaky structures in turbulent channel flows Cess's eddy viscosity model \citep{Cess1958} has proven its prediction efficiency to some extent \citep{hwang2010,morra2019,symon2020,amaral2021}, with a particular need in the logarithmic layer.
It will constitute our comparison model and will be referred as $\nu_t$-resolvent analysis, in contrast with $\nu$-resolvent analysis when no eddy viscosity is considered.
This works well for coherent structures, or waves, where strong production occurs.
However, as argued by \citet{symon2021}, since eddy viscosity is mainly diffusive (up to eddy-diffusion gradients \citep{symonarxiv}), it breaks the energy conservation over the whole spectrum.
Then, it is not well adapted for waves receiving energy from other scales by backscattering.
A dedailed study of the discrepancy of $\nu$ and $\nu_t$-resolvent analysis for a turbulent channel flow in terms of low-rank property, projection onto SPOD modes and energy transfers can be found in \citet{symonarxiv}.
Other attempts to improve the modelling have been proposed.
The embedding of covariance informations of the forcing has been for instance proposed in \citep{nogueira2021,morra2021}.
However, this strategy was considered for diagnostic purposes only and has not been considered for predictions since a fine knowledge of the non-linear term is in that case required.
An estimator has been proposed by \citet{gupta2021} considering together eddy diffusion and a model of stochastic forcing.
As an alternative in the temporal domain, \citet{zare2017,zare2019} have devised a stochastic modelling based on control theory, which incorporates a coloured-in-time noise.

In \citet{tissotJFM2021}, a modelling strategy based on stochastic transport, so-called stochastic linear modes (SLM), has been proposed and will be considered in the present paper.
It starts from a stochastic version of the Navier--Stokes equations, originally introduced by \citet{memin2014}, which is based on the stochastic transport of conserved quantities.
The formalism has been successfully employed to perform large eddy simulations \citep{chandramouli2018}, geophysical flow modelling \citep{resseguier2017b,resseguier2017c,resseguier2017d,chapron2018,bauer2020a,bauer2020b}, near-wall flow modelling \citep{pinier2019}, data assimilation \citep{yang2017,yang2018,chandramouli2020} and reduced-order modelling \citep{resseguier2017,resseguier2021ROM}.
An advantage of the approach is the formulation of closure by defining statistics of a stochastic unresolved time-decorrelated (with respect to the time scales of the resolved part) velocity field.
The associated random perturbation ensues then from a stochastic transport operator.
This stochastic transport involves in addition a stochastic diffusion, and an effective drift velocity similar to the turbophoresis effect.
An exact energy balance is obtained between the stochastic diffusion and the energy backscattering induced by the stochastic transport \citep{resseguier2017b}.
Linearising this model and expressing it in the Fourier domain leads to the so-called \emph{stochastic linear modes} (SLM).

In SLM, the non-linear term, interpreted as the wave-wave interactions, has been neglected, relying on the stochastic transport of the solution by the incoherent turbulence to obtain a physically relevant model.
In the present paper, we come back to this strong assumption.
The generation of streamwise vortices likely involves non-linear interactions between large-scale coherent structures.
As will be detailed further, a close analysis of SLM for these elongated structures shows a poor prediction of the rolls, despite a good prediction of the streamwise velocity fluctuations.
This is consistent with the fact that coherent wave-wave interactions are neglected in SLM.
In order to recover the right roll properties, we propose in this paper to study the response of SLM to a \enquote{non-linear} forcing similarly to what is done in resolvent analysis for modelling non-linear effects through an input-output formalism.
We name this enhanced solution \emph{forced stochastic linear modes} (FSLM).

In addition to adding the aforementioned forcing, we propose some enhancements of the noise definition compared to \citet{tissotJFM2021}.
We propose an iterative procedure enforcing the noise to be incoherent with the solution.
Moreover, the stochastic diffusion tensor is defined by root-mean-square velocity profiles, in order to ensure an approximated consistency between stochastic diffusion and noise expressed in the Fourier domain.
We propose as well a decorrelation time definition based on an inertial range scaling.
Finally, SLM/FSLM numerical computation is improved by the reformulation of the equations as an SVD problem.

With this more complete model which incorporates the effect of decorrelated turbulence on the coherent structure, we will explore the prediction abilities of stochastic modelling in the buffer and logarithmic layer of three turbulent channel flows at friction Reynolds number $Re_\tau=180$, $Re_\tau=550$ and $Re_\tau=1000$.
In particular, we will explore the ability of FSLM to predict coherent structures in the logarithmic layer.

In section~\ref{sec:notations}, notations used along the paper are introduced.
In section~\ref{sec:SLMFSLM}, we present the stochastic model.
In section~\ref{sec:results} we explore the ability of these models to predict buffer and logarithmic layer structures in turbulent channel flows.
Conclusions are provided in section~\ref{sec:conclusion}.
Presentation of the resolvent analysis, numerical details and complementary results are given in Supplementary Material in order to have a more complete view by varying Reynolds number and sweeping the wave-number space.

\section{Notations and preliminaries}
\label{sec:notations}
We consider three turbulent channel flows at the friction Reynolds numbers $Re_\tau=180$, $Re_\tau=550$ and $Re_\tau=1000$ with the Cartesian coordinates $\bm{x}=(x,y,z)$ of the streamwise, wall-normal and spanwise directions of the domain $\Omega$, respectively.
The domain sizes $(L_x,L_y,L_z)$ in outer units are respectively $(4\pi,2,2\pi)$, $(2\pi,2,\pi)$ and $(2\pi,2,\pi)$.
Details and validations of the flow simulations can be found in \cite{amaral2021}, and additional details at $Re_\tau=550$ are present in \cite{morra2021}.
The time-dependent ($t$) state variable $\bm{q}(x,y,z,t)=(\bm{u},p)^T$ is composed of the velocity vector $\bm{u}=(u,v,w)^T$ and the pressure $p$.
The velocity field is decomposed in its time-average and fluctuation $\bm{u}=\bar{\bm{u}}+\bm{u}'$ with $\bar{\bm{u}}=(U(y),0,0)^T$.
By periodicity in the streamwise ($x$) and spanwise ($z$) directions, the space-time Fourier coefficient of the state variable with the sign convention $e^{i(\alpha x+\beta z - \omega t)}$ is noted $\hat{\bm{q}}_{\alpha,\beta,\omega}(y)$.
Variables $\alpha$, $\beta$, $\omega$ refer respectively to streamwise wavenumber, spanwise wavenumber and angular frequency.
In the wall-normal direction, we define a diagonal matrix $\mathsfbi{W}$ of quadrature coefficients.
Finally, we note $\bcdot^H$ the transpose-conjugate operation.

\section{Stochastic linear model and non-linear forcing}
\label{sec:SLMFSLM}
\subsection{Stochastic linear modes}
In \citet{tissotJFM2021}, a modelling strategy for coherent structures in turbulent flows has been proposed.
The formalism relies on the stochastic transport of conserved quantities by a time-differentiable velocity component perturbed by the variation of a Brownian motion.
Under these assumptions a stochastic version of the Navier--Stokes equations under location uncertainty \citep{memin2014} can be written.
In this section, we recall how the stochastic model can be expressed in the frequency-wavenumber domain to predict coherent structures.
More details can be found in \citet{tissotJFM2021}.

The displacement $\bm{X}(\bm{x},t)$ of a particle is written in a differential form
\begin{equation}
    \mathrm{d}\bm{X}(\bm{x},t)=\bm{u}(\bm{x},t)\mathrm{d}t+\boldsymbol{\sigma} \mathrm{d}\bm{B}_t,
    \label{eq:stdisplacement}
\end{equation}
where $\bm{u}$ is a time-differentiable velocity component, and $\dBt$ is the increment of a Brownian motion.
It can be remarked that equation~\eqref{eq:stdisplacement} has to be understood as a time integral over an infinitesimal time increment $\dd t$.
The operator $\boldsymbol{\sigma}$ is an integral operator which hides a spatial convolution in the domain $\Omega$ with a user-defined kernel $\boldsymbol{\check{\sigma}}$
\begin{equation}
   \left(\boldsymbol{\sigma} \mathrm{d}\bm{B}_t\right)_{\bm{x}}^i=\int_\Omega\boldsymbol{\check{\sigma}}^{ij}(\bm{x},\bm{x}',t)\mathrm{d}\bm{B}_t^j(\bm{x}')\,\mathrm{d}\bm{x}'.
\end{equation}

Defined as such, $\sdBt$ is the displacement induced by a velocity component that is smooth in space, but decorrelated in time.
This term aims at representing a time decorrelated (with respect to the time scale of the considered physical processes) turbulent velocity component.
In the general framework, $\boldsymbol{\sigma}$ can be smoothly time-dependent, but for the application of the present paper in statistically stationary turbulent flows, we will assume it constant in time.

Associated with $\boldsymbol{\sigma}$, we define the variance tensor $\mathsfbi{a}$ such that
\begin{equation}
    \mathsfbi{a}_{ij}(\bm{x})\dd t=\mathbb{E}\left(\left(\boldsymbol{\sigma} \mathrm{d}\bm{B}_t\right)_{\bm{x}}^i\left(\boldsymbol{\sigma} \mathrm{d}\bm{B}_{t}\right)_{\bm{x}}^j\right),
    \label{eq:a}
\end{equation}
with $\left(\sdBt\right)_{\bm{x}}^i$ the $i^{\text{th}}$ component of $\sdBt$ at position $\bm{x}$ and $\mathbb{E}$ the expectation operator.

Using the It\={o}-Wentzell formula, conservation of mass and momentum subject to a stochastic transport leads to a stochastic version of the incompressible Navier--Stokes equations \citep{memin2014,resseguier2017b}, referred to as \emph{under location uncertainty}:
\begin{equation}
    \begin{split}
        &\dd_t \bm{u} + \left(\bm{u}_d\bcdot \nabla\right)\bm{u}\,\dd t+\left(\sdBt\bcdot \nabla\right)\bm{u}=-\nabla \left(p_t\, \dd t + \dd p_t\right)
       \\&
       +\frac{1}{Re}\nabla\bcdot \left( \nabla \bm{u} \right)\, \dd t + \nabla\bcdot \left(\frac{1}{2}\mathsfbi{a}\nabla\bm{u}\right)\,\dd t+\frac{1}{Re}\nabla\bcdot \left( \nabla \sdBt \right) \\
        &\nabla\bcdot \bm{u}_d=0
        ;\quad
        \nabla\bcdot \boldsymbol{\sigma}=0,
        \\&\bm{u}_d=\bm{u}-\frac{1}{2}\nabla\bcdot \mathsfbi{a}.
    \end{split}
    \label{eq:stNS}
\end{equation}
In system~\eqref{eq:stNS}, $Re$ is the Reynolds number.
Compared to the deterministic case, the transport of $\bm{u}$ by $\sdBt$ is introduced.
This term brings energy (backscatter) to the system, which is exactly compensated by the stochastic diffusion $\nabla\bcdot \left(\frac{1}{2}\mathsfbi{a}\nabla\bm{u}\right)\,\dd t$ \citep{resseguier2017b}.
The variable $\bm{u}_d$ is called \textit{drift velocity}.
It takes into account that, in average, particles tend to be transported from highly turbulent regions towards low-turbulence regions \citep[see][and references therein]{resseguier2017}.
Mass conservation leads to a divergence-free condition on $\bm{u}_d$, and on $\boldsymbol{\sigma}$.
Finally, a random pressure term $\dd p_t$ corresponding to the small-scale velocity component is involved.
This force balances the martingale part (proportional to $\dBt$) of the system.

In \citet{tissotJFM2021}, system~\eqref{eq:stNS} is linearised around a mean velocity profile $U(y)$ and written in the Fourier domain (see details of the derivation in the latter reference), leading to
    \begin{equation}
       \begin{split}
          &-i\omega \hat{u}_{\alpha,\beta,\omega}+i\alpha U_d\hat{u}_{\alpha,\beta,\omega}+\hat{v}_{\alpha,\beta,\omega}\ddt{U}{y}+i\alpha \hat{p}_{\alpha,\beta,\omega}
           +\widetilde{D}(\hat{u}_{\alpha,\beta,\omega})=-(\dot{\bm{\xi}}_{\alpha,\beta,\omega})_y\ddt{U}{y}+\frac{1}{Re}\Delta(\dot{\bm{\xi}}_{\alpha,\beta,\omega})_x \\
          &-i\omega \hat{v}_{\alpha,\beta,\omega}+i\alpha U_d\hat{v}_{\alpha,\beta,\omega} +\ddt{\hat{p}_{\alpha,\beta,\omega}}{y}+\widetilde{D}(\hat{v}_{\alpha,\beta,\omega})
          =\frac{1}{Re}\Delta(\dot{\bm{\xi}}_{\alpha,\beta,\omega})_y\\
          &-i\omega \hat{w}_{\alpha,\beta,\omega}+i\alpha U_d\hat{w}_{\alpha,\beta,\omega} +i\beta \hat{p}_{\alpha,\beta,\omega}+\widetilde{D}(\hat{w}_{\alpha,\beta,\omega})
           =\frac{1}{Re}\Delta(\dot{\bm{\xi}}_{\alpha,\beta,\omega})_z\\
          &i\alpha \hat{u}_{\alpha,\beta,\omega}+\ddt{\hat{v}_{\alpha,\beta,\omega}}{y}+i\beta \hat{w}_{\alpha,\beta,\omega}=0
          \quad;\quad
          \ddt{\sigma_{x y}}{y}= \ddt{\sigma_{yy}}{y}= \ddt{\sigma_{z y}}{y}=0,
       \end{split}
       \label{eq:stLST}
    \end{equation}
with the modified diffusion operator
\begin{equation}
    \begin{split}
        \widetilde{D}(\bcdot)=&-\frac{1}{Re}\left(-\alpha^2 +\dddt{\bcdot}{y} -\beta^2 \right)
        \\&
        -{\frac{1}{2}\left(-\alpha^2a_{xx}+i\alpha a_{xy}\ddt{\bcdot}{y}-\alpha\beta a_{xz}\right.}
        \\&
        \hspace{30pt}
        {\left.+i\alpha \ddt{a_{yx}\bcdot}{y}+\ddt{}{y}\left(a_{yy}\ddt{\bcdot}{y}\right)+i\beta\ddt{a_{yz}\bcdot}{y} \right.}
        \\&
        \hspace{30pt}
        {\left.-\alpha\beta a_{zx}+i\beta a_{zy}\ddt{\bcdot}{y}-\beta^2a_{zz}\right)}.
   \end{split}
\end{equation}
The drift mean flow is $U_d(y)=U(y)-\frac{1}{2}{\partial a_{xy}}/{\partial y}$. 
The Fourier transform of $\sdBt$ is noted $\dd \bm{\xi}_{\alpha,\beta,\omega}$, and the associated velocity Fourier component $\dot{\bm{\xi}}_{\alpha,\beta,\omega}={\dd \bm{\xi}_{\alpha,\beta,\omega}}/{\dd t}$ is a standard centered Gaussian white noise convolved with the space-Fourier transform of $\boldsymbol{\sigma}$.
As the mean flow is parallel, the random right-hand-side term reduces to $-(\dot{\bm{\xi}}_{\alpha,\beta,\omega})_y{\partial U}/{\partial y}$, which is the strain induced by extraction of energy of the mean flow by the turbulence.
This term is central in the lift-up mechanism \citep{brandt2014} and it is the main actor in the role of incoherent turbulence in the streaks of streamwise velocity $u$.
The choices of stochastic parameters ($\boldsymbol{\sigma}$, $\mathsfbi{a}$, $\tau$) are detailed in section~\ref{sec:param}.

The main added-value of SLM in wall-bounded flows is to model the impact of decorrelated turbulence on the lift-up mechanism and its associated momentum mixing by stochastic diffusion.
Due to stochastic transport, equation~\eqref{eq:stLST} is a stochastic equation for $\hat{\bm{u}}_{\alpha,\beta,\omega}$, Fourier transform of $\bm{u}'$.
As a consequence, $\hat{\bm{u}}_{\alpha,\beta,\omega}$ is a random variable, whose variability will allow us to extract purely coherent components through the estimation of the cross spectral density (CSD) matrix $\mathbb{E}(\hat{\bm{u}}_{\alpha,\beta,\omega}\hat{\bm{u}}_{\alpha,\beta,\omega}^*)$ and its eigenvectors.
The leading eigenvector is called stochastic linear mode, or SLM.
Our objective is to extract the dominant coherent component by SLM, which is compared to the leading spectral proper orthogonal decomposition (SPOD) mode \citep{towne2018}. 
In \citet{tissotJFM2021}, ensemble method is employed for the estimation and we present in Supplementary Material a reformulation of the problem as a singular value decomposition, to improve computational efficiency.


\subsection{Interactions between coherent structures}
Equation~\eqref{eq:stLST} ensues from linearisation of equation~\eqref{eq:stNS}.
Coming back to the ground assumptions in stochastic modelling, a triple decomposition is performed on the displacement
\begin{equation}
    \mathrm{d}\bm{X}(\bm{x},t)=\bar{\bm{u}}(\bm{x})\mathrm{d}t+\bm{u}'(\bm{x},t)\mathrm{d}t+\boldsymbol{\sigma} \mathrm{d}\bm{B}_t.
    \label{eq:stdisplacementtriple}
\end{equation}
The first term $\bar{\bm{u}}(\bm{x})\mathrm{d}t=\left(U(y)\ 0\ 0 \right)^T\mathrm{d}t$ is the time-average displacement.
The fluctuation is split in a time-differentiable component and an incoherent turbulent field, perceived as time-decorrelated compared to the time scale of the coherent structure, and modelled by a Brownian motion.
Even if the time-average is non-ambiguous, the splitting of the fluctuation is less obvious in general and is often performed through a phase/ensemble average operator \citep{reynolds1972,yim2019}.
The formulation~\eqref{eq:stdisplacementtriple} is a way to perform the triple decomposition in a unique manner through time differentiability of the variable (more precisely  this decomposition is unique through the  Bichteler--Dellacherie decomposition of stochastic processes \citep{plotterbook}).
Let us remark that the Brownian part $\sdBt$ is modelled, while the time-differentiable part $\bm{u}'$ is solution of the system.
Moreover, contrary to a splitting based on phase-averaging, $\bm{u}'$ contains coherent and incoherent contributions.

With this decomposition in mind, it can be seen that stochastic diffusion can be interpreted as a generalised eddy diffusion (since a full tensor $\mathsfbi{a}$ is involved) induced by the noise.
In this case, the diffusion comes directly from the time decorrelation assumption and stems from the It\=o-Wentzel formula, where It\=o quadratic variations can be viewed as providing local averaging coefficients.
The diffusion does not come from a Boussinesq hypothesis.
This diffusion term accounts for the effect of time-decorrelated component, and not for nonlinear interactions between time-differentiable components.

In system~\eqref{eq:stLST}, the neglected term (written as a right-hand-side term in the momentum equation) is
\begin{equation}
   \mathcal{F}\left(\overline{(\bm{u}'\bcdot\nabla)\bm{u}'}-(\bm{u}'\bcdot\nabla)\bm{u}' \right),
   \label{eq:wwterm}
\end{equation}
where $\mathcal{F}\left(\bcdot\right)$ stands for space-time Fourier transform.
As in resolvent analysis (presented in Supplementary Material), this term is a convolution over all frequencies and wavenumbers, which renders an explicit expression difficult to obtain.
A major difference compared to $\nu$-resolvent analysis is that it represents non-linear interactions between smooth-in-time structures carrying coherent wave contributions and does not include time-decorrelated turbulent fluctuations.
In that sense, its interpretation is closer to the forcing term in $\nu_t$-resolvent analysis.
We call the term \eqref{eq:wwterm} \emph{wave-wave interactions}.
The contribution of turbulent noise is already taken into account in the stochastic formulation.

We propose to treat the term~\eqref{eq:wwterm} similarly to resolvent analysis, and to model it as a Gaussian white noise forcing term.
The addition of a forcing term to equation~\eqref{eq:stLST} leads to
    \begin{equation}
       \small
       \begin{split}
           &
           \begin{pmatrix}
            -i\omega+i\alpha U_d+{\widetilde{D}(\bcdot)} & \ddt{U}{y}            & 0 & i\alpha         \\
            0                                    & -i\omega+i\alpha U_d+{\widetilde{D}(\bcdot)} & 0 & \ddt{\bcdot}{y}  \\
            0                                    & 0 & -i\omega+i\alpha U_d+{\widetilde{D}(\bcdot)} & i\beta          \\
            i\alpha                              & \ddt{\bcdot}{y}                                          & i\beta & 0          \\
           \end{pmatrix}
           \begin{pmatrix}
           \hat{u}_{\alpha,\beta,\omega}\\
           \hat{v}_{\alpha,\beta,\omega}\\
           \hat{w}_{\alpha,\beta,\omega}\\
           \hat{p}_{\alpha,\beta,\omega}\\
           \end{pmatrix}
           =
           \\&
            \begin{pmatrix}
              -(\dot{\bm{\xi}}_{\alpha,\beta,\omega})_y\ddt{U}{y}+\frac{1}{Re}\Delta(\dot{\bm{\xi}}_{\alpha,\beta,\omega})_x\\
              \frac{1}{Re}\Delta(\dot{\bm{\xi}}_{\alpha,\beta,\omega})_y \\
              \frac{1}{Re}\Delta(\dot{\bm{\xi}}_{\alpha,\beta,\omega})_z \\
              0    
           \end{pmatrix}
           +
           b(y)
           \begin{pmatrix}
              \widetilde{f}^{NL}_x\\
              \widetilde{f}^{NL}_y\\
              \widetilde{f}^{NL}_z\\
              0    
           \end{pmatrix}
           .
       \end{split}
       \label{eq:stLSTmat}
    \end{equation}
The linear operator in the left hand side of~\eqref{eq:stLSTmat} can be written $\widetilde{\mathsfbi{A}}_{\alpha,\beta,\bar{\bm{q}}}-i\omega\mathsfbi{E}$.
The parameter $b(y)$ is an amplitude parameter of the non-linear forcings whose choice is based on the turbulent fluctuation level observed in the data.
Its choice is described in section~\ref{sec:param}.
The vector $(\widetilde{f}^{NL}_x,\widetilde{f}^{NL}_y,\widetilde{f}^{NL}_z)^T$ carries independent standard centered Gaussian white noises.
The above model will lead to forced stochastic linear modes, referred to as FSLM.

In the system~\eqref{eq:stLSTmat}, two stochastic right-hand side terms come from distinct physical mechanisms: the first term function of $\dot{\bm{\xi}}_{\alpha,\beta,\omega}$ is related to stochastic transport by incoherent small scale turbulence, while the second forcing term accounts for the non-linear interactions between coherent structures.

\subsection{Choice of parameters in FSLM}
\label{sec:param}
We recall that we focus on coherent structures perturbed by turbulent flows in the buffer and logarithmic layers at scales where production exceeds dissipation, for which, therefore, a forward energy cascade is expected~\citep{symon2021}.
In the logarithmic layer, we focus at energetic scales, and as highlighted in \citet{jimenez2013nearwallturbulence}, dissipation takes place at a smaller scale.
We expect in this region an energy cascade draining energy from large to small scales through an inter-scale energy flux.
For large energetic scales in the logarithmic layer, we expect as well a larger influence of incoherent turbulence onto the wave compared to the buffer layer; we aim at modelling such influence by FSLM.

The two-point statistics of the noise, carried by $\boldsymbol{\sigma}$, have to represent time decorrelated turbulent velocity field fluctuations.
Its definition is an open question and relies on an \textit{a priori} knowledge of the fluctuating velocity field.
Our strategy is to use few parameters, preferably with quantities available in standard simulation data or well documented in the literature.
Moreover, we need to respect the ground hypothesis that the noise is decorrelated from resolved coherent field (at the large-scale characteristic time scale).

We propose to set the variance tensor, $\mathsfbi{a}$, defined in equation~\eqref{eq:a}, from root-mean-square (RMS) velocity profiles, and variances of velocity fluctuations, which are quantities often available in databases accompanying the mean flow profile:
\begin{equation}
    \mathsfbi{a}(y)=
    \tau
    \begin{pmatrix}
     \langle u'^2(y)\rangle & \langle u'(y)v'(y) \rangle & 0 \\
     \langle u'(y)v'(y)\rangle & \langle v'^2(y) \rangle & 0 \\
     0                   & 0                   & \langle w'^2(y) \rangle 
    \end{pmatrix}
    .
    \label{eq:RMS}
\end{equation}
The underlying hypothesis to use RMS profiles is that the contribution of the single coherent wave we are trying to predict is small compared to the whole time-domain solution.
Thus, the RMS, which contains all contributions of the turbulent velocity field is a fair estimate of the turbulence which impacts the wave.
The decorrelation time $\tau$, necessary for dimensional consistency, represents the time scale necessary for the Brownian motion to perform mixing by stochastic diffusion.
This parameter is crucial for obtaining relevant results since it controls the level of diffusion.
The time scale $\tau$ should represent, at a given wavelength, the time scale necessary for the turbulence to affect the wave by a transport mechanism.
For this, we rely on an inertial scaling $\tau=\tau_0\left({l}/{l_0}\right)^{\frac{2}{3}}$ proposed in \citep{kadriharouna2017}, assuming that the wave length lies within the inertial range of an energy cascade under Kolmogorov hypotheses.
The time $\tau_0=l_0/U_0$ is the outer time scale, $l_0=2$ is the channel height, $U_0$ is the velocity averaged over the wall-normal direction; the scale of the wave is $l={2\pi}/{\sqrt{k_x^2+k_z^2}}$ with $k_x={2\pi}/{\lambda_x}$ and $k_z={2\pi}/{\lambda_z}$ \citep{popebook}.
This scaling is valid for scales such that $l<l_0$.
We do not expect our scaling to be valid for $l$ larger than the channel height leading to structures living in the outer region.
It can be noticed that the structure of the model allows a scale-dependent stochastic diffusion through the decorrelation time $\tau$, which we determine by a physical scaling.
In \citet{gupta2021}, a similar scale dependence of the eddy diffusion has been observed to produce accurate results.

%
The noise $\dd \bm{\xi}_{\alpha,\beta,\omega}$ is the space-time Fourier transform of $\sdBt$.
It is white in time, and its covariance should match the Fourier transform of the tensor $\mathsfbi{a}$ since both are the CSD of $\sdBt$ at a given location.
Indeed, the cross spectral density becomes
\begin{equation}
    \mathcal{F}\left(\mathbb{E}\left(\left(\boldsymbol{\sigma} \mathrm{d}\bm{B}_t\right)_{\bm{x}}^i\left(\boldsymbol{\sigma} \mathrm{d}\bm{B}_{t'}\right)_{\bm{x'}}^j\right)\right)
    =
    \mathbb{E}\left(\dd \bm{\xi}_{\alpha,\beta,\omega}^i \dd \bm{\xi}_{\alpha,\beta,\omega}^j \right)
    .
    \label{eq:ffta}
\end{equation}
In order to obtain an approximate consistency with the diffusion tensor $\mathsfbi{a}$ equation~\eqref{eq:RMS}, and to ensure that the noise is decorrelated from the wave, we define the noise in a specific manner.
First, we express it as an expansion onto an orthonormal basis 
\begin{equation}
     \dd \bm{\xi}_{\alpha,\beta,\omega}=\sum_{k=1}^{N_\sigma}c_k\,\bm{\Phi}^{\sigma}_k\eta_k.
    \label{eq:noise}
\end{equation}
We then propose a first guess by defining $\bm{\Phi}^\sigma_k=\bm{\Phi}^{\text{\tiny $\nu$-resolvent}}_{k+1}$ and 
$c_k={\sqrt{\lambda^{\text{\tiny SPOD}}_{1}}}s_{k+1}^{\text{\tiny $\nu$-resolvent}}/{s_1^{\text{\tiny $\nu$-resolvent}}}$, with $k\in[1,\cdots,N_\sigma]$, $(s^{\text{\tiny $\nu$-resolvent}}_k,\bm{\Phi}^{\text{\tiny $\nu$-resolvent}}_{k})$ the $k^{\text{th}}$ singular value and optimal response mode of $\nu$-resolvent analysis and $\lambda^{\text{\tiny SPOD}}_{1}$ the first SPOD eigenvalue.
This guess rescales the noise spanned by $\nu$-resolvent suboptimal modes in such a way that the energy of the first mode matches the first SPOD mode.
Doing this, we define an orthonormal family of vectors orthogonal to the dominant resolvent mode.
The amplitude rescaling aims at obtaining an approximate consistency between $\boldsymbol{\sigma}$ and the definition of $\mathsfbi{a}$ by the RMS profiles.
The use of resolvent modes as a first guess frees the modelling from the data.
Only the first SPOD eigenvalue is required, but this single parameter can be replaced by a free parameter fixed by some physical knowledge to obtain a fully model-based procedure.

In a second step, we correct the definition of~\eqref{eq:noise} in order to ensure that the noise is decorrelated from the first FSLM.
For that, we choose $\bm{\Phi}^\sigma_k=\bm{\Phi}^{\text{\tiny FSLM}}_{k+1}$ and 
$c_k=\sqrt{\lambda^{\text{\tiny FSLM}}_{k+1}}$, with $k\in[1,\cdots,N_\sigma]$,
where $(\lambda_k^{\text{\tiny FSLM}},\bm{\Phi}^{\text{\tiny FSLM}}_k)$ are eigen-elements of CSD matrix $\mathsfbi{S}$ of FSLM solutions.
This choice is motivated by the fact that in such a way, the first FSLM is by construction decorrelated from the noise, since the noise is spanned by the other eigenfunctions of the CSD (as explained in \citet{towne2018} for the SPOD modes).
The procedure is cyclic since FSLM are mandatory to predict FSLM, but it is possible to compute it iteratively through a fixed point procedure initialised with the first guess, as summarised in algorithm~\ref{algo}.
In practice, calculations converge quickly in few (less than 10) iterations with a relative tolerance on the Frobenius norm $\|\bcdot\|_F$ of the CSD equal to $\epsilon=10^{-3}$.
An example of convergence is shown in Supplementary Material.
\begin{algorithm}
\caption{Iterative procedure for FSLM}\label{algo}
Compute $\nu$-resolvent: $\mapsto(s_{k+1}^{\text{\tiny $\nu$-resolvent}}, \bm{\Phi}^{\text{\tiny $\nu$-resolvent}})$\\
 $\bm{\Phi}^{\sigma,(1)}_k\gets\bm{\Phi}^{\text{\tiny $\nu$-resolvent}}_{k+1}$, $c_k^{(1)}=s_{k+1}^{\text{\tiny $\nu$-resolvent}}\frac{\sqrt{\lambda^{\text{\tiny SPOD}}_{1}}}{s_1^{\text{\tiny $\nu$-resolvent}}}$, $\mathsfbi{S}^{(1)}\gets\text{real}(L_{\alpha,\beta,\omega}L_{\alpha,\beta,\omega}^*)$;
\\
$n\gets1$;\\
\While{not converged}{
    Compute FSLM by SVD procedure (see Suppl. Material):
    $\mapsto (\widetilde{L}_{\alpha,\beta,\omega}, \lambda^{\text{\tiny FSLM}}, \bm{\Phi}^{\text{\tiny FSLM}})$;
    \\
    $\mathsfbi{S}^{(n+1)}\gets\text{real}(\widetilde{L}_{\alpha,\beta,\omega}\widetilde{L}_{\alpha,\beta,\omega}^*)$;
    \\
    $\bm{\Phi}^{\sigma,(n+1)}_k\gets\bm{\Phi}^{\text{\tiny FSLM}}_{k+1}$,
    $c_k^{(n+1)}\gets\sqrt{\lambda^{\text{\tiny FSLM}}_{k+1}}$;
    \\
    \If{$\|\mathsfbi{S}^{(n+1)}-\mathsfbi{S}^{(n)}\|_F/\|\mathsfbi{S}^{(1)}\|_F<\epsilon$}{
        $converged \gets True$;
    }
}
\end{algorithm}

To summarise, the proposed procedure uses the RMS profiles to define the diffusion tensor $\mathsfbi{a}$, which is the one defined in the time-domain in equation~(4) of the main document.
The noise $\dd \bm{\xi}_{\alpha,\beta,\omega}$, space-time Fourier transform of $\sdBt$, is expanded on an orthonormal basis, which is estimated by an iterative procedure ensuring decorrelation between the noise and the solution.
An initial guess is defined by resolvent modes rescaled using the first SPOD eigenvalue in order to obtain consistency with the definition of $\mathsfbi{a}$ with a minimum of data.

Finally, in FSLM, the non-linear forcing amplitude has to be given.
In order to obtain a physically relevant order of magnitude, the non-linear forcing amplitude $b(y)$ is chosen as $\left({\sqrt{\lambda^{\text{\tiny SPOD}}_1}}/{s^{\text{\tiny $\nu$-resolvent}}_1}\right)\left({\text{TKE}(y)}/{\max(\text{TKE})}\right)$, with $\text{TKE}(y)=\langle u'^2\rangle +\langle v'^2\rangle + \langle w'^2\rangle $ the turbulent kinetic energy.
This scaling allows us to define a profile of non-linear forcing in the wall-normal direction consistent with the turbulent activity, and such that the response of the deterministic linearised system (without eddy viscosity) to this forcing has an amplitude comparable with SPOD.
\section{Application to turbulent channel flow}
\label{sec:results}

\subsection{Numerical simulation}
\begin{figure}
   \centering
    \subfigure[Mean profiles for the three Reynolds numbers.]{
   \includegraphics[width=0.47\textwidth]{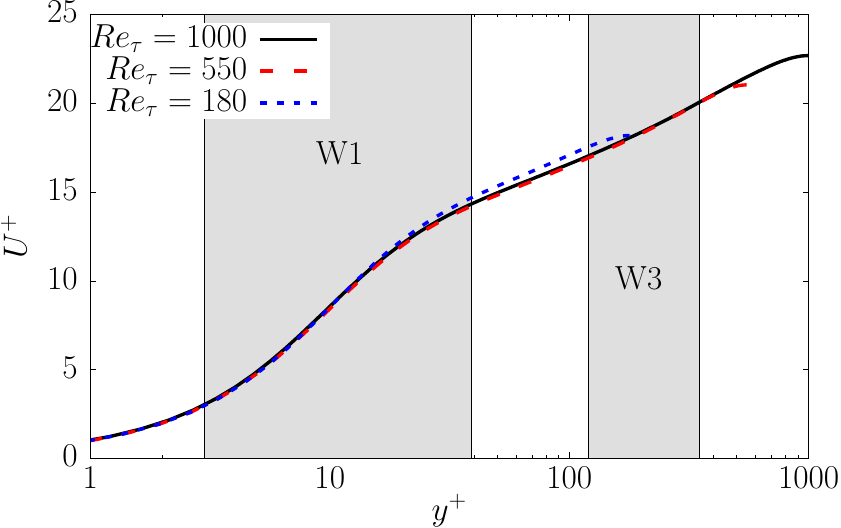}
   \label{fig:u}
   }
   \hfill
    \subfigure[Root-mean-square profiles, $Re_\tau=1000$.]{
   \includegraphics[width=0.47\textwidth]{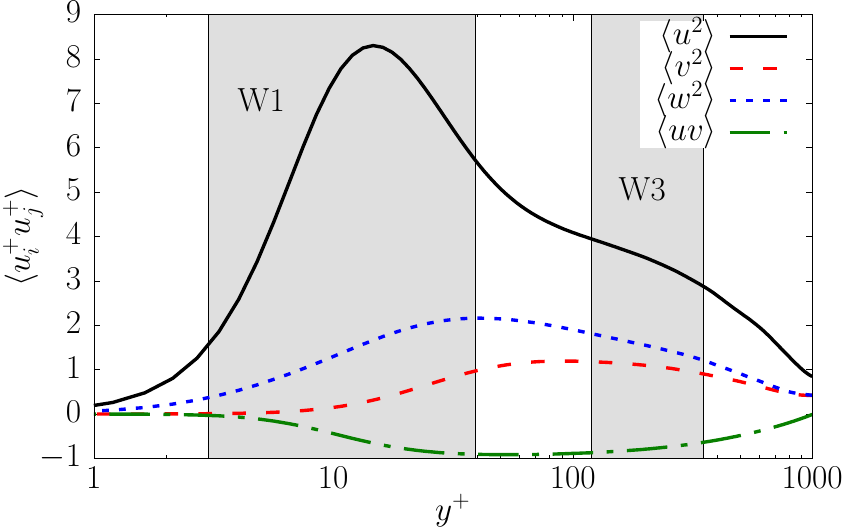}
   \label{fig:rms}
   }
   \caption{Mean and root-mean-square profiles. Grey areas indicate the spatial supports of W1 and W3.}
   \label{fig:flow}
\end{figure}

Databases of direct numerical simulation of turbulent channels were obtained with the pseudospectral code Channelflow~2.0 \citep{channelflow}.
Periodic boundary conditions are enforced in the streamwise ($x$) and spanwise ($z$) directions and Chebyshev polynomials are used in the wall-normal direction ($y$).
Parameters are given in table~\ref{tab:simu} and additional numerical details, including validation results, can be found in \citet{amaral2021}.
\begin{table*}
    \centering
    \begin{tabular}{cccccccccc}
    $Re_\tau$  & $Re$   & $N_x$ & $N_y$ & $N_z$ & $\Delta x^+$ & $\Delta y^+_{\min}$ & $\Delta y^+_{\max}$ & $\Delta z^+$ & $\Delta t^+$  \\
    180 (179)  &  2800  & 192   & 129   & 192   & 11.7         & $5.4\cdot10^{-2}$   & 4.4                 & 5.9          & 5.7           \\
    550 (543)  &  10000 & 384   & 257   & 384   & 8.9          & $4.1\cdot10^{-2}$   & 6.7                 & 4.4          & 3.0           \\
    1000 (996) &  20000 & 484   & 385   & 484   & 12.9         & $3.3\cdot10^{-2}$   & 8.2                 & 6.5          & 2.5           \\
    \end{tabular}
    \caption{Numerical parameters for the simulations.}
    \label{tab:simu}
\end{table*}
Mean flow profiles for the three Reynolds numbers and root-mean-square (RMS) profiles at $Re_\tau=1000$ are presented in figure~\ref{fig:flow}.
Results are presented using non-dimensional quantities using viscous (wall) scaling, denoted with a $+$ superscript.

SPOD has been computed as a reference to which $\nu_t$-resolvent analysis and FSLM will be compared.
They represent the most energetic structure for a given frequency-wavenumber combination, which can be meaningfully compared to most amplified responses of resolvent and FSLM. 
Numerical details are given in Supplementary Material.

Resolvent modes are known \citep{mckeon2010,moarref2013} to show large responses around the critical layer $y_c^+$, \textit{i.e.} where the phase speed $c^+=\lambda_x^+/\lambda_t^+$ matches the mean flow $U^+(y_c^+)$.
SPOD modes follow the same trend, which is consistent with the fact that these modes are equivalent if the non-linear term behaves as a Gaussian white noise \citep{cavalieri2019}.
Two waves have been selected: one denoted W1, typical of the streaks structures in the buffer layer chosen at the lowest Reynolds number $Re_\tau=180$; and one denoted W3, evolving within the logarithmic layer at the highest Reynolds number $Re_\tau=1000$.
In Supplementary Material, the robustness of the method is shown by varying Reynolds number for W1, and two other waves (W2 and W4) are presented in order to vary the wall-distance of the wave spatial support.
%
%
As in \citet{tissotJFM2021} we consider modes that are odd in $u$ and $w$ (and thus even in $v$) around the channel centerline.

\begin{figure}
    \centering
    \includegraphics[width=0.48\textwidth]{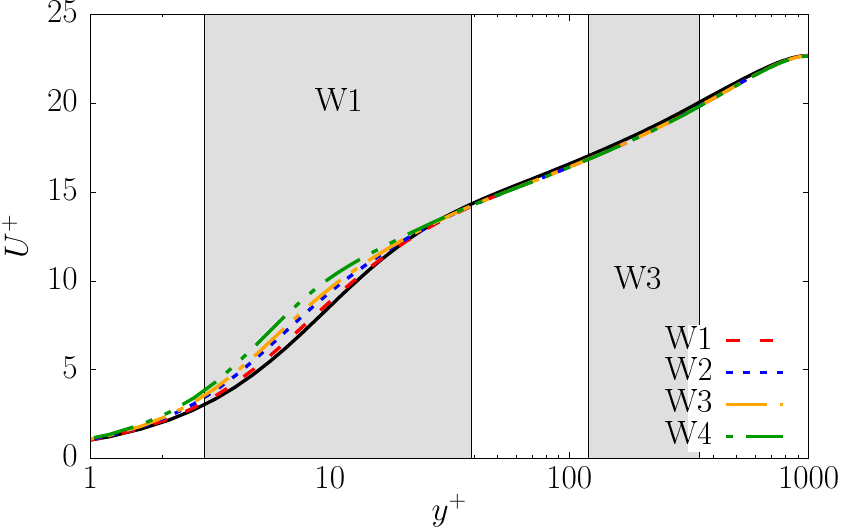}
    \caption{Mean velocity (plain black line) and mean velocity with the corrective drift associated with W1 to W4 at $Re_\tau=1000$ (see Supplementary Material for W2 and W4).}
    \label{fig:drift}
\end{figure}

The spatial supports of the waves W1 and W3 are reported in figures~\ref{fig:flow}.
In figure~\ref{fig:drift}, the drift velocity (in wall units) associated with W1 to W4 are displayed at $Re_\tau=1000$. It shows that the corrective drift $-\frac{1}{2}{\partial a_{xy}}/{\partial y}$ plays essentially a role in the buffer region.
This confirms the observations  and modelling of \citet{pinier2019}.
Additionally, we can see that with our definition, the effect of drift velocity is more pronounced for waves evolving at higher wall-normal distance due to larger decorrelation times $\tau$.
As a matter of fact, for such long waves there is a more substantial contribution of the stochastic transport, which occurs with a longer decorrelation time.

\subsection{Buffer layer}
\label{sec:buffer}
In the buffer layer, we present the results at $Re_\tau=180$, and complementary results at $Re_\tau=550$ and $Re_\tau=1000$ are given in the Supplementary Material.
Figure~\ref{fig:W1SPOD} shows a velocity field cross section of the SPOD for W1.
It shows a typical streaky structure of $u$ with streamwise vortices (rolls), which highlights the lift-up mechanism: in regions of high streamwise velocity high-speed streak are emerging.
They are associated with negative $v$ components, which transport fluid with high streamwise velocity to a region with lower mean flow; the opposite happens for low-speed streaks, which are associated with positive $v$ components.
Predictions by $\nu$-resolvent and $\nu_t$-resolvent analysis are shown in figures~\ref{fig:W1res} and~\ref{fig:W1reseddy} respectively.
We can see a relevant prediction, with an improvement when eddy viscosity is added.
This is consistent with \citet{morra2019}. 
Figure~\ref{fig:W1SLM} shows the solution of the proposed stochastic model, but omitting the non-linear forcing (SLM).
It can be seen that the streaks are well predicted, but the rolls are absent.
However, taking into account wave-wave interactions by a non-linear forcing (FSLM) enables us to recover the rolls, and to obtain accurate predictions.
This can be explained by the fact that stochastic transport models the effect of the incoherent part of the velocity field, thus leading to good predictions of the $u$ profiles.
However, since near-wall streamwise vortices are thought to arise from a non-linear interaction of coherent structures \citep{hamilton1995}, a non-linear forcing is mandatory to predict them.
Resolvent analysis with eddy viscosity leads to good predictions since it takes into account this non-linear forcing and incorporates eddy diffusion.
These predictions are significantly enhanced by the stochastic model since it explicitly modifies lift-up by incoherent turbulent motions through three complementary terms: the transport by the noise, a diffusion tensor with non-zero off-diagonal terms and a drift velocity active in the buffer region \citep{tissotJFM2021}.
\begin{figure}
    \centering
      \subfigure[SPOD.]{
      \includegraphics[width=0.31\textwidth]{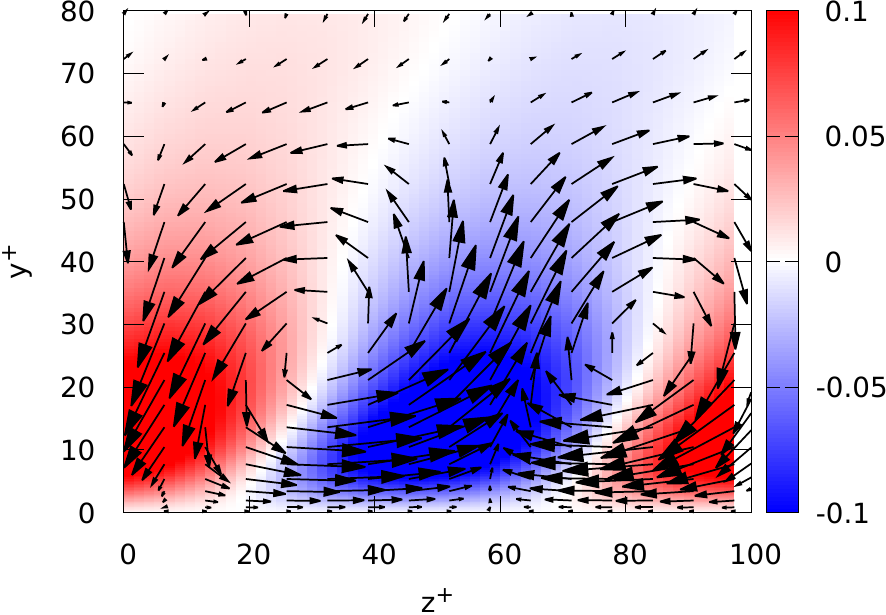}
      \label{fig:W1SPOD}
      }
      \hfill
      \subfigure[$\nu$-resolvent.]{
      \includegraphics[width=0.31\textwidth]{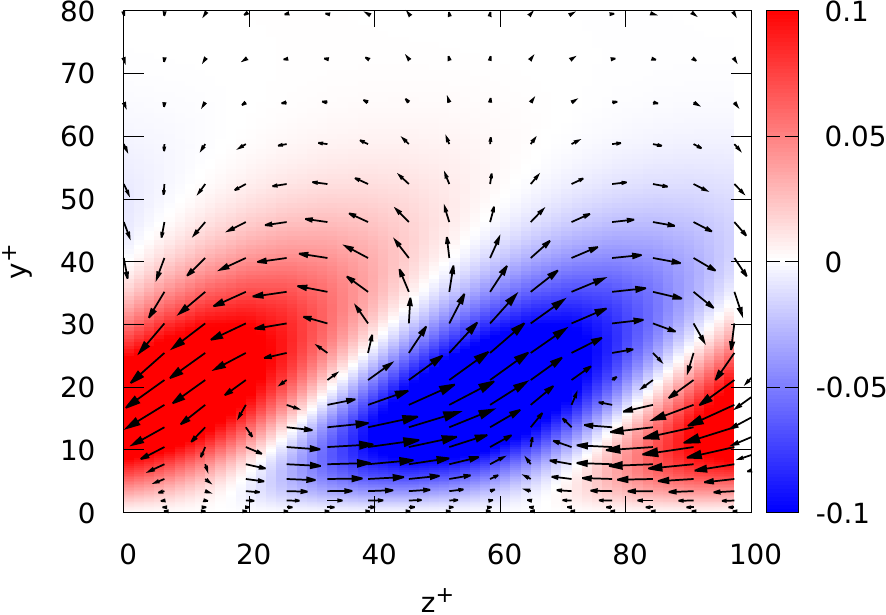}
      \label{fig:W1res}
      }
      \hfill
      \subfigure[$\nu_t$-resolvent.]{
      \includegraphics[width=0.31\textwidth]{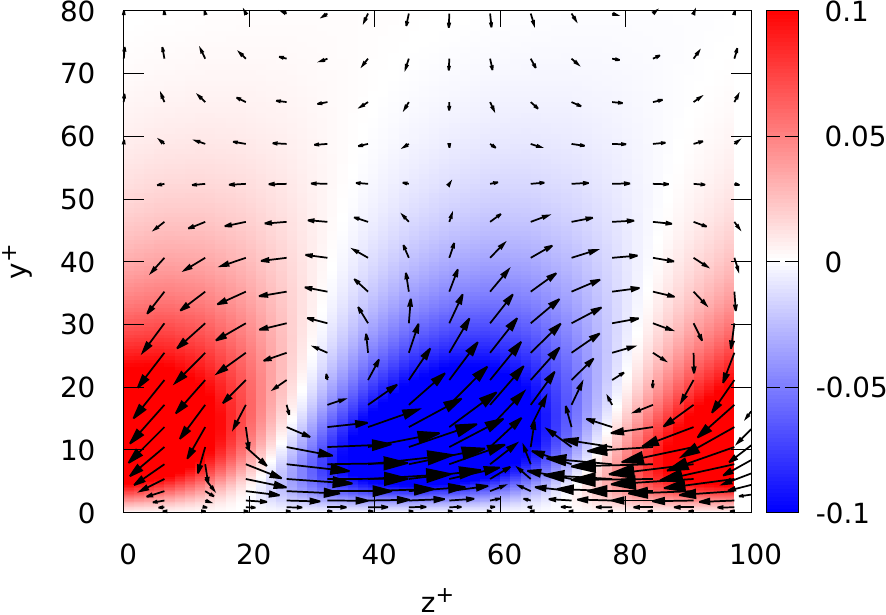}
      \label{fig:W1reseddy}
      }
      \\
      \subfigure[SLM.]{
      \includegraphics[width=0.31\textwidth]{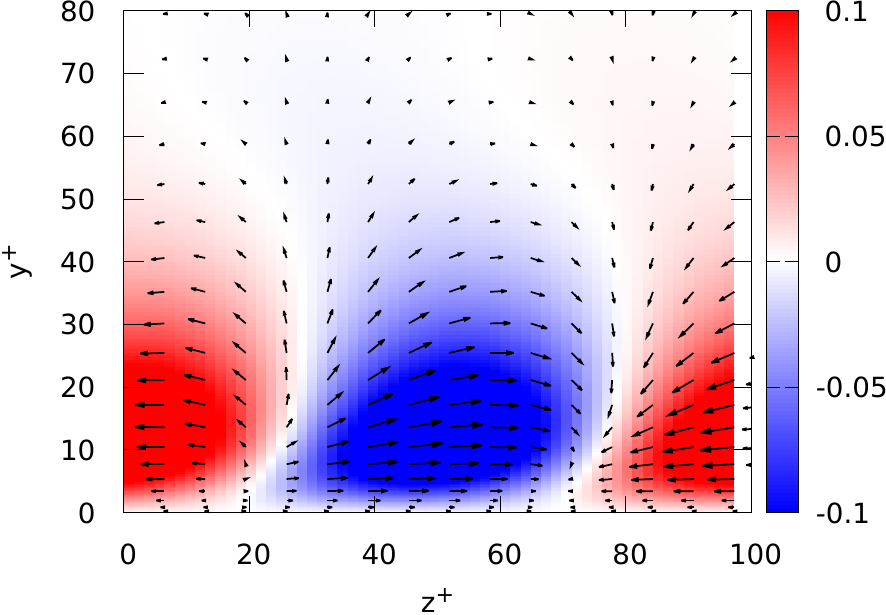}
      \label{fig:W1SLM}
      }
      \hfill
      \subfigure[FSLM.]{
      \includegraphics[width=0.31\textwidth]{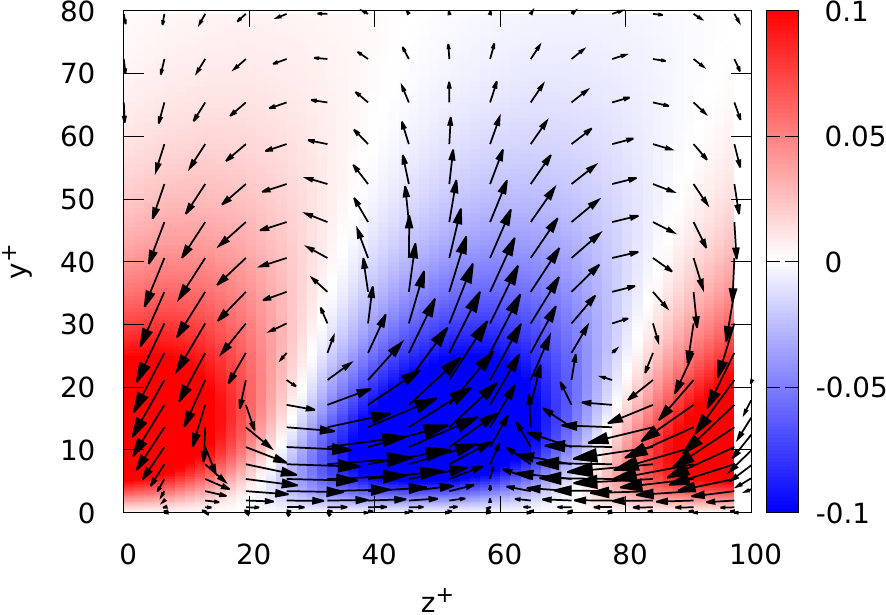}
      \label{fig:W1FSLM}
      }
      \caption{Reconstructions of W1 at $Re_\tau=180$. Colors are streamwise velocities, arrows are in-plane velocity fields.}
     \label{fig:W1Re180}
\end{figure}

\begin{figure}
    \centering
    \subfigure[Streamwise velocity $|\hat{u}|^2$. ]{\includegraphics[width=0.48\textwidth]{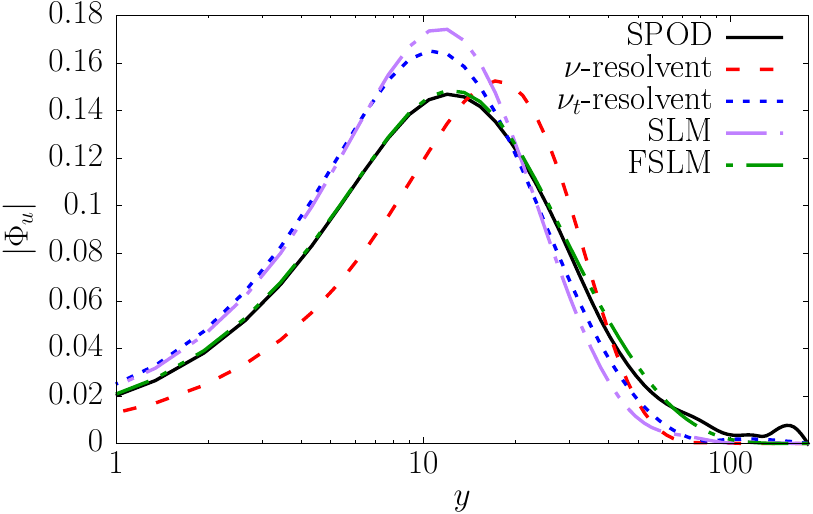}}
    \\                                        
    \subfigure[Wall-normal velocity $|\hat{v}|^2$.]{\includegraphics[width=0.48\textwidth]{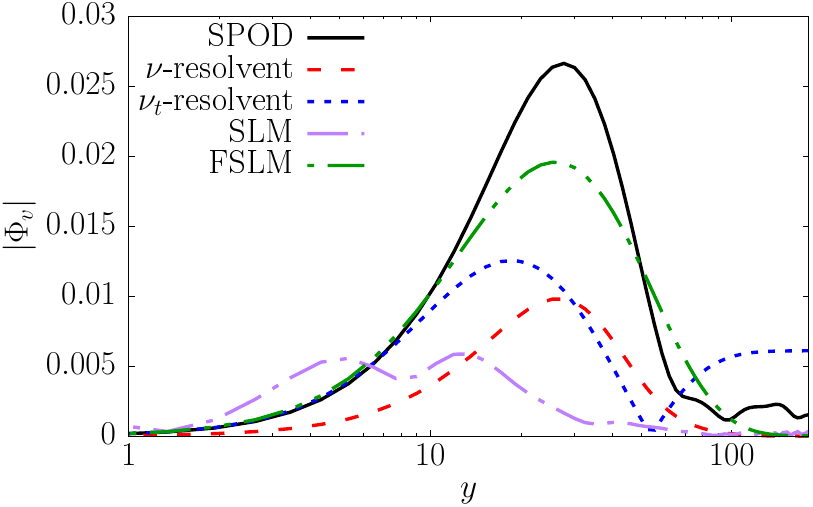}}
    \hfill                                        
    \subfigure[Spanwise velocity $|\hat{w}|^2$.   ]{\includegraphics[width=0.48\textwidth]{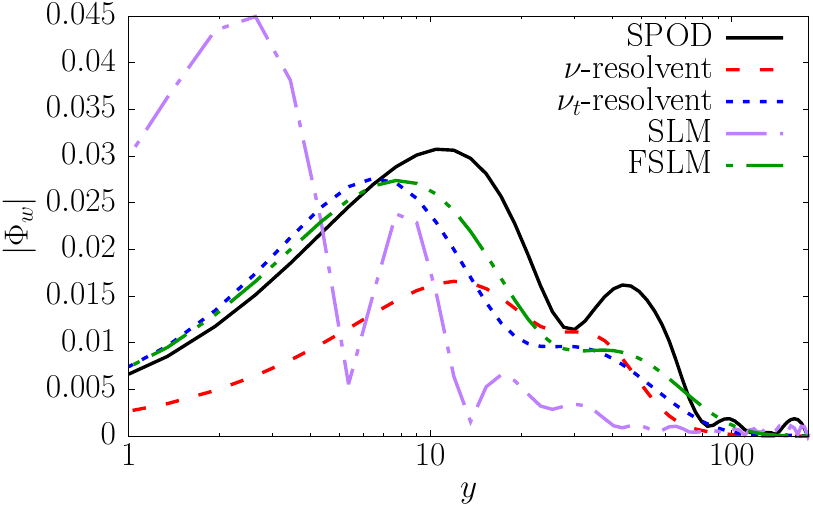}}
    \caption{PSD velocity profiles of W1 at $Re_\tau=180$.}
    \label{fig:W1profiles}
\end{figure}

Profiles of power spectral density (PSD) of the three velocity components are shown in figure~\ref{fig:W1profiles}.
They confirm that streamwise velocity ($u$) profiles are similarly captured by $\nu_t$-resolvent and by SLM.
The agreement of FSLM with SPOD data is significantly improved.
Moreover, the streamwise vortices signing on the ($v$, $w$) profiles are not captured by SLM but strongly intensified in FSLM.
The wall-normal velocity ($v$) is especially affected by the stochastic transport leading to the best agreement with SPOD.

To demonstrate the robustness of the procedure, buffer layer modes at other Reynolds numbers are shown in Supplementary Material.
Despite a slight overall deterioration of agreement between all models and SPOD when the Reynolds number increases, the trend is maintained and FSLM shows systematically a better agreement.

\subsection{Logarithmic layer}
\begin{figure}
 \centering
   \subfigure[SPOD.]{
   \includegraphics[width=0.30\textwidth]{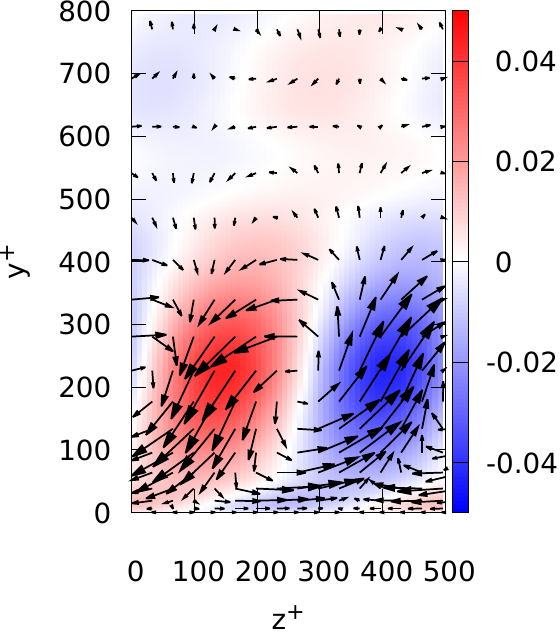}
   \label{fig:W3SPOD}
   }
   \subfigure[$\nu$-resolvent.]{
   \includegraphics[width=0.30\textwidth]{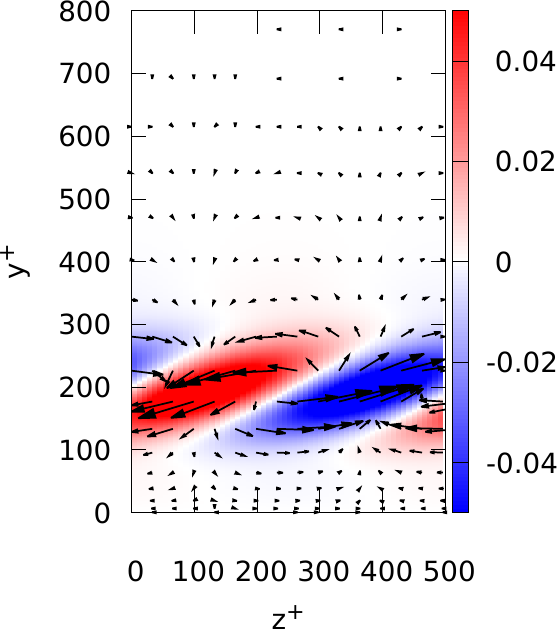}
   \label{fig:W3res}
   }
   \subfigure[$\nu_t$-resolvent.]{
   \includegraphics[width=0.30\textwidth]{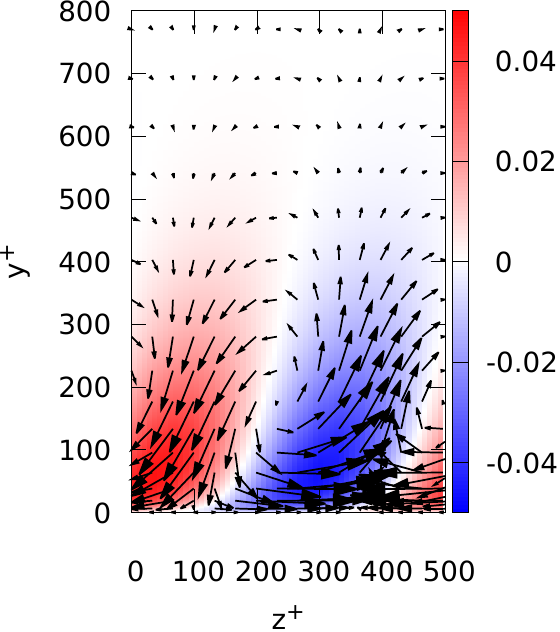}
   \label{fig:W3reseddy}
   }
   \subfigure[FSLM.]{
   \includegraphics[width=0.30\textwidth]{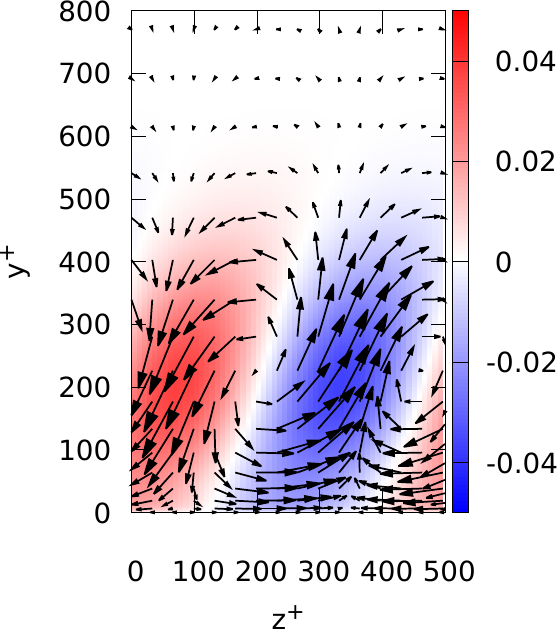}
   \label{fig:W3FSLM}
   }
   \caption{Reconstructions of W3 at $Re_\tau=1000$.}
\end{figure}

We now select a wave named W3 evolving within the logarithmic layer by setting the phase speed $c^+=18.1$ associated with a critical level $y_c^+=180$. The wave-number $(\lambda_x^+=2087,\lambda_z^+=522)$ has been chosen in the energy peak \citep[taken from ref.][]{amaral2021} of the premultiplied powerspectra of the streamwise velocity.
It can be extracted by SPOD, as shown in figure~\ref{fig:W3SPOD}.
The $\nu$-resolvent analysis (figure~\ref{fig:W3res}) extracts typical critical layer modes, with a narrow spatial support located at the critical layer, \textit{i.e.} at the wall-normal position $y_c^+$ where the phase speed $c^+$ matches the mean velocity.
Incorporating eddy viscosity (figure~\ref{fig:W3reseddy}) leads to wider spatial support, more similar to SPOD modes, which is again consistent with previous studies \citep{morra2019}.
However, there is room for improvements, since SPOD modes show a structure that peaks further from the wall than what is predicted by $\nu_t$-resolvent.
FSLM in figure~\ref{fig:W3FSLM} improves significantly this prediction with the streamwise velocity structure further from the wall and a more accurate shape of the rolls.

\begin{figure}
    \centering
    \subfigure[Streamwise velocity $|\hat{u}|^2$. ]{\includegraphics[width=0.48\textwidth]{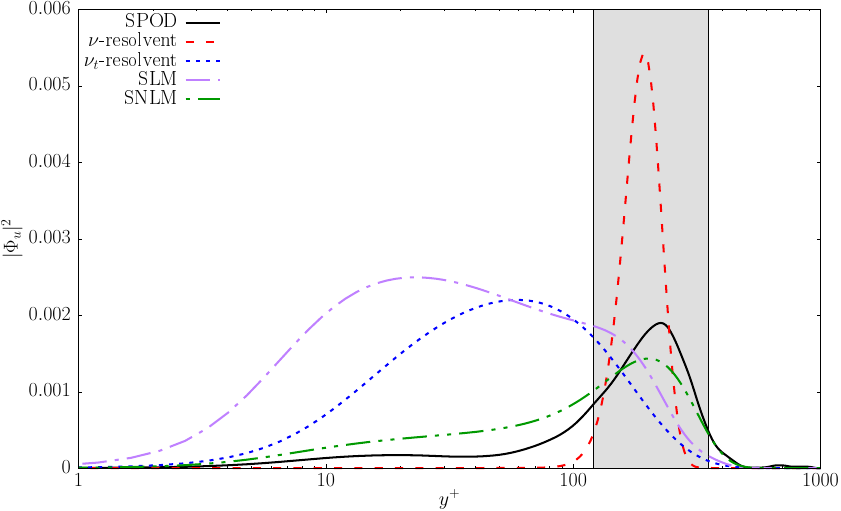}\label{fig:W3profileu}}
    \\
    \subfigure[Wall-normal velocity $|\hat{v}|^2$.]{\includegraphics[width=0.48\textwidth]{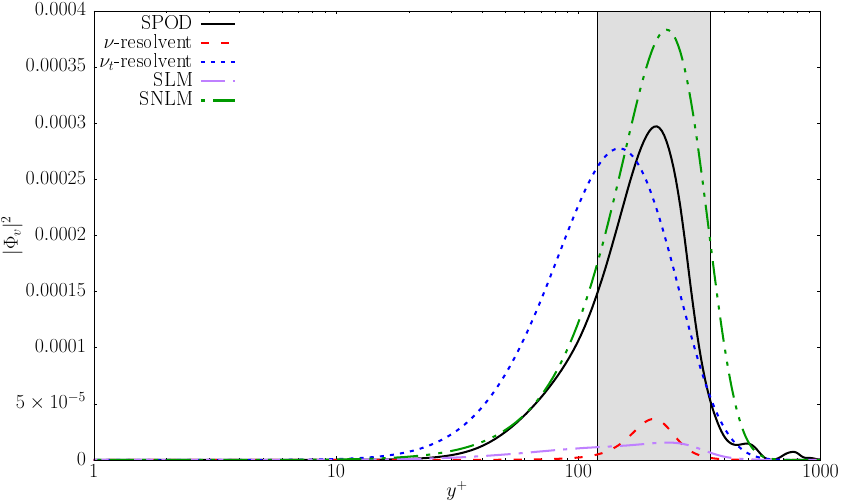}\label{fig:W3profilev}}
    \hfill
    \subfigure[Spanwise velocity $|\hat{w}|^2$.   ]{\includegraphics[width=0.48\textwidth]{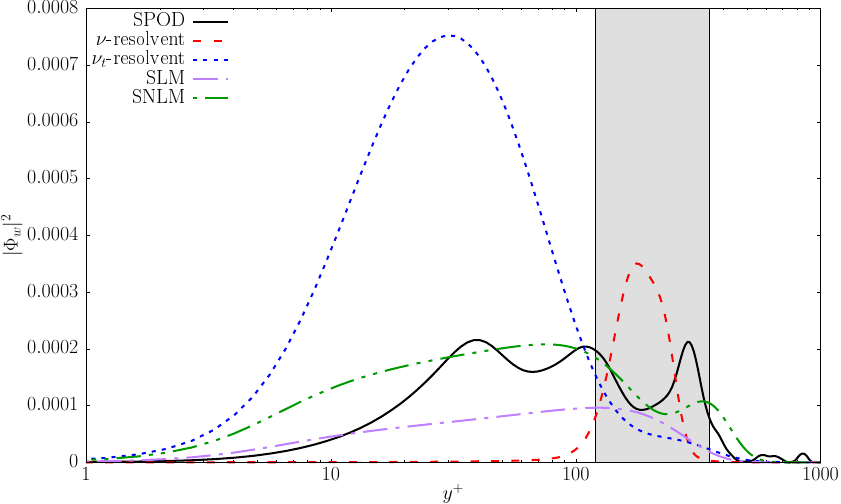}\label{fig:W3profilew}}
    \caption{PSD velocity profiles of W3 at $Re_\tau=1000$.}
    \label{fig:W3profiles}
\end{figure}

Figure~\ref{fig:W3profiles} showing the PSD profiles highlights quantitatively this improvement.
Concerning the $u$ component, figure~\ref{fig:W3profileu} shows that differently from the other models, the spatial support is very well captured.
As for the wall-normal $v$ velocity, the shape of the profile is better predicted but with a high relative amplitude.
The spanwise $w$ velocity is better captured as well.

The lower accuracy of $\nu_t$-resolvent predictions can be understood by the fact that the effect of the incoherent turbulent field on the wave is modelled only by a diffusive mechanism.
On the contrary, FSLM incorporates through stochastic transport some driving mechanisms induced by the incoherent motions existing at the same scale.
The success of FSLM suggests that in the logarithmic region, where the turbulence is developed, taking into account the stochastic nature of the log-layer structures is central to perform accurate predictions.

In addition, profiles of SLM, \textit{i.e.} neglecting the non-linear forcing, are shown to produce poor predictions.
This suggests again that coherent non-linear wave-wave interactions are crucial for self-sustaining process for log-layer.
It corroborates hypotheses in \citet{flores2010} and \citet{cossu2017} that a coherent large scale self-sustaining process is in action for large log-layer structures.

\begin{figure}
    \centering
    \subfigure[$c^+=10$; $y_c^+=13$. The green cross locates W1.]{\includegraphics[width=0.48\textwidth]{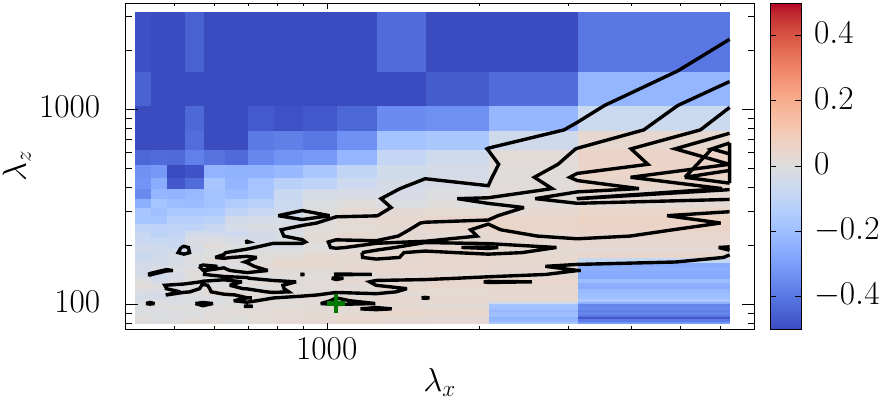}\label{fig:betac10}}
    \subfigure[$c^+=15$; $y_c^+=53$. The green cross locates W2.]{\includegraphics[width=0.48\textwidth]{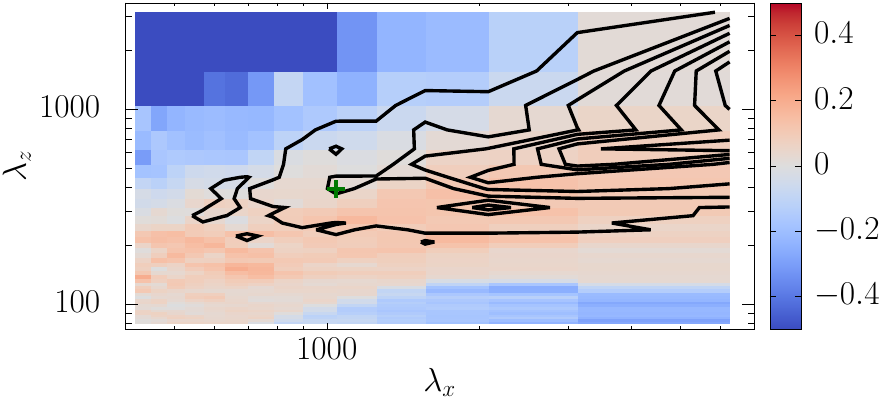}\label{fig:betac15}}
    \subfigure[$c^+=18.3$; $y_c^+=200$. The green cross locates W3.]{\includegraphics[width=0.48\textwidth]{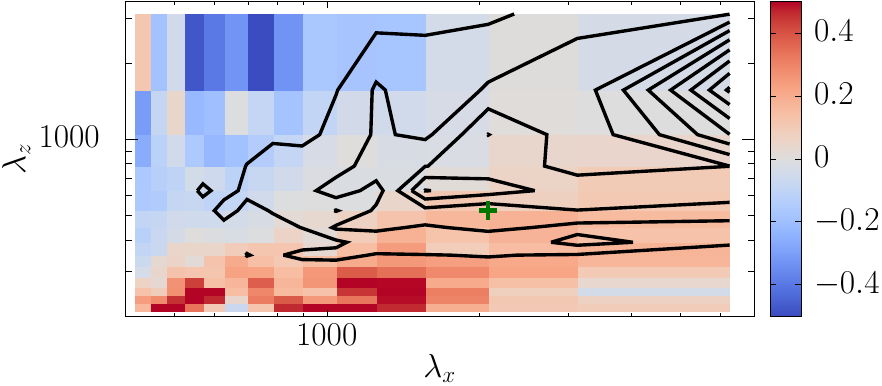}\label{fig:betac18}}
    \subfigure[$c^+=19$; $y_c^+=250$. The green cross locates W4.]{\includegraphics[width=0.48\textwidth]{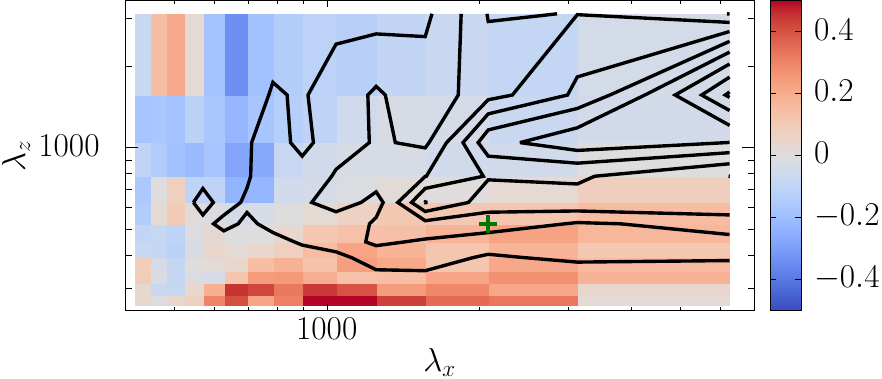}\label{fig:betac19}}
    \caption{Metric $\gamma_{\alpha,\beta,\omega}=\log(\beta_{FSLM}/\beta_{\nu_t-\text{res}})$ of improvement ($>0$) or deterioration ($<0$) of collinearity between FSLM and SPOD compared to $\nu_t$-resolvent analysis as a function of $\lambda_x,\lambda_z$ for various critical layer positions. Isocontours are the pre-multiplied value of the first SPOD $\alpha \beta\lambda_1^{\text{\tiny SPOD}}$.}
    \label{fig:ratio_beta}
\end{figure}

Frequency-wavenumber space has been swept, and colinearity metrics
\begin{equation}
\beta^{\text{\tiny model}}_{\alpha,\beta,\omega}=\frac{\left|\left(\bm{\Phi}^{\text{\tiny model}}_{1,\alpha,\beta,\omega},\bm{\Phi}_{1,\alpha,\beta,\omega}^{\text{\tiny SPOD}}\right)\right|}{\left\|\bm{\Phi}^{\text{\tiny model}}_{1,\alpha,\beta,\omega}\right\|\left\|\bm{\Phi}_{1,\alpha,\beta,\omega}^{\text{\tiny SPOD}}\right\|}
\end{equation}
have been computed (metric used for instance in \citep{cavalieri2013}).
This metric is a normalised inner-product between the dominant SPOD mode $\bm{\Phi}_{1,\alpha,\beta,\omega}^{\text{\tiny SPOD}}$ and the mode issued from a given model $\bm{\Phi}^{\text{\tiny model}}_{1,\alpha,\beta,\omega}$.
A value of 1 means exact collinearity between modes, while 0 happens when the modes are orthogonal.
Then, we compute the metric $\gamma_{\alpha,\beta,\omega}=\log\left({\beta^{\text{\tiny FSLM}}_{\alpha,\beta,\omega}}/{\beta^{\text{\tiny {$\nu_t$-resolvent}}}_{\alpha,\beta,\omega}}\right)$ which represents the improvement ($\gamma>0$) or deterioration ($\gamma<0$) of colinearity with SPOD compared to the $\nu_t$-resolvent model.
Figure~\ref{fig:ratio_beta} shows the value of $\gamma_{\alpha,\beta,\omega}$ at four critical layer positions as a function of streamwise and spanwise wavenumbers.
We can see that in the buffer and logarithmic layer, a wide range of streamwise elongated structures are improved with FSLM compared to $\nu_t$-resolvent analysis.
The improvement is more pronounced further from the wall since agreement is more difficult to obtain.
Complementary maps of $\beta^{\text{\tiny model}}_{\alpha,\beta,\omega}$ are given in Supplementary Material.
Isocontour of the pre-multiplied first SPOD eigenvalue $\alpha \beta\lambda_1^{\text{\tiny SPOD}}$ are superimposed, and show that deterioration happens at scales where less energy is present.
Finally, in figure~\ref{fig:betac19} almost in the outer-region, we can see that FSLM provide slightly less good performances than $\nu_t$-resolvent for large $\lambda_x$ and $\lambda_z$. We explain this discrepancy by the choice of decorrelation time $\tau$ which is designed based on inertial-range scalings (see section~\ref{sec:param}).
These maps prove a wide range of validity of the proposed model.

\section{Conclusion}
\label{sec:conclusion}
In this paper we have proposed a stochastic modelling strategy of coherent structures in turbulent channel flows.
By adding a stochastic non-linear forcing term, we obtain a refinement of the model proposed in \citet{tissotJFM2021}.
This forced model aims at improving consistency with the physical processes involved in these flows while maintaining the mathematical assumptions of the stochastic formulation.
We used this model to explore the prediction abilities in the buffer and logarithmic layers in channels with friction Reynolds number equal to 180, 550 and 1000.

A central ingredient is the incorporation of a non-linear forcing representing coherent wave-wave interactions, which are essential in the self-sustaining processes of wall-bounded turbulence to generate streamwise vortices for large-scale structures in the logarithmic layer.
In the model predictions, such forcing is central in the buffer layer, consistently with the Hamilton-Kim-Waleffe scenario \citep{hamilton1995}.
Moreover we have shown that it is central as well for large-scale structures in the logarithmic layer, which is in line with a large body of evidence in the literature supporting that self-sustaining processes are in action in the logarithmic layer.
In addition to this, the model predicts a significant effect of incoherent turbulence on large log-layer structures, showing that it is crucial to model this effect in order to obtain good predictions.
The model that we propose takes into account stochastic transport by this incoherent velocity field leading to an improvement compared to the state-of-the-art $\nu_t$-resolvent analysis.
The present model requires a knowledge of the statistics of the turbulent field, and there are some possible open directions concerning its specification.

The modelling ingredients act on several physical mechanisms. First the stochastic diffusion induced by incoherent velocity field, unlike an eddy viscosity, has the shape of a full tensor with in particular $\langle u' v'\rangle$ off-diagonal components, which are defined consistently with the RMS profiles.
A mean drift velocity takes into account the turbophoresis effect (effective transport from high to low turbulence regions), which is active in the buffer layer.
A stochastic term representing the lift-up induced by the random incoherent velocity field is explicitly taken into account.

In addition, we have brought technical improvements by ensuring the decorrelation of the incoherent component with the solution through an iterative procedure, and we have proposed an efficient computation of FSLM by reformulating it as a singular value decomposition problem.
With these effects taken into account in the model, we obtained an agreement between model predictions and turbulent fluctuations at various wall-normal positions.

In summary, the proposed model incorporates features of resolvent analysis, via the forcing resulting from non-linear wave-wave interactions, to the stochastic formalism introduced in our previous work \citep{tissotJFM2021}, combining hence the benefits of the two approaches.
The stochastic framework can be seen as a refined model of incoherent turbulence on large-scale structures, which not only includes the standard additional diffusion present in eddy-viscosity models \citep{cossu2009,pickering2021}, but also involves all aspects of stochastic transport at the lengthscale of interest.
The study shows that FSLM seems to carry advantageous features for reduced-order modelling of coherent structures in turbulent flows.

\section*{Acknowledgements}
The authors acknowledge the support of the ERC EU project 856408 STUOD. 

\section*{Supplementary Material}
\appendix
In this Supplementary Material section, complementary elements on theoretical formalisms, numerical details and additional results are provided.
In section~\ref{sec:resolvent}, resolvent analysis is presented.
In section~\ref{sec:computation}, details concerning numerical computations of the (forced) stochastic linear modes (SLM and FSLM) are given.
In section~\ref{sec:computation_SPOD_modes} numerical details related to the spectral proper orthogonal decomposition (SPOD) are explained and in section~\ref{sec:results}, additional complementary results are shown.
We show the convergence of the iterative procedure (sec.~\ref{sec:convergence}), for waves in the buffer layer at various friction Reynolds numbers $Re_\tau$ (sec.~\ref{sec:buffer}) and in the logarithmic layer at $Re_\tau=1000$ at various critical layer positions (sec.~\ref{sec:log}).
In section~\ref{sec:beta} we show complementary collinearity metrics.

\section{Resolvent analysis}
\label{sec:resolvent}
\subsection{General formulation}
Resolvent analysis is a tool to explore the response $\hat{\bm{q}}_{\alpha,\beta,\omega}$ of a linearised operator $\mathsfbi{A}_{\alpha,\beta,\bar{\bm{q}}}$ to a harmonic forcing $\hat{\bm{f}}_{\alpha,\beta,\omega}$.
It is today a formalism widely used to extract coherent structures in turbulent flows \citep[for instance]{moarref2014,lesshafft2019,abreu2021}.
The considered linear system is the Navier--Stokes equations linearised around the mean flow $\bar{\bm{q}}$, in which one may incorporate an eddy-viscosity model to take into account a part of energy transfers toward smaller scales.
The forcing is interpreted as the Fourier transform of the remaining non-linear term, which is \textit{a priori} unknown since it stems from a convolution over all frequencies and wave-numbers.
The response can be compactly written as the solution of the linear system
\begin{equation}
    \left( \mathsfbi{A}_{\alpha,\beta,\bar{\bm{q}}}-i\omega\mathsfbi{E} \right)\hat{\bm{q}}_{\alpha,\beta,\omega} =\hat{\bm{f}}_{\alpha,\beta,\omega}.
\end{equation}
The problem in discretised such that the operators become matrices, and detailed expression of the matrices are given in section~\ref{sec:res}.
We define an output matrix $\mathsfbi{H}$ such that $\bm{u}=\mathsfbi{H}\bm{q}$, and an input matrix $\mathsfbi{B}=\mathsfbi{H}^T$.
The input matrix allows to restrict the forcing space to momentum equation, in accordance to the structure of the non-linear term.
The output matrix targets the kinetic energy to be the quantity to optimise.
We define the transfer function operator $\mathsfbi{L}_{\alpha,\beta,\omega}=\mathsfbi{H}\mathsfbi{R}_{\alpha,\beta,\omega}\mathsfbi{B}$, with $\mathsfbi{R}_{\alpha,\beta,\omega}=\left( \mathsfbi{A}_{\alpha,\beta,\bar{\bm{q}}}-i\omega\mathsfbi{E} \right)^{-1}$ the resolvent operator.
Singular value decomposition of $\mathsfbi{W}^{\frac{1}{2}}\mathsfbi{L}_{\alpha,\beta,\omega}\mathsfbi{W}^{-\frac{1}{2}}=\mathsfbi{U}_r\mathsfbi{\Sigma_r}\mathsfbi{V}^H$ gives access to optimal forcing modes $\bm{\Psi}_i^{\text{resolvent}}=\mathsfbi{W}^{-\frac{1}{2}}\bm{V}_{r,i}$ and associated response modes $\bm{\Phi}_i^{\text{resolvent}}=\mathsfbi{W}^{-\frac{1}{2}}\bm{U}_{r,i}$, where $\bm{U}_{r,i}$ (resp. $\bm{V}_{r,i}$) denotes the $i^{\text{th}}$ column of $\mathsfbi{U}_r$ (resp. $\mathsfbi{V}_r$), and $\mathsfbi{W}$ is the diagonal matrix containing quadrature coefficients associated with integrals in the wall-normal direction. The diagonal elements $s_i$ of $\mathsfbi{\Sigma}_r$ are the amplification gains.
We call this method $\nu$-resolvent analysis.

Cess's eddy viscosity \citep{Cess1958} can be incorporated in the linearised system.
The associated resolvent analysis \citep{hwang2010,morra2019,symon2020,amaral2021}, noted \emph{$\nu_t$-resolvent}, constitutes the reference case used for comparisons.

\subsection{System used in resolvent analysis}
\label{sec:res}
In this section, we detail the operators used in resolvent analysis.
We have
\begin{equation}
   \mathsfbi{A}_{\alpha,\beta,\bar{\bm{q}}}
   =
   \begin{pmatrix}
             i\alpha U+D(\bcdot) & \ddt{U}{y}                   & 0                            & i\alpha         \\
    0                            &          i\alpha U+D(\bcdot) & 0                            & \ddt{\bcdot}{y} \\
    0                            & 0                            &          i\alpha U+D(\bcdot) & i\beta          \\
    i\alpha                      & \ddt{\bcdot}{y}              & i\beta                       & 0               \\
   \end{pmatrix}
   ,
   \label{eq:A}
\end{equation}
with the diffusion operator $$D(\bcdot)=-\frac{1}{Re}\left(-\alpha^2 +\dddt{\bcdot}{y} -\beta^2 \right)$$ in the case of $\nu$-resolvent analysis and
$$
D(\bcdot)=-\left(\frac{1}{Re}+\nu_t\right)\left(-\alpha^2 +\dddt{\bcdot}{y} -\beta^2 \right)-\ddt{\nu_t}{y}\ddt{\bcdot}{y}
$$
with
\begin{equation}
    \nu_t=\frac{1}{Re}\left(\frac{1}{2}\left(1+\frac{\kappa^2Re_\tau^2}{9}(1-y^2)^2(1+2y^2)^2(1-e^{-\frac{y^+}{A}})^2\right)^{\frac{1}{2}}-\frac{1}{2}\right),
\end{equation}
in the case of $\nu_t$-resolvent analysis,
where $y^+=Re_\tau (1-|y|)$, $\kappa=0.426$ is the von K\'arm\'an constant and $A=25.5$ is a constant chosen consistently following \citet{pujals2009}. 
Moreover, we have
\begin{equation}
   \mathsfbi{E}
   =
   \begin{pmatrix}
    \mathbb{I} & 0 & 0 & 0 \\
    0 & \mathbb{I} & 0 & 0 \\
    0 & 0 & \mathbb{I} & 0 \\
    0 & 0 & 0 & 0          \\
   \end{pmatrix}
   ,
   \label{eq:E}
\end{equation}
with $\mathbb{I}$ the identity matrix.

\section{Numerical computation of the forced stochastic linear modes formalism}
\label{sec:computation}
In \citet{tissotJFM2021}, an ensemble of solutions are computed to obtain an empirical cross spectral density (CSD) matrix.
This procedure turns out to be more expensive than a singular value decomposition (SVD) for small size problems (one-dimensional in the $y$ direction).
We can note that for large scale problems, advanced ensemble-based techniques \citep{moarref2013,ribeiro2020} or time-domain formulations \citep{martini2021} can be employed.
We propose here to write FSLM as a SVD problem.
Starting from the system~(9) of the main document and similarly as in resolvent analysis, we define $\widetilde{\mathsfbi{L}}_{\alpha,\beta}=\mathsfbi{H}\left( \widetilde{\mathsfbi{A}}_{\alpha,\beta,\bar{\bm{q}}}-i\omega\mathsfbi{E} \right)^{-1}\widetilde{\mathsfbi{B}}$ and perform the singular value decomposition
\begin{equation}
    \mathsfbi{W}^{\frac{1}{2}}\mathsfbi{H}\left( \widetilde{\mathsfbi{A}}_{\alpha,\beta,\bar{\bm{q}}}-i\omega\mathsfbi{E} \right)^{-1}\widetilde{\mathsfbi{B}}\mathsfbi{W}_f^{-\frac{1}{2}}=\mathsfbi{U}_{\text{\tiny FSLM}}\boldsymbol{\Sigma}_{\text{\tiny FSLM}}\mathsfbi{V}_{\text{\tiny FSLM}}^*,
\end{equation}
with
\begin{equation}
    \widetilde{\mathsfbi{B}}
    =
    \begin{pmatrix}
        -\boldsymbol{\Phi}_y^\sigma \mathbb{D}^\sigma \frac{\partial U}{\partial y}+\frac{1}{Re}\Delta \boldsymbol{\Phi}_x^\sigma \mathbb{D}^\sigma  & b(y)\mathbb{I} & 0 & 0 \\
       \frac{1}{Re}\Delta \boldsymbol{\Phi}_y^\sigma \mathbb{D}^\sigma & 0 & b(y)\mathbb{I} & 0\\
       \frac{1}{Re}\Delta \boldsymbol{\Phi}_z^\sigma \mathbb{D}^\sigma & 0 & 0 & b(y)\mathbb{I} \\
       0 & 0 & 0 & 0
    \end{pmatrix}
    ,
    \label{eq:B}
\end{equation}
and
\begin{equation}
   \mathsfbi{W}_f=
   \begin{pmatrix}
   \mathbb{I} & 0 \\
   0 & \mathsfbi{W}
   \end{pmatrix}
   .
\end{equation}
The matrix $\boldsymbol{\Phi}^\sigma$ gathers in columns $\bm{\Phi}^\sigma_k$ for $k\in[1,N_\sigma]$, expansion basis of the noise $\dd \bm{\xi}_{\alpha,\beta,\omega}$ defined in eq.~(12) of the main document.
The diagonal matrix $\mathbb{D}^\sigma$ contains the associated amplitude coefficients $c_k$. 
The matrix $\widetilde{\mathsfbi{B}}$ applies on the vector of random variables $$\widetilde{\bm{f}}=(\eta_1,\cdots,\eta_{N_\sigma},\widetilde{f}^{NL}_x,\widetilde{f}^{NL}_y,\widetilde{f}^{NL}_z)^T.$$

Following the same notations than in resolvent analysis, FSLM are defined by $\bm{\Phi}^{\text{\tiny FSLM}}_i=\mathsfbi{W}^{-\frac{1}{2}}\bm{V}_{\text{\tiny FSLM},i}$, where the first mode is the predicted coherent structure and the higher order modes are used to define the noise at the next iteration (see section~III.3 of the main document).

\section{Computation of SPOD modes}
\label{sec:computation_SPOD_modes}
SPOD has been computed for the three numerical simulations.
Numerical parameters are given in table~\ref{tab:SPOD} with $N_{\text{\tiny FFT}}$ denoting the window size for performing the discrete Fourier transform, $T$ the duration of the simulation in outer units and $N_t$ the associated number of time steps.
A weighting window $w_j=2\sin^2\left(\pi\frac{j-1}{N_{\text{\tiny FFT}}}\right)$ is used with an overlap region of $N_{\text{\tiny overlap}}$.
\begin{table}
    \centering
    \begin{tabular}{ccccc}
                   & $N_{\text{\tiny FFT}}$ & $N_{\text{\tiny overlap}}$ & $N_t$ & $T$ \\
    $Re_\tau=180$  &  256                   & 192                        & 2000  & 1000\\
    $Re_\tau=550$  &  256                   & 224                        & 3000  & 300 \\
    $Re_\tau=1000$ &  512                   & 448                        & 4000  & 200 \\
    \end{tabular}
    \caption{Numerical parameters for the SPOD.}
    \label{tab:SPOD}
\end{table}

For the waves in the logarithmic layer, the wave-numbers $(\lambda_x^+,\lambda_z^+)$ have been chosen in the energy peak \citep[taken from][]{amaral2021} computed at the critical layer $y_c^+$.
Values of the wave numbers are given in table~\ref{tab:lambda}.
\begin{table}
    \centering
    \begin{tabular}{ccccccc}
                              & & $\lambda_x^+$ & $\lambda_z^+$ & $\lambda_t^+$ & $c^+$ & $y_c^+$\\
         W1 at $Re_\tau=180$  & & 1124        & 102         & 100               & 11.2  & 16   \\
         W1 at $Re_\tau=550$  & & 1137        & 100         & 100               & 11.2  & 16    \\
         W1 at $Re_\tau=1000$ & & 1043        & 101         & 100               & 10.4  & 16    \\
         W2 at $Re_\tau=1000$ & & 1043        & 391         & 66                & 15.8  & 70    \\
         W3 at $Re_\tau=1000$ & & 2087        & 522         & 115               & 18.1  & 180   \\
         W4 at $Re_\tau=1000$ & & 2087        & 522         & 110               & 19    & 250
    \end{tabular}
    \caption{Values of wave length and time period of the selected modes.}
    \label{tab:lambda}
\end{table}

\section{Complementary results}
\label{sec:results}
\subsection{Convergence}
\label{sec:convergence}
\begin{figure}
   \centering
   \includegraphics[width=0.4\textwidth]{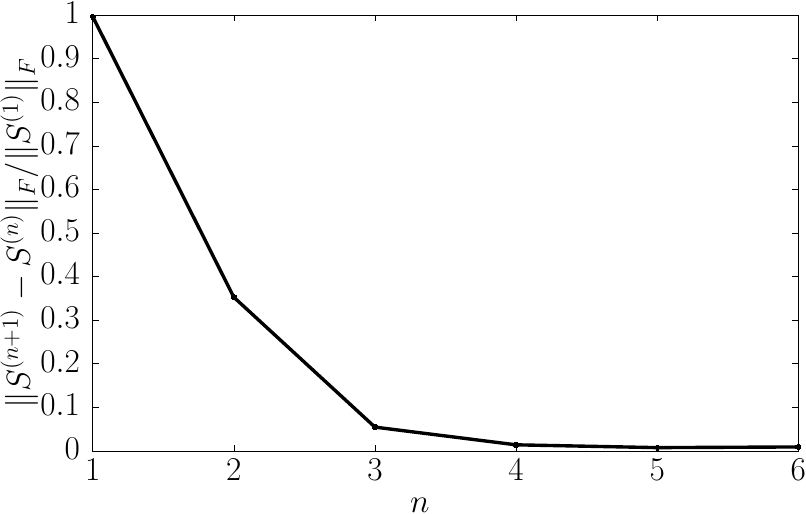}
   \caption{Convergence criteria of W3 $Re_\tau=1000$. Relative variation of the CSD $\mathsfbi{S}^{(n)}$ based on the Frobenius norm.}
   \label{fig:convergence-crit}
\end{figure}

In this section, we show the effect of the iterative procedure to define the operator $\boldsymbol{\sigma}$ described in section~\ref{sec:computation_SPOD_modes}.
Figure~\ref{fig:convergence-crit} shows the decay of the convergence criteria $\frac{\|\mathsfbi{S}^{(n+1)}-\mathsfbi{S}^{(n)}\|_F}{\|\mathsfbi{S}^{(1)}\|_F}$ in the case of W3 at $Re_\tau=1000$, with $\mathsfbi{S}^{(n)}$ being the CSD of the FSLM solution at iteration $n$ and $\|\bcdot\|_F$ the Frobenius norm.
The procedure converges in practice in few iterations.

\begin{figure*}
   \centering
    \subfigure[$n=1$.]{
    \includegraphics[width=0.22\textwidth]{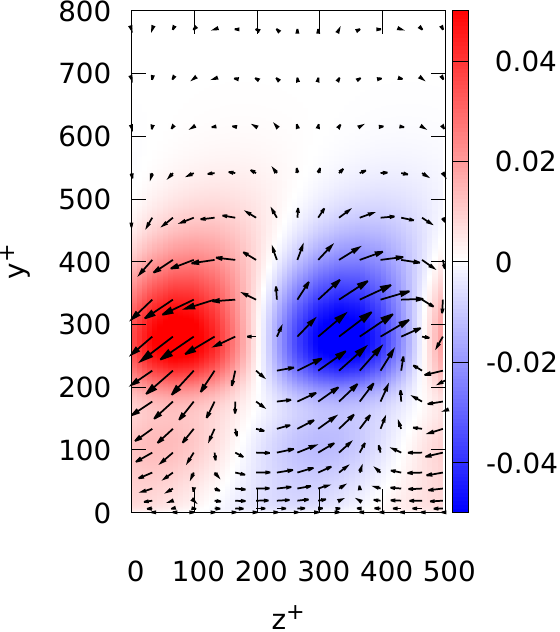}
    }
    \hfill
    \subfigure[$n=2$.]{
    \includegraphics[width=0.22\textwidth]{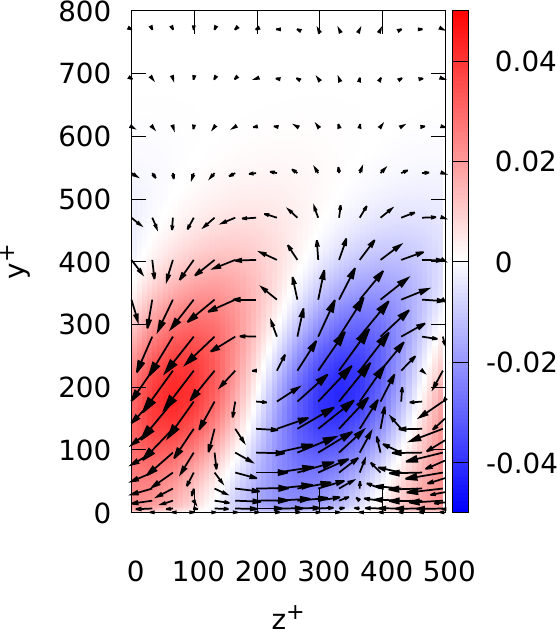}
    }
    \hfill
    \subfigure[$n=3$.]{
    \includegraphics[width=0.22\textwidth]{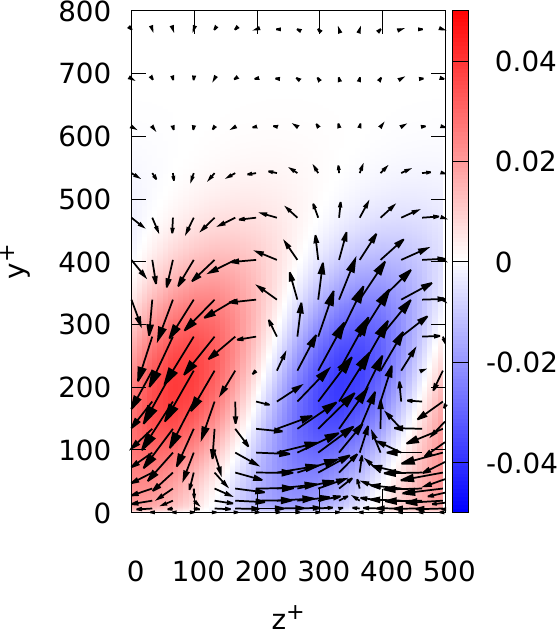}
    }
    \hfill
    \subfigure[Converged.]{
    \includegraphics[width=0.22\textwidth]{images/Re1000/W3_st_iter.pdf}
    }
    \caption{Convergence of W3 at $Re=1000$.}
   \label{fig:convergence-rec}
\end{figure*}

Figure~\ref{fig:convergence-rec} shows reconstructions of W3 during the iteration process.
At the first iteration, the solution is concentrated around the critical layer, which is likely a consequence of the thin spatial support of $\nu$-resolvent modes.
Even if the first guess leads to irrelevant solutions, in few iterations good agreement compared to SPOD are obtained. This indicates the relevance and the robustness of the procedure.
Very similar behaviours can be observed for the other waves.

\subsection{Behaviour in the buffer layer at $Re_\tau=550$ and $Re_\tau=1000$}
\label{sec:buffer}
\begin{figure*}
    \centering
   \subfigure[SPOD]{
   \includegraphics[width=0.22\textwidth]{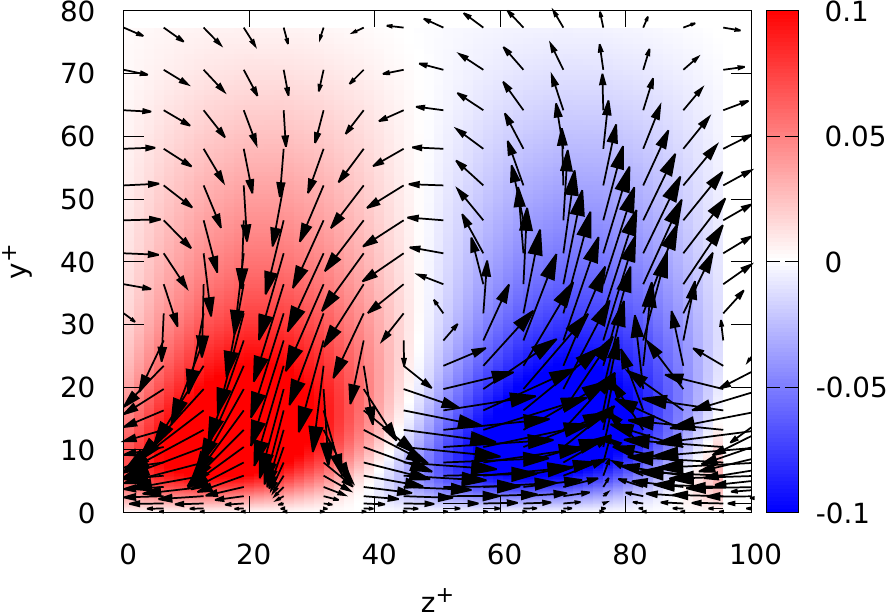}
   }
   \hfill
   \subfigure[$\nu$-resolvent]{
   \includegraphics[width=0.22\textwidth]{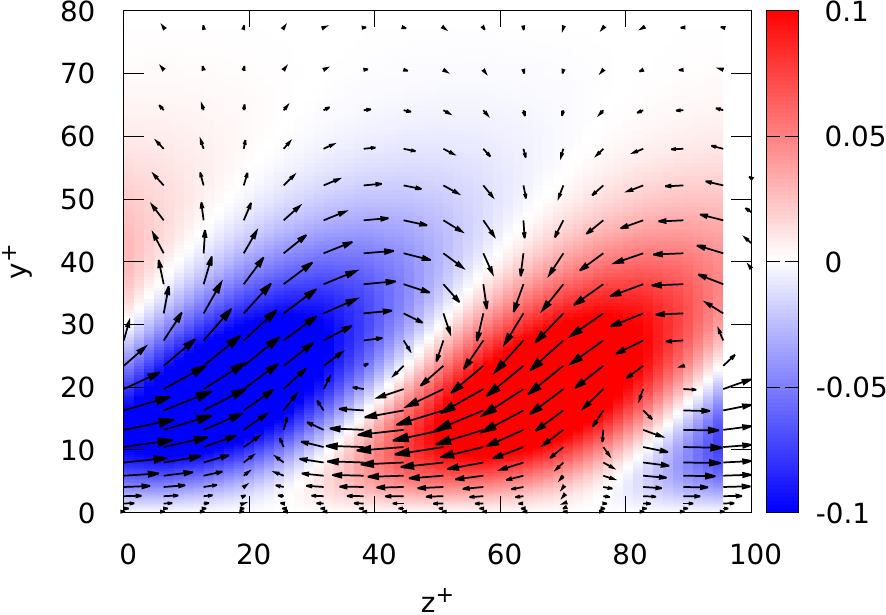}
   }
   \hfill
   \subfigure[$\nu_t$-resolvent]{
   \includegraphics[width=0.22\textwidth]{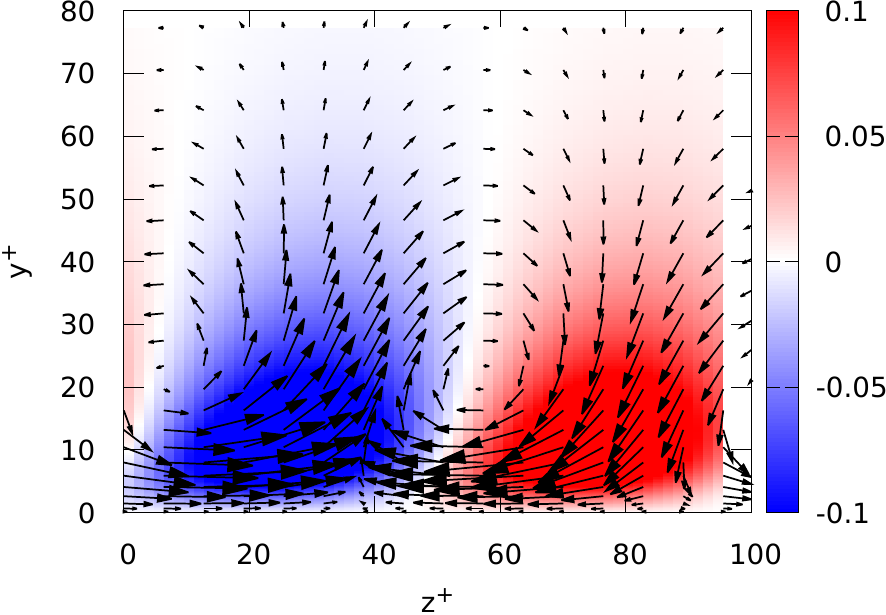}
   }
   \hfill
   \subfigure[FSLM.]{
   \includegraphics[width=0.22\textwidth]{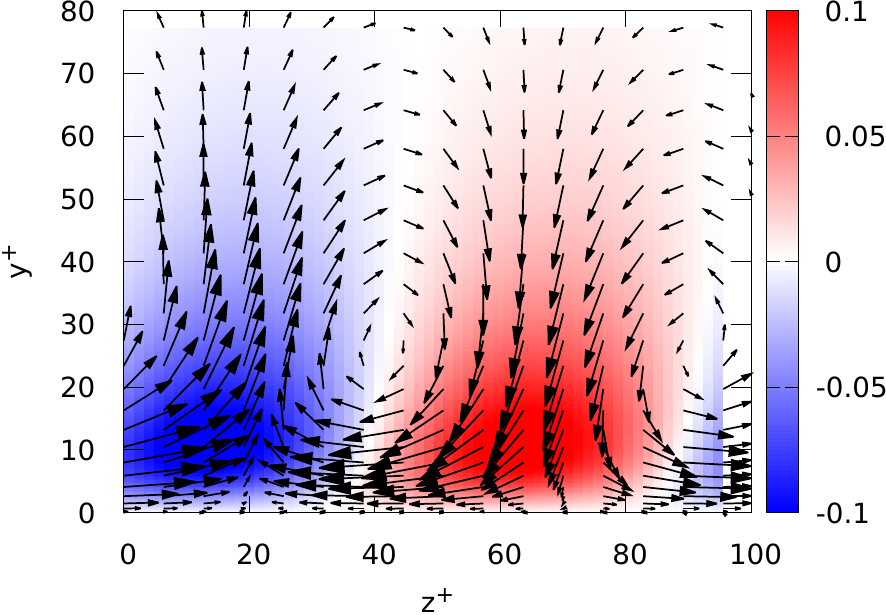}
   }
   \caption{Reconstructions of W1 at $Re_\tau=550$. Colors are streamwise velocities, arrows are in-plane velocity fields.}
   \label{fig:W1Re550}
\end{figure*}

\begin{figure*}
    \centering
    \subfigure[Streamwise velocity $|\hat{u}|^2$. ]{\includegraphics[width=0.32\textwidth]{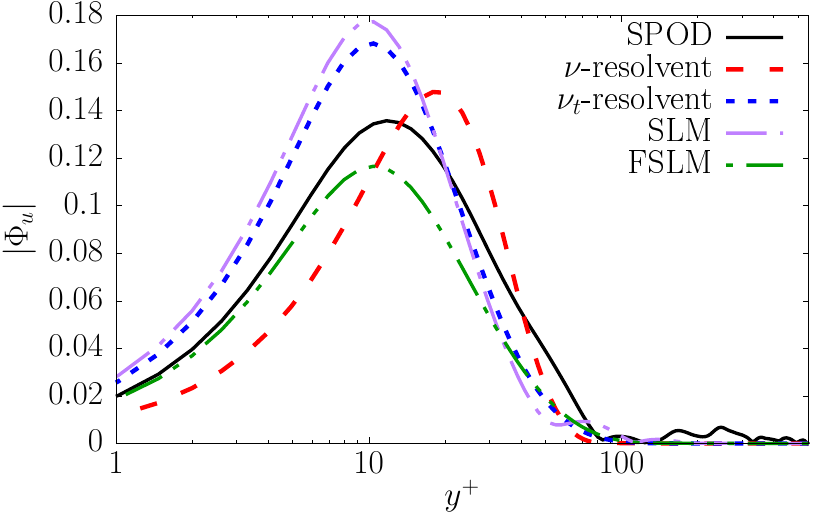}}
    \hfill                                                                                                  
    \subfigure[Wall-normal velocity $|\hat{v}|^2$.]{\includegraphics[width=0.32\textwidth]{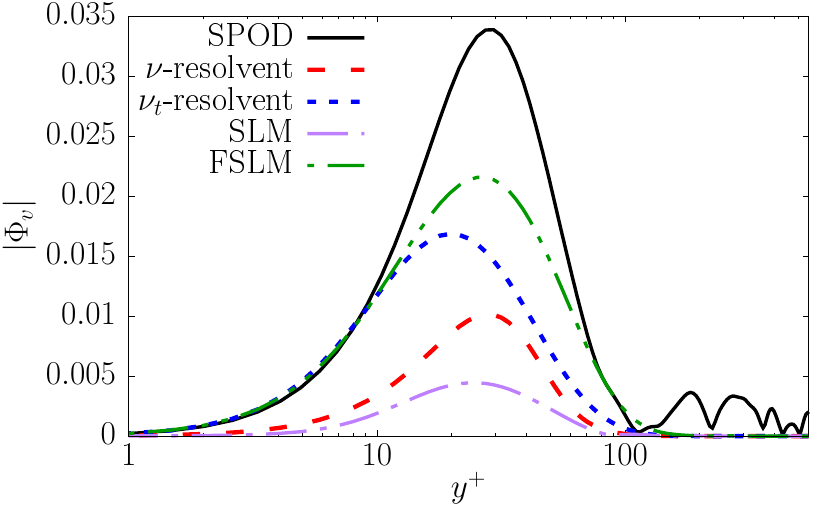}}
    \hfill                                                                                              
    \subfigure[Spanwise velocity $|\hat{w}|^2$.   ]{\includegraphics[width=0.32\textwidth]{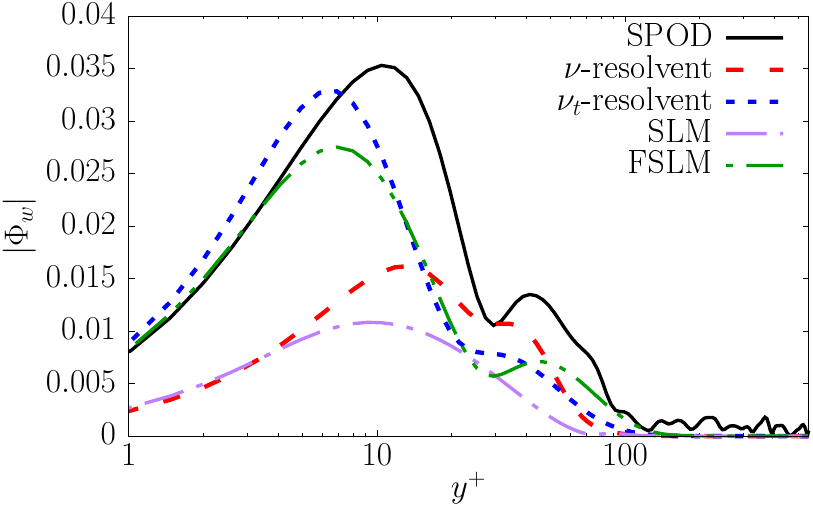}}
    \caption{PSD velocity profiles of W1 at $Re_\tau=550$.}
   \label{fig:profileW1Re550}
\end{figure*}

\begin{figure*}
    \centering
   \subfigure[SPOD]{
   \includegraphics[width=0.22\textwidth]{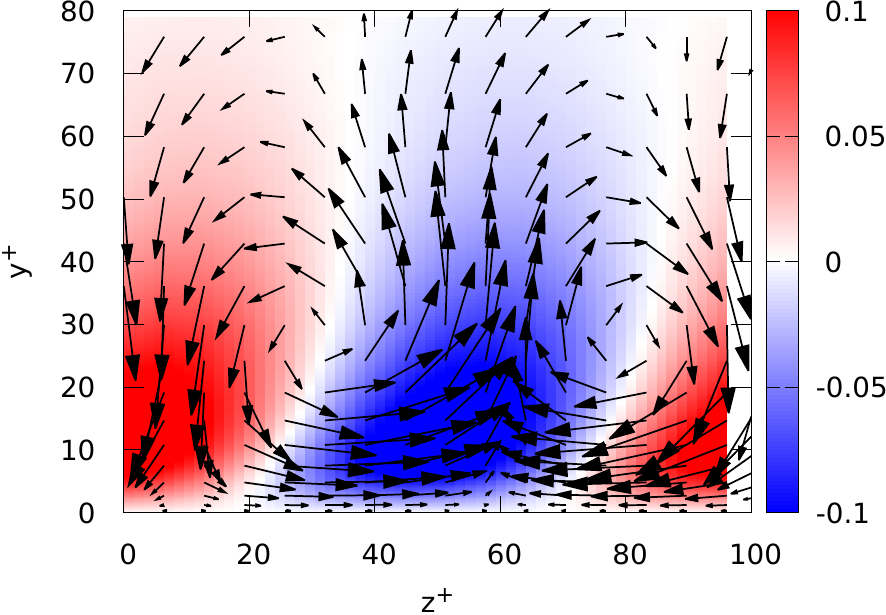}
   }
   \hfill
   \subfigure[$\nu$-resolvent]{
   \includegraphics[width=0.22\textwidth]{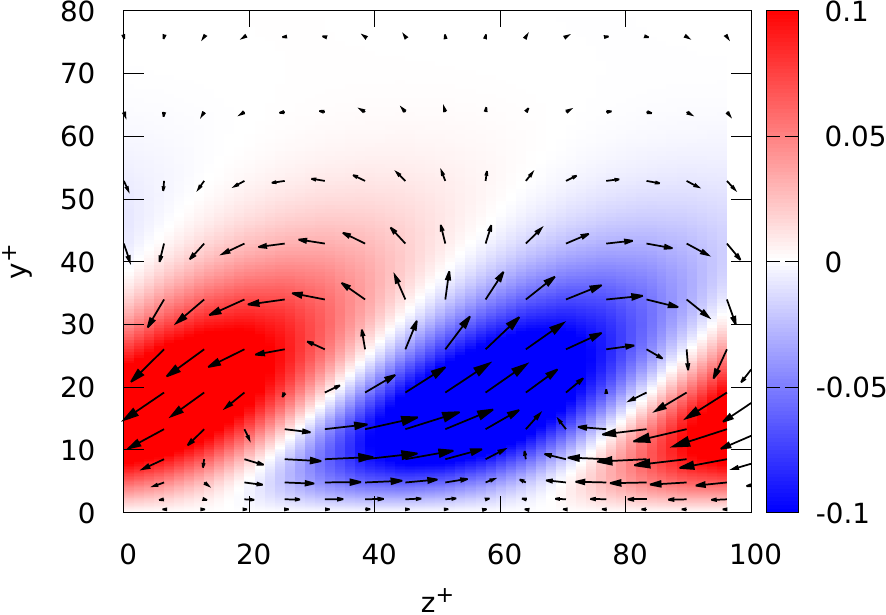}
   }
   \hfill
   \subfigure[$\nu_t$-resolvent]{
   \includegraphics[width=0.22\textwidth]{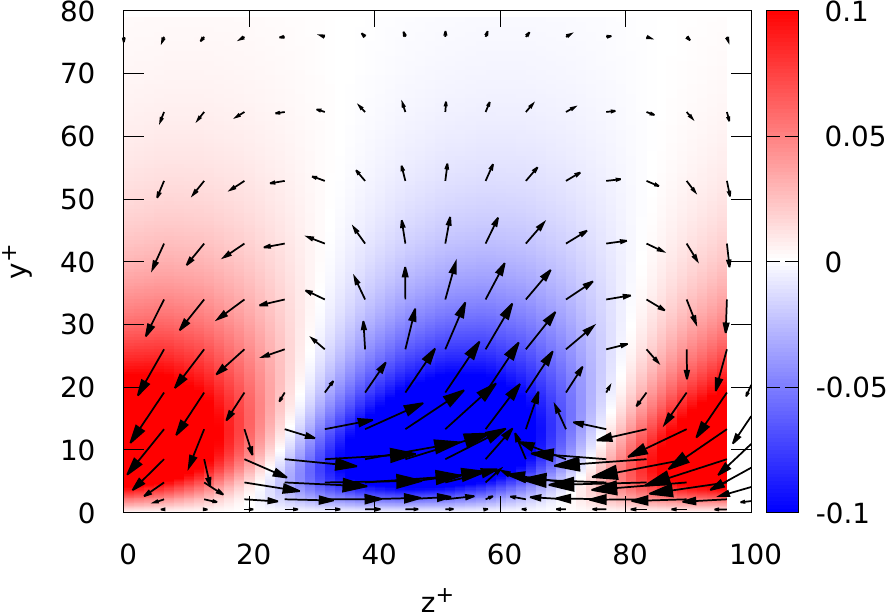}
   }
   \hfill
   \subfigure[FSLM.]{
   \includegraphics[width=0.22\textwidth]{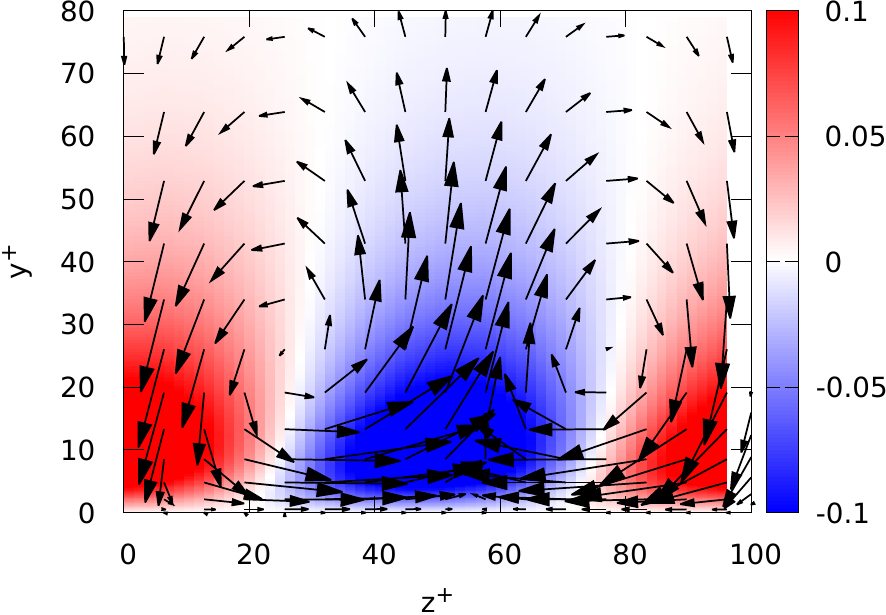}
   }
   \caption{Reconstructions of W1 at $Re_\tau=1000$. Colors are streamwise velocities, arrows are in-plane velocity fields.}
   \label{fig:W1Re1000}
\end{figure*}

\begin{figure*}
    \centering
    \subfigure[Streamwise velocity $|\hat{u}|^2$. ]{\includegraphics[width=0.32\textwidth]{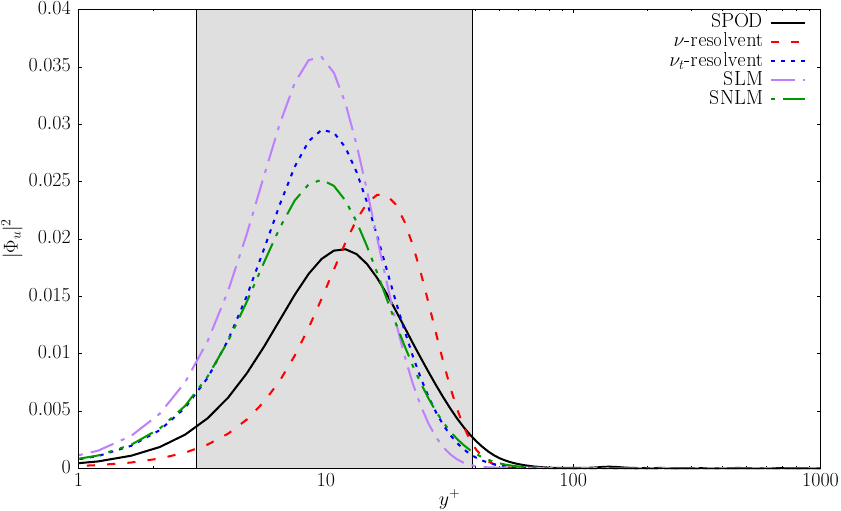}}
    \hfill                                                                                            
    \subfigure[Wall-normal velocity $|\hat{v}|^2$.]{\includegraphics[width=0.32\textwidth]{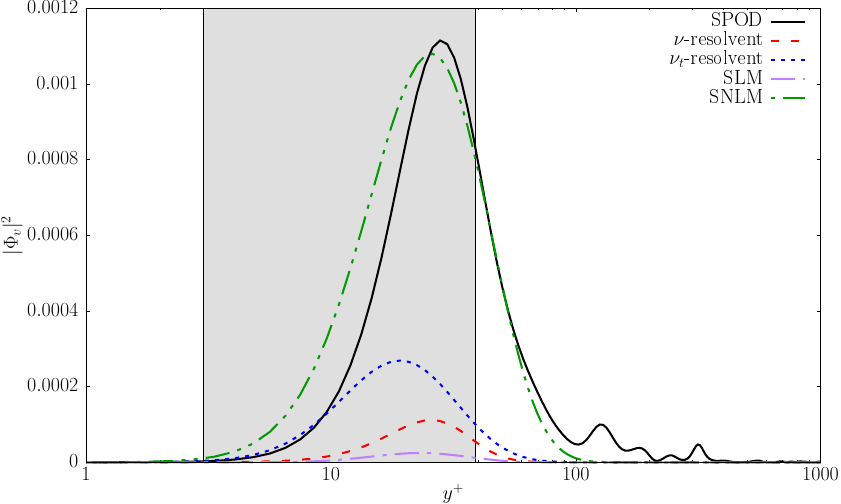}}
    \hfill                                                                                              
    \subfigure[Spanwise velocity $|\hat{w}|^2$.   ]{\includegraphics[width=0.32\textwidth]{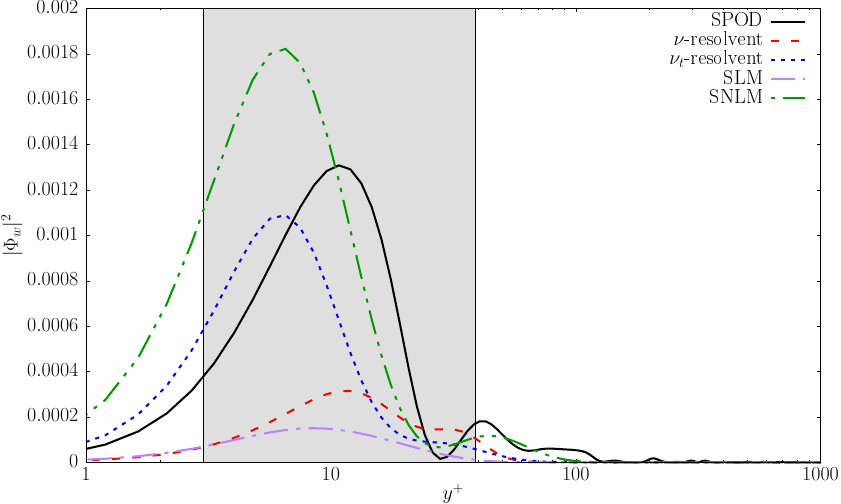}}
    \caption{PSD velocity profiles of W1 at $Re_\tau=1000$.}
   \label{fig:profileW1Re1000}
\end{figure*}

In order to show the robustness of the procedure, waves W1 evolving within the buffer layer are computed at $Re_\tau=550$ in figure~\ref{fig:W1Re550} and at $Re_\tau=1000$ in figure~\ref{fig:W1Re1000}.
Associated PSD profiles are shown in figures~\ref{fig:profileW1Re550} and~\ref{fig:profileW1Re1000}.
The results are quite similar to the ones described in section~\ref{sec:buffer}.
We can note a worsening of the agreement with SPOD when the Reynolds number increases for all models.
However, FSLM consistently displays the best predictions of near-wall structures.

\subsection{Two other waves in the logarithmic layer}
\label{sec:log}
\begin{figure*}
   \centering
   \includegraphics[width=0.45\textwidth]{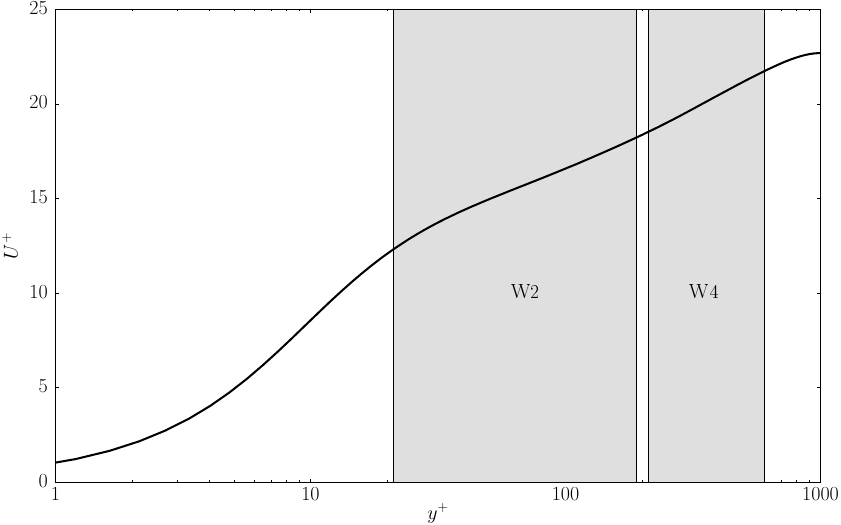}
   \hfill
   \includegraphics[width=0.45\textwidth]{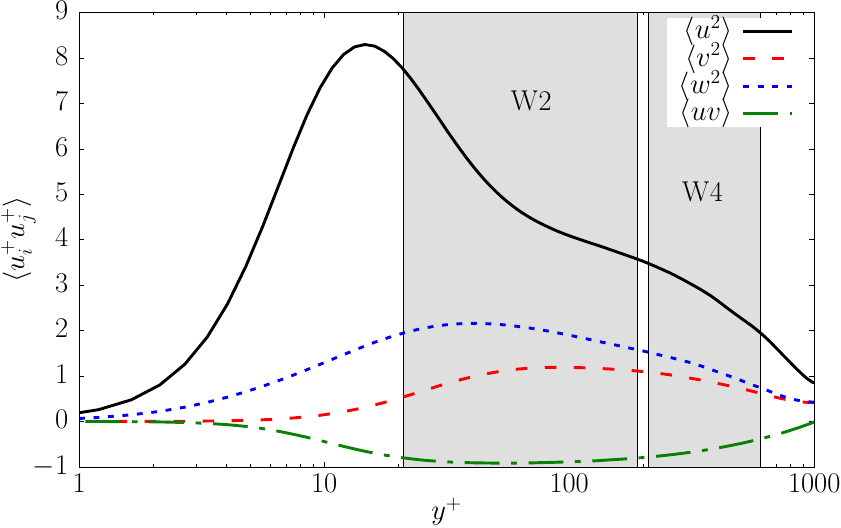}
   \caption{Spatial support of W2 and W4 superimposed with the mean and RMS profiles.}
   \label{fig:flow2}
\end{figure*}

To observe the robustness of the procedure in the logarithmic layer, waves labelled as W2 and W4 are considered at $Re_\tau=1000$.
This is shown only at $Re_\tau=1000$ since only for this Reynolds number one may study an indicial logarithmic layer with some separation of scales.
They have a spatial support respectively in the low and high region of the logarithmic layer.
Figure~\ref{fig:flow2} shows their spatial supports superimposed with the mean flow and RMS profiles.

\begin{figure*}
    \centering
      \subfigure[SPOD]{
      \includegraphics[width=0.22\textwidth]{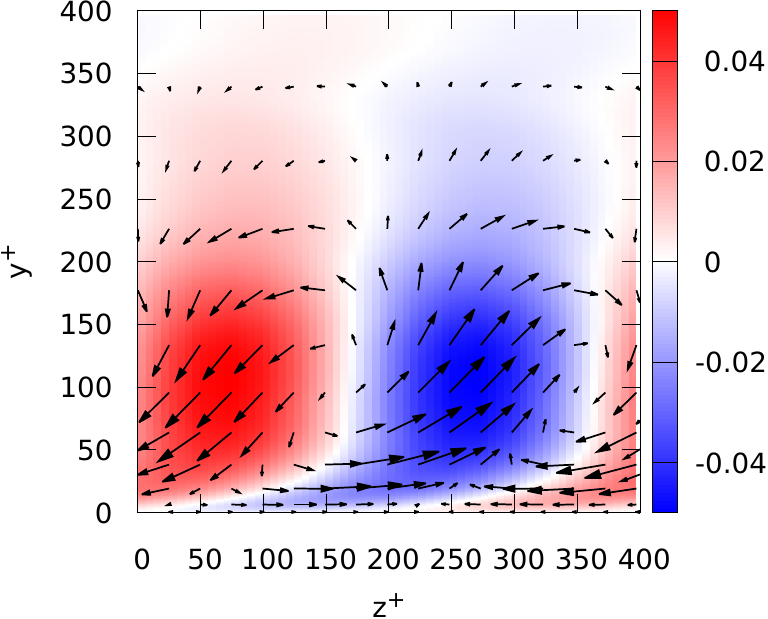}
      }
      \hfill
      \subfigure[$\nu$-resolvent]{
      \includegraphics[width=0.22\textwidth]{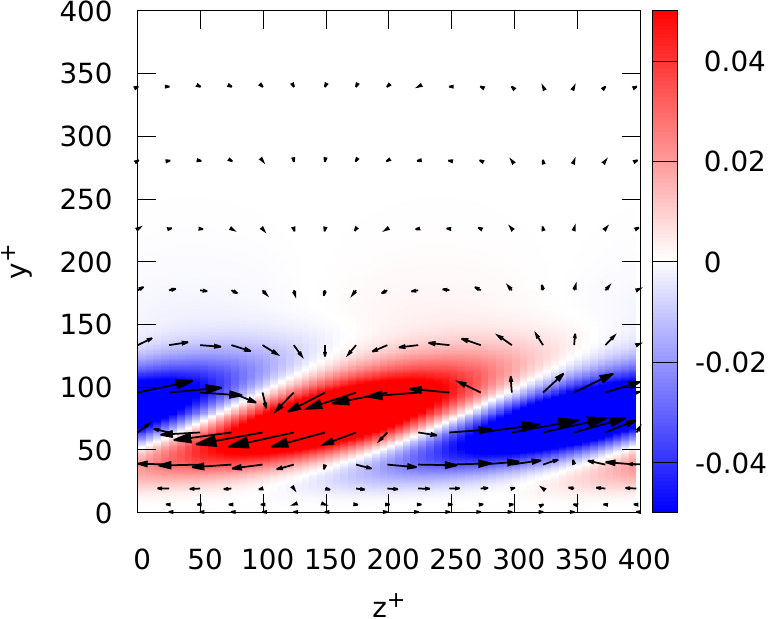}
      }
      \hfill
      \subfigure[$\nu_t$-resolvent]{
      \includegraphics[width=0.22\textwidth]{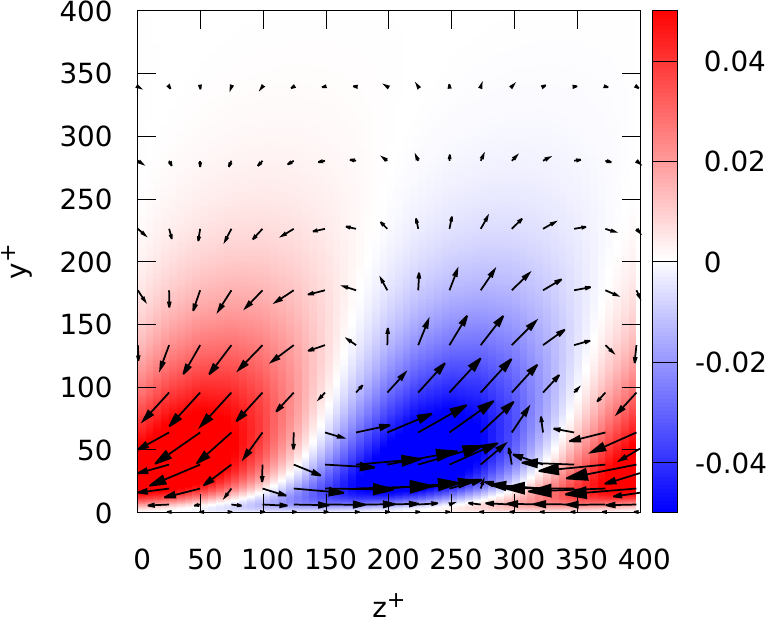}
      }
      \hfill
      \subfigure[FSLM.]{
      \includegraphics[width=0.22\textwidth]{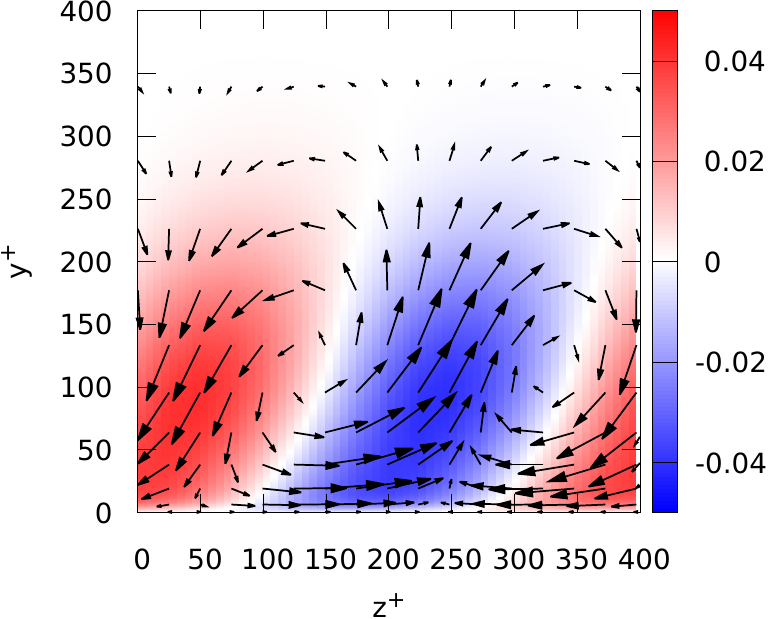}
      }
      \caption{Reconstructions W2 $Re_\tau=1000$. Colors are streamwise velocities, arrows are in-plane velocity fields.}
      \label{fig:reconstructW2}
   \end{figure*}

\begin{figure*}
    \centering
    \subfigure[Streamwise velocity $|\hat{u}|^2$. ]{\includegraphics[width=0.32\textwidth]{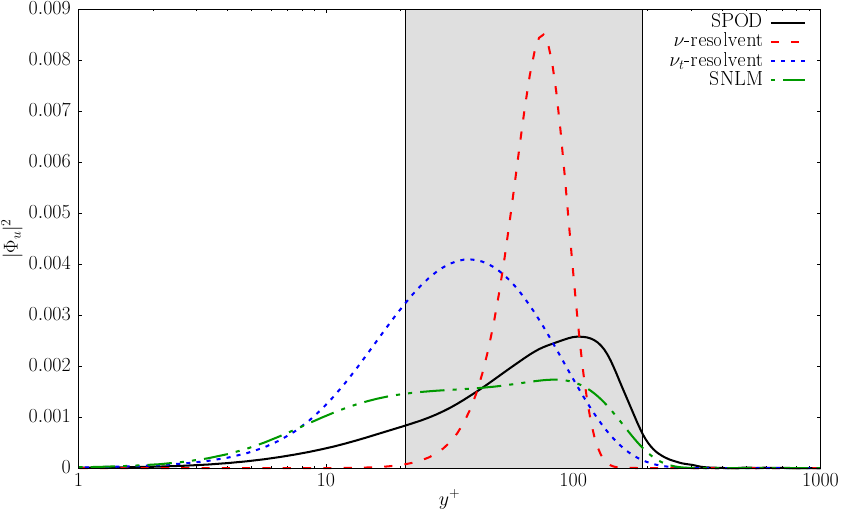}}
    \hfill 
    \subfigure[Wall-normal velocity $|\hat{v}|^2$.]{\includegraphics[width=0.32\textwidth]{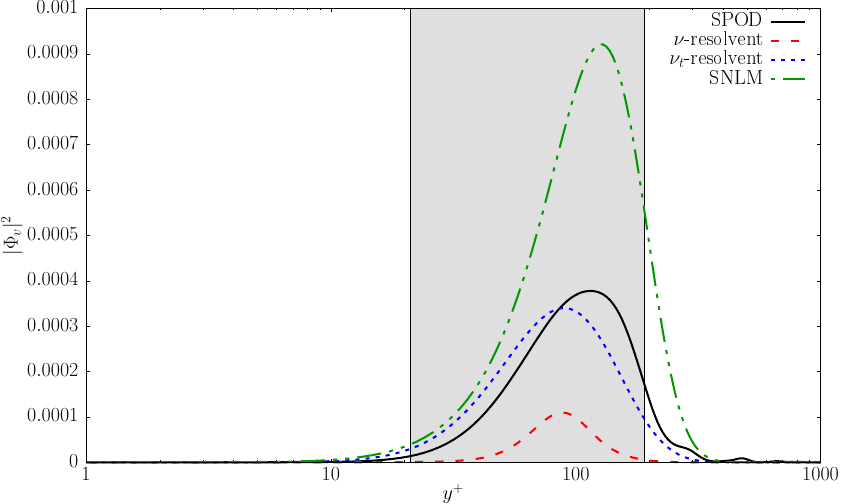}}
    \hfill                                        
    \subfigure[Spanwise velocity $|\hat{w}|^2$.   ]{\includegraphics[width=0.32\textwidth]{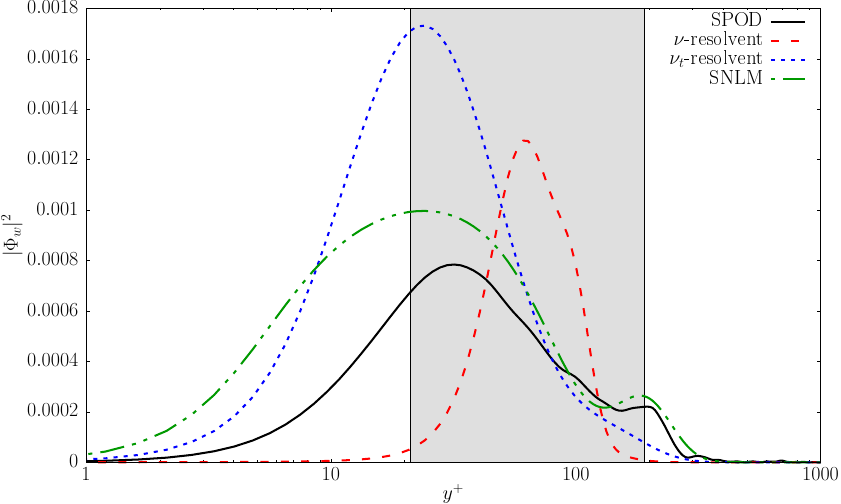}}
    \caption{PSD velocity profiles of W2 at $Re_\tau=1000$.}
    \label{fig:profilesW2}
\end{figure*}

Figures~\ref{fig:reconstructW2} and \ref{fig:profilesW2} show reconstructions and profiles for W2.
We can observe very good reconstructions and accurate velocity profiles for FSLM.
The results are similar to the ones of W3.

\begin{figure*}
    \centering
      \subfigure[SPOD]{
      \includegraphics[width=0.22\textwidth]{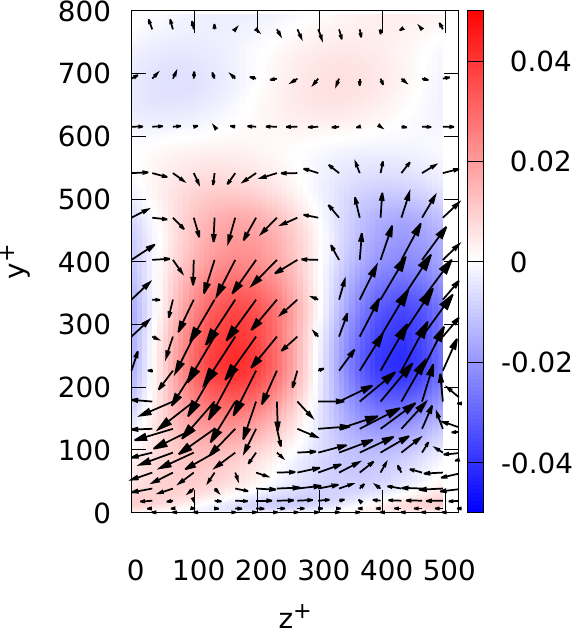}
      }
      \hfill
      \subfigure[$\nu$-resolvent]{
      \includegraphics[width=0.22\textwidth]{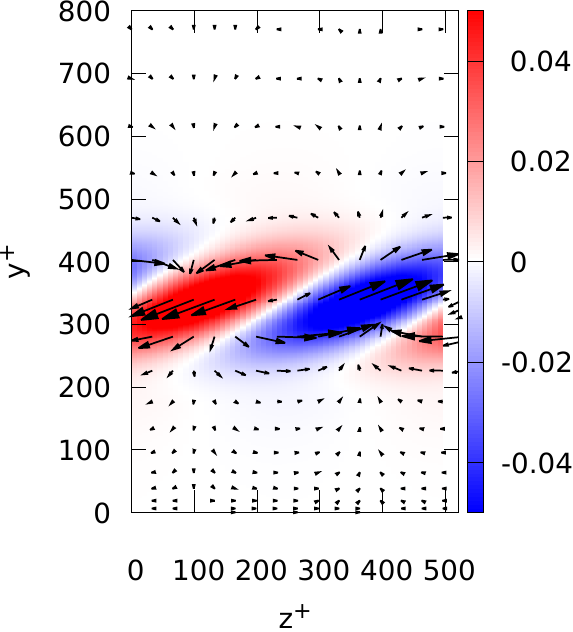}
      }
      \hfill
      \subfigure[$\nu_t$-resolvent]{
      \includegraphics[width=0.22\textwidth]{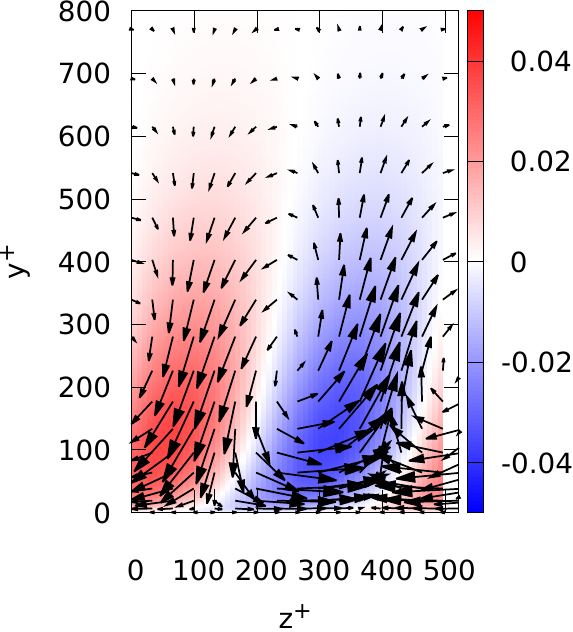}
      }
      \hfill
      \subfigure[FSLM.]{
      \includegraphics[width=0.22\textwidth]{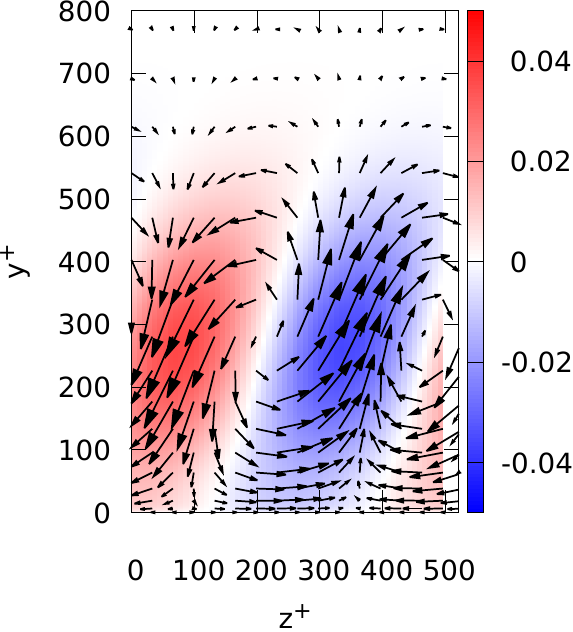}
      }
      \caption{Reconstructions W4 $Re_\tau=1000$. Colors are streamwise velocities, arrows are in-plane velocity fields.}
      \label{fig:reconstructW4}
   \end{figure*}

\begin{figure*}
    \centering
    \subfigure[Streamwise velocity $|\hat{u}|^2$. ]{\includegraphics[width=0.32\textwidth]{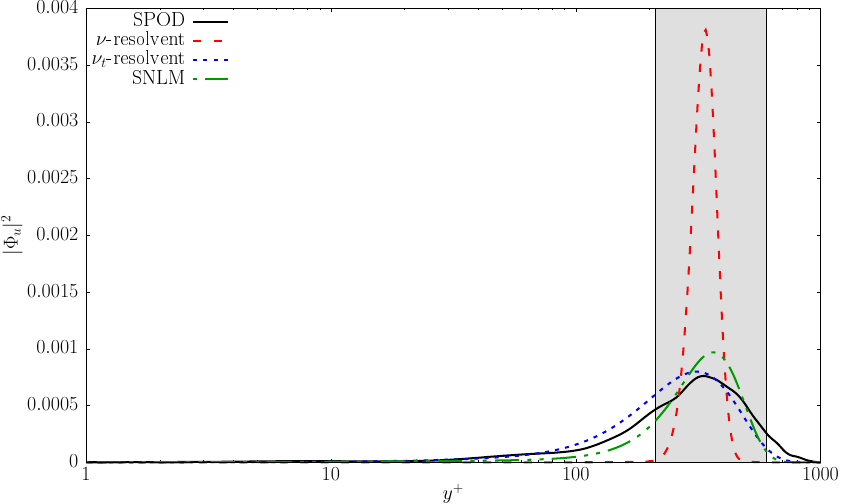}}
    \hfill 
    \subfigure[Wall-normal velocity $|\hat{v}|^2$.]{\includegraphics[width=0.32\textwidth]{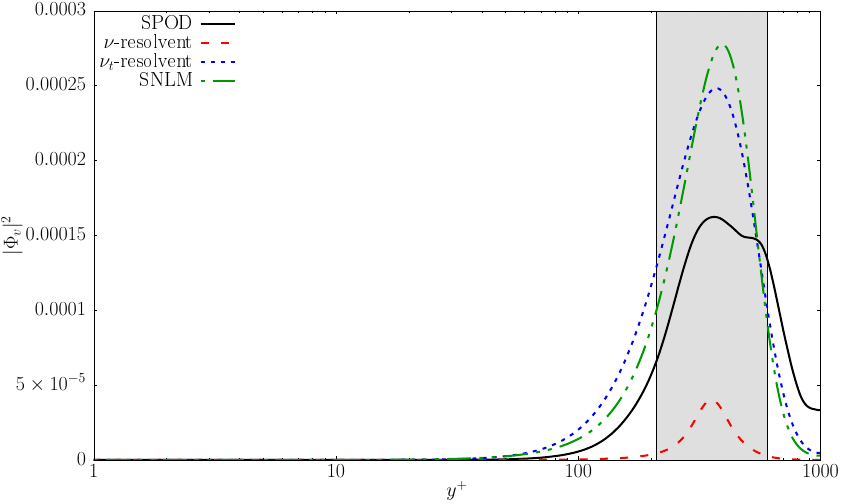}}
    \hfill                                        
    \subfigure[Spanwise velocity $|\hat{w}|^2$.   ]{\includegraphics[width=0.32\textwidth]{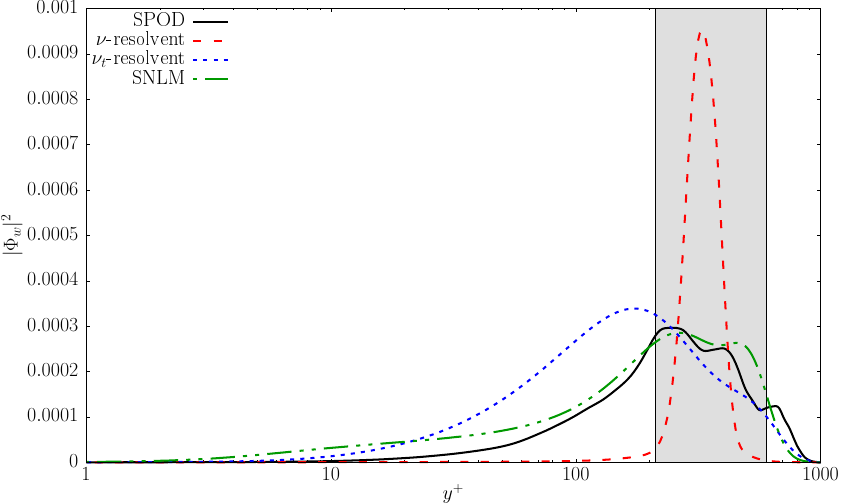}}
    \caption{PSD velocity profiles of W4 at $Re_\tau=1000$.}
    \label{fig:profilesW4}
\end{figure*}

Figures~\ref{fig:reconstructW4} and \ref{fig:profilesW4} show reconstructions and profiles for W4.
Again, a similar behaviour is observed.
As a caveat, reconstructions of the upper part, reaching the outer layer, become less accurate.
Since the correlation times are defined based on a intertial scaling, they are not valid in the outer region.
We interpret this discrepancy by an inconsistency of decorrelation times $\tau$, leading to a wrong stochastic diffusion intensity.
Despite this, we note a close agreement of $w$ components.

\subsection{Complementary metrics}
\label{sec:beta}
\begin{figure*}
    \centering
    \subfigure[$\alpha\beta\lambda^{\text{\tiny SPOD}}_1$, $y_c^+=13$.]{\includegraphics[width=0.23\textwidth]{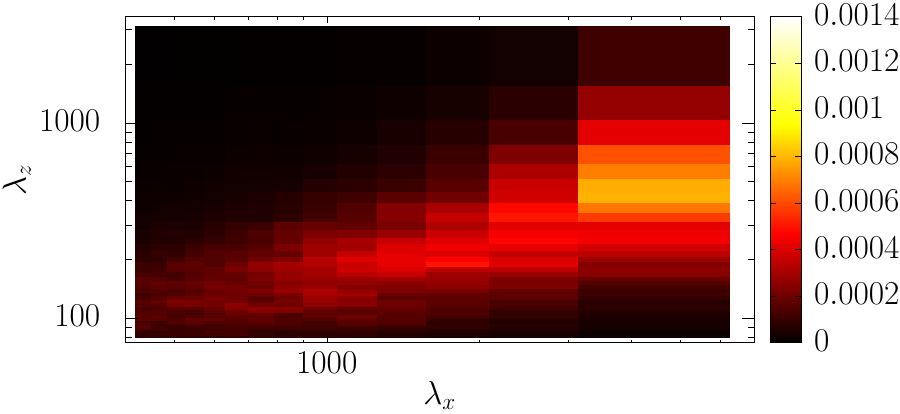}\label{fig:lambda_SPOD_c10}}
    \hfill
    \subfigure[$\beta_{\alpha,\beta,\omega}^{\text{\tiny $\nu$-resolvent}}$, $y_c^+=13$.]{\includegraphics[width=0.23\textwidth]{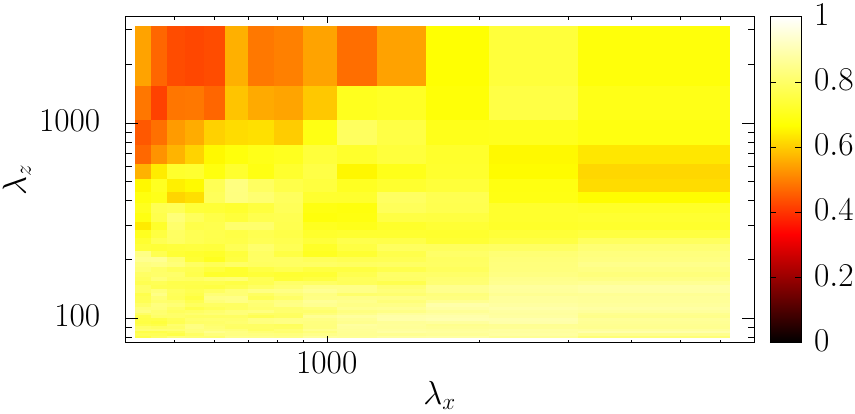}\label{fig:beta_resolvent_c10}}
    \hfill
    \subfigure[$\beta_{\alpha,\beta,\omega}^{\text{\tiny $\nu_t$-resolvent}}$, $y_c^+=13$.]{\includegraphics[width=0.23\textwidth]{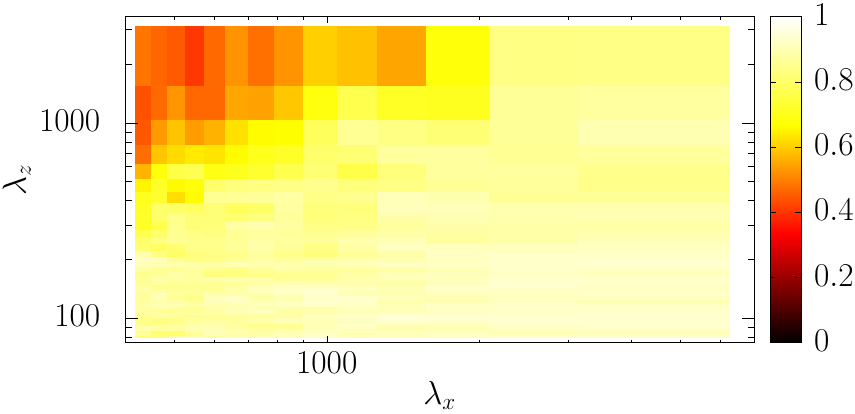}\label{fig:beta_resolvent_eddy_c10}}
    \hfill
    \subfigure[$\beta_{\alpha,\beta,\omega}^{\text{\tiny FSLM}}$, $y_c^+=13$.]{\includegraphics[width=0.23\textwidth]{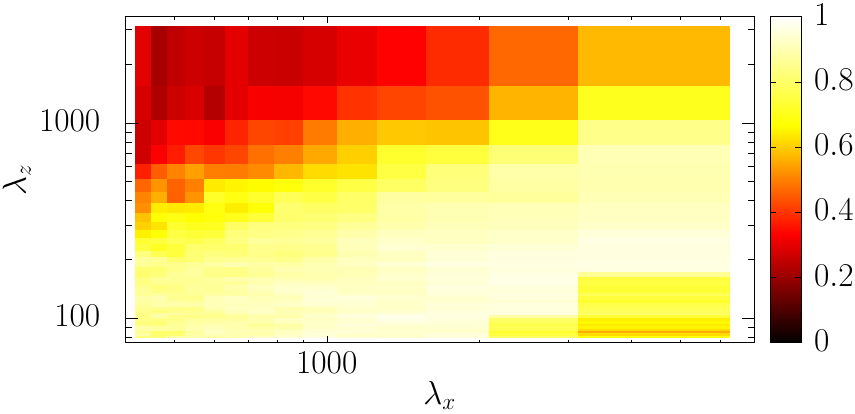}\label{fig:beta_FSLM_c10}}
    \\
    \subfigure[$\alpha\beta\lambda^{\text{\tiny SPOD}}_1$, $y_c^+=53$.]{\includegraphics[width=0.23\textwidth]{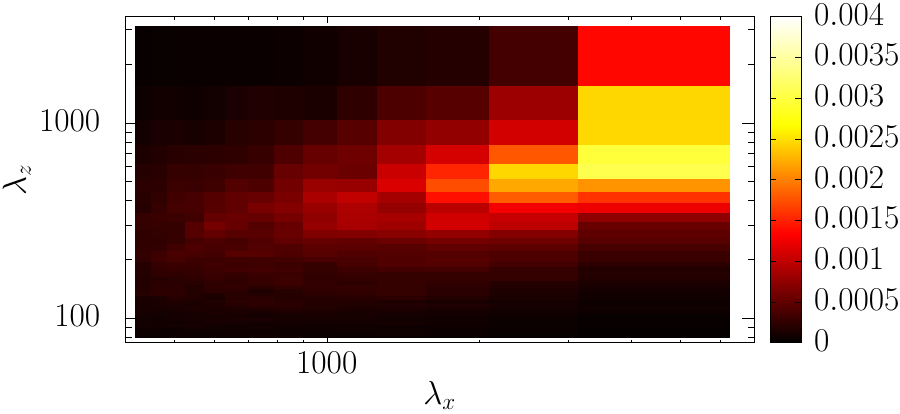}\label{fig:lambda_SPOD_c15}}
    \hfill
    \subfigure[$\beta_{\alpha,\beta,\omega}^{\text{\tiny $\nu$-resolvent}}$, $y_c^+=53$.]{\includegraphics[width=0.23\textwidth]{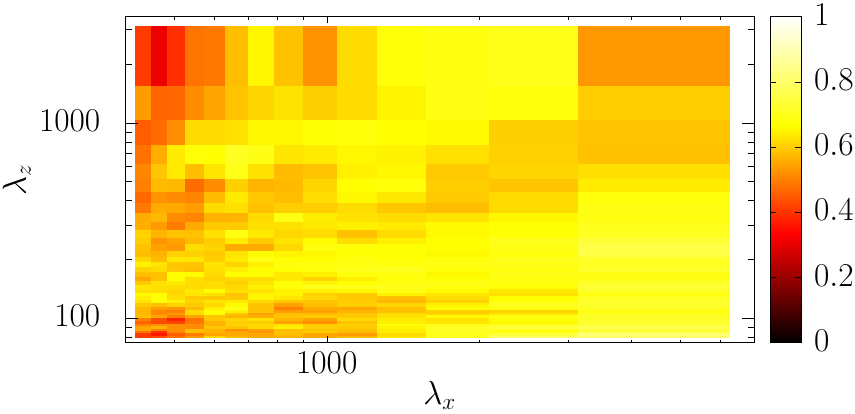}\label{fig:beta_resolvent_c15}}
    \hfill
    \subfigure[$\beta_{\alpha,\beta,\omega}^{\text{\tiny $\nu_t$-resolvent}}$, $y_c^+=53$.]{\includegraphics[width=0.23\textwidth]{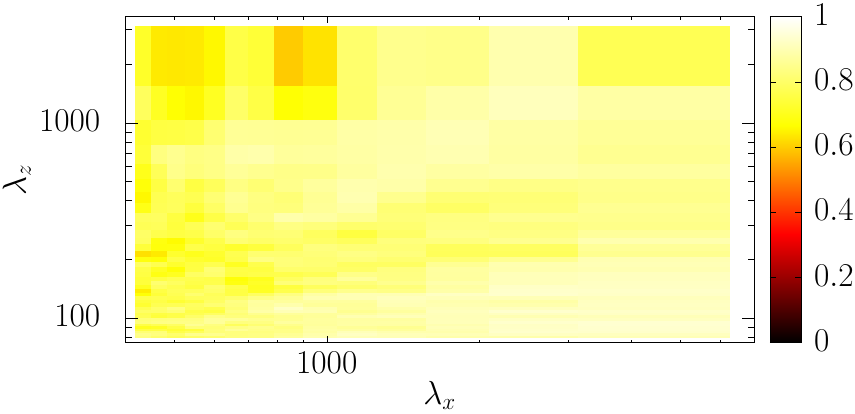}\label{fig:beta_resolvent_eddy_c15}}
    \hfill
    \subfigure[$\beta_{\alpha,\beta,\omega}^{\text{\tiny FSLM}}$, $y_c^+=53$.]{\includegraphics[width=0.23\textwidth]{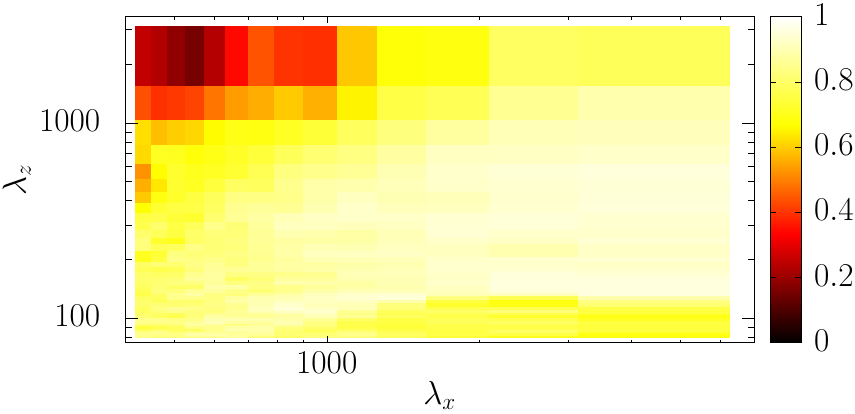}\label{fig:beta_FSLM_c15}}
    \\
    \subfigure[$\alpha\beta\lambda^{\text{\tiny SPOD}}_1$, $y_c^+=200$.]{\includegraphics[width=0.23\textwidth]{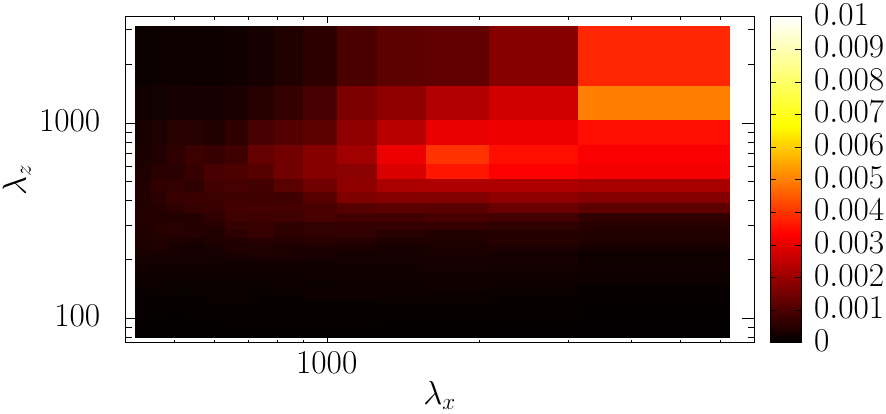}\label{fig:lambda_SPOD_c18.3}}
    \hfill
    \subfigure[$\beta_{\alpha,\beta,\omega}^{\text{\tiny $\nu$-resolvent}}$, $y_c^+=200$.]{\includegraphics[width=0.23\textwidth]{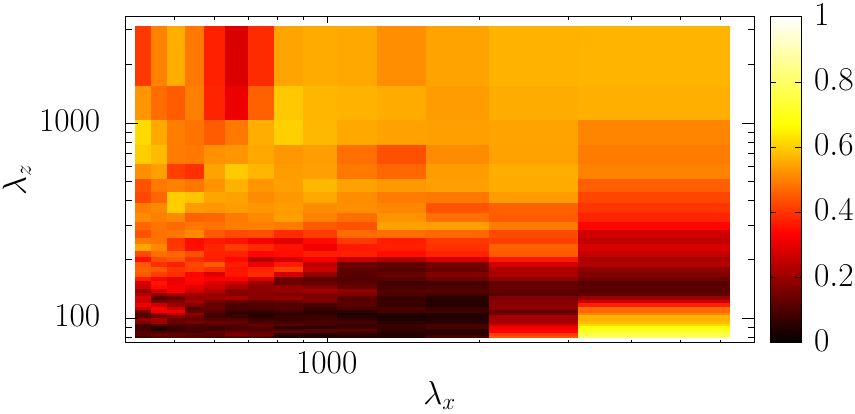}\label{fig:beta_resolvent_c18.3}}
    \hfill
    \subfigure[$\beta_{\alpha,\beta,\omega}^{\text{\tiny $\nu_t$-resolvent}}$, $y_c^+=200$.]{\includegraphics[width=0.23\textwidth]{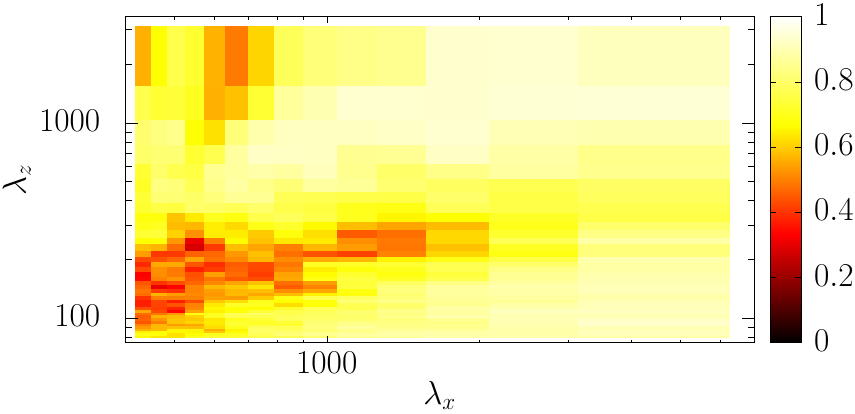}\label{fig:beta_resolvent_eddy_c18.3}}
    \hfill
    \subfigure[$\beta_{\alpha,\beta,\omega}^{\text{\tiny FSLM}}$, $y_c^+=200$.]{\includegraphics[width=0.23\textwidth]{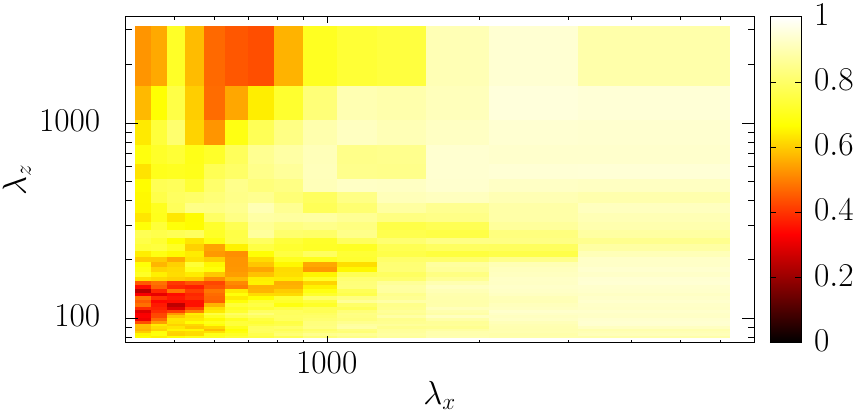}\label{fig:beta_FSLM_c18.3}}
    \\
    \subfigure[$\alpha\beta\lambda^{\text{\tiny SPOD}}_1$, $y_c^+=250$.]{\includegraphics[width=0.23\textwidth]{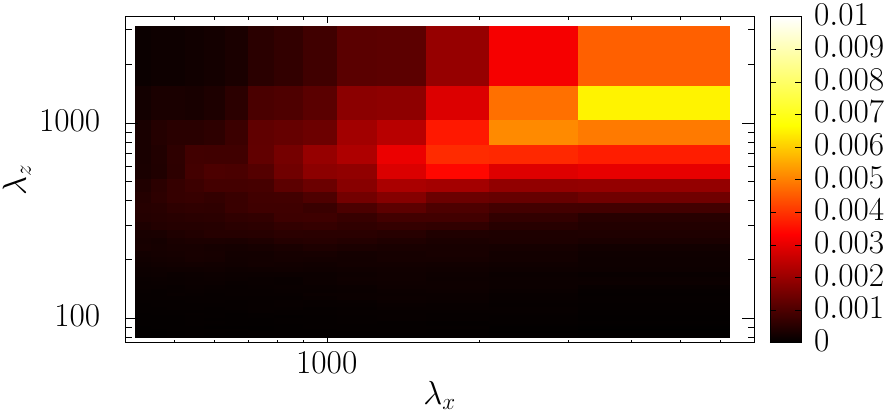}\label{fig:lambda_SPOD_c19}}
    \hfill
    \subfigure[$\beta_{\alpha,\beta,\omega}^{\text{\tiny $\nu$-resolvent}}$, $y_c^+=250$.]{\includegraphics[width=0.23\textwidth]{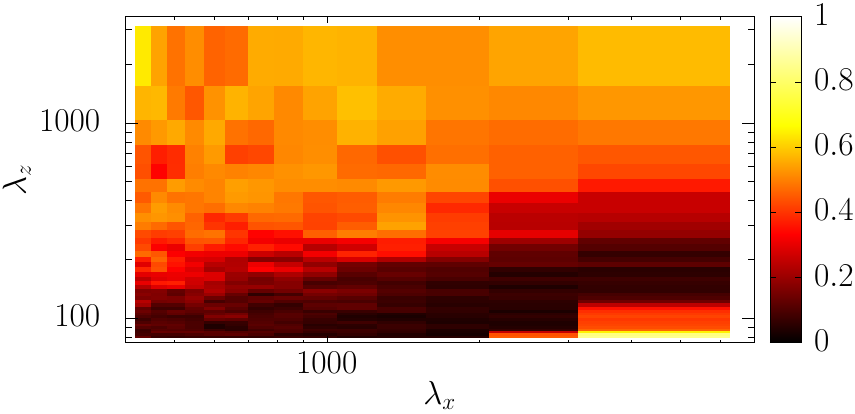}\label{fig:beta_resolvent_c19}}
    \hfill
    \subfigure[$\beta_{\alpha,\beta,\omega}^{\text{\tiny $\nu_t$-resolvent}}$, $y_c^+=250$.]{\includegraphics[width=0.23\textwidth]{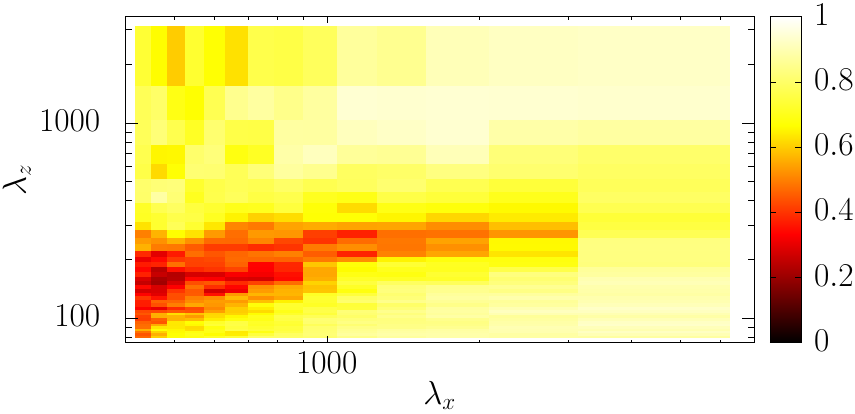}\label{fig:beta_resolvent_eddy_c19}}
    \hfill
    \subfigure[$\beta_{\alpha,\beta,\omega}^{\text{\tiny FSLM}}$, $y^+=250$.]{\includegraphics[width=0.23\textwidth]{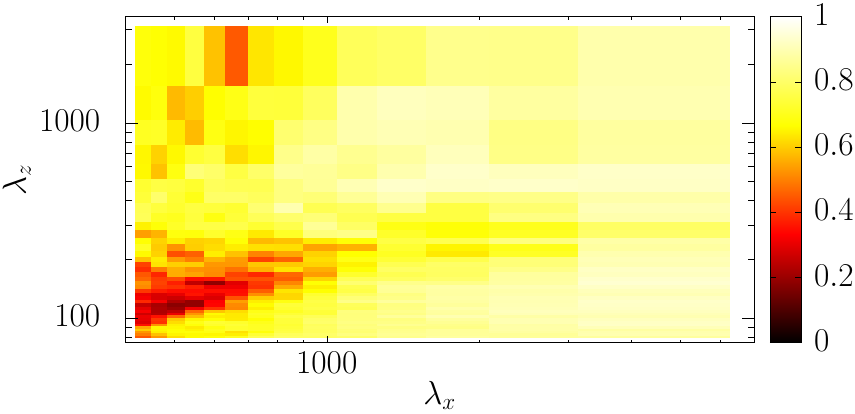}\label{fig:beta_FSLM_c19}}
    \caption{Comparison of collinearity agreement with SPOD $\beta_{\alpha,\beta,\omega}^{\text{\tiny model}}$ function of $(\lambda_x,\lambda_z)$ at various critical levels $y_c^+$. The pre-multiplied spectra of the first SPOD eigenvalue $\alpha\beta\lambda_1^{\text{\tiny SPOD}}$ is displayed.}
\end{figure*}

The metric $\gamma_{\alpha,\beta,\omega}$ defined in the main document represents the improvement of collinearity with SPOD compared to the $\nu_t$-resolvent model.
This comparison is relative to the reference model and this section provides the maps of collinearity $\beta_{\alpha,\beta,\omega}^{\text{\tiny model}}$ of the various models with SPOD at various critical layer positions.
Figure~\ref{fig:lambda_SPOD_c10}, \ref{fig:lambda_SPOD_c15}, \ref{fig:lambda_SPOD_c18.3} and \ref{fig:lambda_SPOD_c19} show the pre-multiplied spectrum of the first SPOD eigenvalue $\alpha\beta\lambda_1^{\text{\tiny SPOD}}$ to indicate the energetic scales.
Then, the $\beta_{\alpha,\beta,\omega}^{\text{\tiny model}}$ metrics are provided for $\nu$-resolvent (figures~\ref{fig:beta_resolvent_c10}, \ref{fig:beta_resolvent_c15}, \ref{fig:beta_resolvent_c18.3} and \ref{fig:beta_resolvent_c19}), $\nu_t$-resolvent (figures~\ref{fig:beta_resolvent_eddy_c10}, \ref{fig:beta_resolvent_eddy_c15}, \ref{fig:beta_resolvent_eddy_c18.3} and \ref{fig:beta_resolvent_eddy_c19}). 
First, we can see that $\nu$-resolvent is relevant only in the buffer-layer.
This highlights again the necessity of modelling the effect of small-scale turbulence on the coherent structure in turbulent regions.
The overall performances of $\nu_t$-resolvent and FSLM are good, and it is shown more clearly in the main document that FSLM improves agreement in a wide range of streamwise elongates scales.

We can note a bad performance of FSLM in the buffer layer at $y^+=13$ for small $\lambda_x$ and large $\lambda_z$.
These waves are not energetic (see figure~\ref{fig:lambda_SPOD_c10}), where the convergence of SPOD is difficult to obtain.
Finally, we observe a slight worsening at large scales, with $\lambda_z>\lambda_x$ far from the wall (for W3 and W4).
We interpret this by effects of the outer region, which have not been included in the decorrelation time scale model $\tau$, but present in Cess's eddy viscosity model.
This could constitute a potential improvement of the noise parameters definition.

\bibliographystyle{biblio/unsrtnat}

\begin{thebibliography}{70}
\providecommand{\natexlab}[1]{#1}
\providecommand{\url}[1]{\texttt{#1}}
\expandafter\ifx\csname urlstyle\endcsname\relax
  \providecommand{\doi}[1]{doi: #1}\else
  \providecommand{\doi}{doi: \begingroup \urlstyle{rm}\Url}\fi

\bibitem[Kline et~al.(1967)Kline, Reynolds, Schraub, and Runstadler]{kline1967}
S.~J. Kline, W.~C. Reynolds, F.~A. Schraub, and P.~W. Runstadler.
\newblock The structure of turbulent boundary layers.
\newblock \emph{Journal of Fluid Mechanics}, 30\penalty0 (04):\penalty0
  741--773, 1967.

\bibitem[Smith and Metzler(1983)]{smith1983}
C.~R. Smith and S.~P. Metzler.
\newblock The characteristics of low-speed streaks in the near-wall region of a
  turbulent boundary layer.
\newblock \emph{Journal of Fluid Mechanics}, 129:\penalty0 27–54, 1983.
\newblock \doi{10.1017/S0022112083000634}.

\bibitem[Jim{\'e}nez and Moin(1991)]{jimenez1991}
J.~Jim{\'e}nez and P.~Moin.
\newblock The minimal flow unit in near-wall turbulence.
\newblock \emph{Journal of Fluid Mechanics}, 225\penalty0 (213--240), 1991.

\bibitem[Panton(2001)]{panton2001}
R.~L. Panton.
\newblock Overview of the self-sustaining mechanisms of wall turbulence.
\newblock \emph{Progress in Aerospace Sciences}, 37\penalty0 (4):\penalty0
  341--383, 2001.

\bibitem[Hamilton et~al.(1995)Hamilton, Kim, and Waleffe]{hamilton1995}
J.~M. Hamilton, J.~Kim, and F.~Waleffe.
\newblock Regeneration mechanisms of near-wall turbulence structures.
\newblock \emph{Journal of Fluid Mechanics}, 287:\penalty0 317–348, 1995.
\newblock \doi{10.1017/S0022112095000978}.

\bibitem[Ellingsen and Palm(1975)]{ellingsen1975}
T.~Ellingsen and E.~Palm.
\newblock Stability of linear flow.
\newblock \emph{The Physics of Fluids}, 18\penalty0 (4):\penalty0 487--488,
  1975.
\newblock \doi{10.1063/1.861156}.

\bibitem[Brandt(2014)]{brandt2014}
L.~Brandt.
\newblock The lift-up effect: {T}he linear mechanism behind transition and
  turbulence in shear flows.
\newblock \emph{European Journal of Mechanics-B/Fluids}, 47:\penalty0 80--96,
  2014.

\bibitem[Flores and Jim{\'e}nez(2010)]{flores2010}
O.~Flores and J.~Jim{\'e}nez.
\newblock Hierarchy of minimal flow units in the logarithmic layer.
\newblock \emph{Physics of Fluids}, 22\penalty0 (7):\penalty0 071704, 2010.
\newblock \doi{http://dx.doi.org/10.1063/1.3464157}.
\newblock URL
  \url{http://scitation.aip.org/content/aip/journal/pof2/22/7/10.1063/1.3464157}.

\bibitem[Smits et~al.(2011)Smits, McKeon, and Marusic]{smits2011}
A.~J. Smits, B.~J. McKeon, and I.~Marusic.
\newblock High–{R}eynolds number wall turbulence.
\newblock \emph{Annual Review of Fluid Mechanics}, 43\penalty0 (1):\penalty0
  353--375, 2011.
\newblock \doi{10.1146/annurev-fluid-122109-160753}.

\bibitem[Pope(2000)]{popebook}
S.~B. Pope.
\newblock \emph{Turbulent Flows}.
\newblock Cambridge University Press, 2000.
\newblock \doi{10.1017/CBO9780511840531}.

\bibitem[Hwang and Cossu(2010{\natexlab{a}})]{hwangPRL2010}
Y.~Hwang and C.~Cossu.
\newblock Self-sustained process at large scales in turbulent channel flow.
\newblock \emph{Physical Review Letters}, 105:\penalty0 044505, Jul
  2010{\natexlab{a}}.
\newblock \doi{10.1103/PhysRevLett.105.044505}.
\newblock URL \url{https://link.aps.org/doi/10.1103/PhysRevLett.105.044505}.

\bibitem[Cossu and Hwang(2017)]{cossu2017}
C.~Cossu and Y.~Hwang.
\newblock Self-sustaining processes at all scales in wall-bounded turbulent
  shear flows.
\newblock \emph{Philosophical Transactions of the Royal Society A:
  Mathematical, Physical and Engineering Sciences}, 375\penalty0
  (2089):\penalty0 20160088, 2017.

\bibitem[Lozano-Dur\'an et~al.(2020)Lozano-Dur\'an, Bae, and
  Encinar]{lozano2020}
A.~Lozano-Dur\'an, H.~J. Bae, and M.~P. Encinar.
\newblock Causality of energy-containing eddies in wall turbulence.
\newblock \emph{Journal of Fluid Mechanics}, 882:\penalty0 A2, 2020.
\newblock \doi{10.1017/jfm.2019.801}.

\bibitem[Bae et~al.(2021)Bae, Lozano-Dur\'an, and McKeon]{bae2021}
H.~J. Bae, A.~Lozano-Dur\'an, and B.~J. McKeon.
\newblock Nonlinear mechanism of the self-sustaining process in the buffer and
  logarithmic layer of wall-bounded flows.
\newblock \emph{Journal of Fluid Mechanics}, 914:\penalty0 A3, 2021.
\newblock \doi{10.1017/jfm.2020.857}.

\bibitem[Barkley(2006)]{barkley2006}
D.~Barkley.
\newblock Linear analysis of the cylinder wake mean flow.
\newblock \emph{Europhysics Letters ({EPL})}, 75\penalty0 (5):\penalty0
  750--756, sep 2006.
\newblock \doi{10.1209/epl/i2006-10168-7}.

\bibitem[Schmid and Henningson(2001)]{schmid2001}
P.~J. Schmid and D.~S. Henningson.
\newblock \emph{Stability and transition in shear flows}, volume 142.
\newblock Springer-Verlag, 2001.

\bibitem[Trefethen and Embree(2005)]{trefethenbook}
L.~N. Trefethen and M.~Embree.
\newblock \emph{Spectra and pseudospectra: the behavior of nonnormal matrices
  and operators}.
\newblock Princeton University Press, 2005.

\bibitem[Jovanovi\'c and Bamieh(2005)]{jovanovic2005}
M.~R. Jovanovi\'c and B.~Bamieh.
\newblock Componentwise energy amplification in channel flows.
\newblock \emph{Journal of Fluid Mechanics}, 534:\penalty0 145–183, 2005.
\newblock \doi{10.1017/S0022112005004295}.

\bibitem[McKeon and Sharma(2010)]{mckeon2010}
B.~J. McKeon and A.~S. Sharma.
\newblock A critical-layer framework for turbulent pipe flow.
\newblock \emph{Journal of Fluid Mechanics}, 658:\penalty0 336--382, 2010.

\bibitem[G\'omez et~al.(2016)G\'omez, Sharma, and Blackburn]{gomez2016b}
F.~G\'omez, A.~S. Sharma, and H.~M. Blackburn.
\newblock Estimation of unsteady aerodynamic forces using pointwise velocity
  data.
\newblock \emph{Journal of Fluid Mechanics}, 804:\penalty0 R4, 2016.
\newblock \doi{10.1017/jfm.2016.546}.

\bibitem[Symon et~al.(2019)Symon, Sipp, and McKeon]{symon2019}
S.~Symon, D.~Sipp, and B.~J. McKeon.
\newblock A tale of two airfoils: resolvent-based modelling of an oscillator
  versus an amplifier from an experimental mean.
\newblock \emph{Journal of Fluid Mechanics}, 881:\penalty0 51–83, 2019.
\newblock \doi{10.1017/jfm.2019.747}.

\bibitem[Martini et~al.(2020)Martini, Cavalieri, Jordan, Towne, and
  Lesshafft]{martini2020}
E.~Martini, A.~V.~G. Cavalieri, P.~Jordan, A.~Towne, and L.~Lesshafft.
\newblock Resolvent-based optimal estimation of transitional and turbulent
  flows.
\newblock \emph{Journal of Fluid Mechanics}, 900:\penalty0 A2, 2020.
\newblock \doi{10.1017/jfm.2020.435}.

\bibitem[Towne et~al.(2020)Towne, Lozano-Dur\'an, and Yang]{towne2020}
A.~Towne, A.~Lozano-Dur\'an, and X.~Yang.
\newblock Resolvent-based estimation of space–time flow statistics.
\newblock \emph{Journal of Fluid Mechanics}, 883:\penalty0 A17, 2020.
\newblock \doi{10.1017/jfm.2019.854}.

\bibitem[Amaral et~al.(2021)Amaral, Cavalieri, Martini, Jordan, and
  Towne]{amaral2021}
F.~R. Amaral, A.~V.~G. Cavalieri, E.~Martini, P.~Jordan, and A.~Towne.
\newblock Resolvent-based estimation of turbulent channel flow using wall
  measurements.
\newblock \emph{Journal of Fluid Mechanics}, 927:\penalty0 A17, 2021.
\newblock \doi{10.1017/jfm.2021.764}.

\bibitem[Franceschini et~al.(2021)Franceschini, Sipp, and
  Marquet]{franceschini2021}
L.~Franceschini, D.~Sipp, and O.~Marquet.
\newblock Mean- and unsteady-flow reconstruction with one or two time-resolved
  measurements.
\newblock \emph{arXiv preprint arXiv:2102.03839}, 2021.

\bibitem[Leclercq et~al.(2019)Leclercq, Demourant, Poussot-Vassal, and
  Sipp]{leclercq2019}
C.~Leclercq, F.~Demourant, C.~Poussot-Vassal, and D.~Sipp.
\newblock Linear iterative method for closed-loop control of quasiperiodic
  flows.
\newblock \emph{Journal of Fluid Mechanics}, 868:\penalty0 26–65, 2019.
\newblock \doi{10.1017/jfm.2019.112}.

\bibitem[Reynolds and Hussain(1972)]{reynolds1972}
W.~C. Reynolds and A.~K. M.~F. Hussain.
\newblock The mechanics of an organized wave in turbulent shear flow. {P}art 3.
  {T}heoretical models and comparisons with experiments.
\newblock \emph{Journal of Fluid Mechanics}, 54\penalty0 (2):\penalty0
  263–288, 1972.
\newblock \doi{10.1017/S0022112072000679}.

\bibitem[Cess(1958)]{Cess1958}
R.~D. Cess.
\newblock A survey of the literature on heat transfer in turbulent tubeflow.
\newblock \emph{Research Report No. 8-0529-R24}, 1958.

\bibitem[Hwang and Cossu(2010{\natexlab{b}})]{hwang2010}
Y.~Hwang and C.~Cossu.
\newblock Linear non-normal energy amplification of harmonic and stochastic
  forcing in the turbulent channel flow.
\newblock \emph{Journal of Fluid Mechanics}, 664:\penalty0 51–73,
  2010{\natexlab{b}}.
\newblock \doi{10.1017/S0022112010003629}.

\bibitem[Morra et~al.(2019)Morra, Semeraro, Henningson, and Cossu]{morra2019}
P.~Morra, O.~Semeraro, D.~S. Henningson, and C.~Cossu.
\newblock On the relevance of {R}eynolds stresses in resolvent analyses of
  turbulent wall-bounded flows.
\newblock \emph{Journal of Fluid Mechanics}, 867:\penalty0 969–984, 2019.
\newblock \doi{10.1017/jfm.2019.196}.

\bibitem[Symon et~al.(2020)Symon, Illingworth, and Marusic]{symon2020}
S.~Symon, S.~J. Illingworth, and I.~Marusic.
\newblock Large-scale structures predicted by linear models of wall-bounded
  turbulence.
\newblock \emph{Journal of Physics: Conference Series}, 1522\penalty0
  (1):\penalty0 012006, 2020.

\bibitem[Symon et~al.(2021)Symon, Illingworth, and Marusic]{symon2021}
S.~Symon, S.~J. Illingworth, and I.~Marusic.
\newblock Energy transfer in turbulent channel flows and implications for
  resolvent modelling.
\newblock \emph{Journal of Fluid Mechanics}, 911, 2021.

\bibitem[Symon et~al.(2022)Symon, Madhusudanan, Illingworth, and
  Marusic]{symonarxiv}
S.~Symon, A.~Madhusudanan, S.~J. Illingworth, and I.~Marusic.
\newblock On the use of eddy viscosity in resolvent analysis of turbulent
  channel flow.
\newblock \emph{arXiv}, 2022.
\newblock \doi{10.48550/ARXIV.2205.11216}.
\newblock URL \url{https://arxiv.org/abs/2205.11216}.

\bibitem[Nogueira et~al.(2021)Nogueira, Morra, Martini, Cavalieri, and
  Henningson]{nogueira2021}
P.~A.~S. Nogueira, P.~Morra, E.~Martini, A.~V.~G. Cavalieri, and D.~S.
  Henningson.
\newblock Forcing statistics in resolvent analysis: application in minimal
  turbulent {C}ouette flow.
\newblock \emph{Journal of Fluid Mechanics}, 908:\penalty0 A32, 2021.
\newblock \doi{10.1017/jfm.2020.918}.

\bibitem[Morra et~al.(2021)Morra, Nogueira, Cavalieri, and
  Henningson]{morra2021}
P.~Morra, P.~A.~S. Nogueira, A.~V.~G. Cavalieri, and D.~S. Henningson.
\newblock The colour of forcing statistics in resolvent analyses of turbulent
  channel flows.
\newblock \emph{Journal of Fluid Mechanics}, 907:\penalty0 A24, 2021.
\newblock \doi{10.1017/jfm.2020.802}.

\bibitem[Gupta et~al.(2021)Gupta, Madhusudanan, Wan, Illingworth, and
  Juniper]{gupta2021}
V.~Gupta, A.~Madhusudanan, M.~Wan, S.~J. Illingworth, and M.~P. Juniper.
\newblock Linear-model-based estimation in wall turbulence: improved stochastic
  forcing and eddy viscosity terms.
\newblock \emph{Journal of Fluid Mechanics}, 925:\penalty0 A18, 2021.
\newblock \doi{10.1017/jfm.2021.671}.

\bibitem[Zare et~al.(2017)Zare, Jovanovi\'c, and Georgiou]{zare2017}
A.~Zare, M.~R. Jovanovi\'c, and T.~T. Georgiou.
\newblock Colour of turbulence.
\newblock \emph{Journal of Fluid Mechanics}, 812:\penalty0 636–680, 2017.
\newblock \doi{10.1017/jfm.2016.682}.

\bibitem[Zare et~al.(2019)Zare, Georgiou, and Jovanovi{\'c}]{zare2019}
A.~Zare, T.~T. Georgiou, and M.~R. Jovanovi{\'c}.
\newblock Stochastic dynamical modeling of turbulent flows.
\newblock \emph{Annual Review of Control, Robotics, and Autonomous Systems}, 3,
  2019.

\bibitem[Tissot et~al.(2021)Tissot, Cavalieri, and Mémin]{tissotJFM2021}
G.~Tissot, A.~V.~G. Cavalieri, and E.~Mémin.
\newblock Stochastic linear modes in a turbulent channel flow.
\newblock \emph{Journal of Fluid Mechanics}, 912:\penalty0 A51, 2021.
\newblock \doi{10.1017/jfm.2020.1168}.

\bibitem[M{\'e}min(2014)]{memin2014}
E.~M{\'e}min.
\newblock Fluid flow dynamics under location uncertainty.
\newblock \emph{Geophysical \& Astrophysical Fluid Dynamics}, 108\penalty0
  (2):\penalty0 119--146, 2014.

\bibitem[Chandramouli et~al.(2018)Chandramouli, Heitz, Laizet, and
  M\'emin]{chandramouli2018}
P.~Chandramouli, D.~Heitz, S.~Laizet, and E.~M\'emin.
\newblock Coarse large-eddy simulations in a transitional wake flow with flow
  models under location uncertainty.
\newblock \emph{Computers \& Fluids}, 168:\penalty0 170--189, 2018.
\newblock ISSN 0045--7930.
\newblock \doi{https://doi.org/10.1016/j.compfluid.2018.04.001}.
\newblock URL
  \url{http://www.sciencedirect.com/science/article/pii/S0045793018301890}.

\bibitem[Resseguier et~al.(2017{\natexlab{a}})Resseguier, M{\'e}min, and
  Chapron]{resseguier2017b}
V.~Resseguier, E.~M{\'e}min, and B.~Chapron.
\newblock {Geophysical flows under location uncertainty, Part I Random
  transport and general models}.
\newblock \emph{{Geophysical and Astrophysical Fluid Dynamics}}, 111\penalty0
  (3):\penalty0 149--176, April 2017{\natexlab{a}}.
\newblock \doi{10.1080/03091929.2017.1310210}.
\newblock URL \url{https://hal.inria.fr/hal-01391420}.

\bibitem[Resseguier et~al.(2017{\natexlab{b}})Resseguier, M{\'e}min, and
  Chapron]{resseguier2017c}
V.~Resseguier, E.~M{\'e}min, and B.~Chapron.
\newblock {Geophysical flows under location uncertainty, Part II
  Quasi-geostrophy and efficient ensemble spreading}.
\newblock \emph{{Geophysical and Astrophysical Fluid Dynamics}}, 111\penalty0
  (3):\penalty0 177--208, April 2017{\natexlab{b}}.
\newblock \doi{10.1080/03091929.2017.1312101}.
\newblock URL \url{https://hal.inria.fr/hal-01391476}.

\bibitem[Resseguier et~al.(2017{\natexlab{c}})Resseguier, M{\'e}min, and
  Chapron]{resseguier2017d}
V.~Resseguier, E.~M{\'e}min, and B.~Chapron.
\newblock {Geophysical flows under location uncertainty, Part III SQG and
  frontal dynamics under strong turbulence conditions}.
\newblock \emph{{Geophysical and Astrophysical Fluid Dynamics}}, 111\penalty0
  (3):\penalty0 209--227, April 2017{\natexlab{c}}.
\newblock \doi{10.1080/03091929.2017.1312102}.
\newblock URL \url{https://hal.inria.fr/hal-01391484}.

\bibitem[Chapron et~al.(2018)Chapron, D{\'e}rian, M{\'e}min, and
  Resseguier]{chapron2018}
B.~Chapron, P.~D{\'e}rian, E.~M{\'e}min, and V.~Resseguier.
\newblock {Large scale flows under location uncertainty: a consistent
  stochastic framework}.
\newblock \emph{{Quarterly Journal of the Royal Meteorological Society}},
  144\penalty0 (710):\penalty0 251--260, 2018.
\newblock \doi{10.1002/qj.3198}.
\newblock URL \url{https://hal.inria.fr/hal-01629898}.

\bibitem[Bauer et~al.(2020{\natexlab{a}})Bauer, Chandramouli, Chapron, Li, and
  M{\'e}min]{bauer2020a}
W.~Bauer, P.~Chandramouli, B.~Chapron, L.~Li, and E.~M{\'e}min.
\newblock {Deciphering the role of small-scale inhomogeneity on geophysical
  flow structuration: a stochastic approach}.
\newblock \emph{{Journal of Physical Oceanography}}, February
  2020{\natexlab{a}}.
\newblock \doi{10.1175/JPO-D-19-0164.1}.
\newblock URL \url{https://hal.inria.fr/hal-02398521}.

\bibitem[Bauer et~al.(2020{\natexlab{b}})Bauer, Chandramouli, Li, and
  M{\'e}min]{bauer2020b}
W.~Bauer, P.~Chandramouli, L.~Li, and E.~M{\'e}min.
\newblock {Stochastic representation of mesoscale eddy effects in
  coarse-resolution barotropic models}.
\newblock \emph{{Ocean Modelling}}, 151:\penalty0 1--50, 2020{\natexlab{b}}.
\newblock \doi{10.1016/j.ocemod.2020.101646}.
\newblock URL \url{https://hal.inria.fr/hal-02666147}.

\bibitem[Pinier et~al.(2019)Pinier, M{\'e}min, Laizet, and
  Lewandowski]{pinier2019}
B.~Pinier, E.~M{\'e}min, S.~Laizet, and R.~Lewandowski.
\newblock {A stochastic flow approach to model the mean velocity profile of
  wall-bounded flows}.
\newblock \emph{{Physical Review E }}, 2019.
\newblock URL \url{https://hal.inria.fr/hal-01947662}.

\bibitem[Yang and M{\'e}min(2017)]{yang2017}
Y.~Yang and E.~M{\'e}min.
\newblock {High-resolution data assimilation through stochastic subgrid tensor
  and parameter estimation from 4DEnVar}.
\newblock \emph{{Tellus A}}, page~19, April 2017.
\newblock URL \url{https://hal.inria.fr/hal-01500140}.

\bibitem[Yang and M{\'e}min(2018)]{yang2018}
Y.~Yang and E.~M{\'e}min.
\newblock {Estimation of physical parameters under location uncertainty using
  an Ensemble$^2$-Expectation-Maximization algorithm}.
\newblock \emph{{Quaterly journal of the royal meteorological society}}, pages
  1--47, November 2018.
\newblock \doi{10.1002/qj.3438}.
\newblock URL \url{https://hal.inria.fr/hal-01944730}.

\bibitem[Chandramouli et~al.(2020)Chandramouli, M{\'e}min, and
  Heitz]{chandramouli2020}
P.~Chandramouli, E.~M{\'e}min, and D.~Heitz.
\newblock {4D large scale variational data assimilation of a turbulent flow
  with a dynamics error model}.
\newblock \emph{{Journal of Computational Physics}}, 412, July 2020.
\newblock \doi{10.1016/j.jcp.2020.109446}.
\newblock URL \url{https://hal.inria.fr/hal-02547763}.

\bibitem[Resseguier et~al.(2017{\natexlab{d}})Resseguier, M{\'e}min, Heitz, and
  Chapron]{resseguier2017}
V.~Resseguier, E.~M{\'e}min, D.~Heitz, and B.~Chapron.
\newblock Stochastic modelling and diffusion modes for proper orthogonal
  decomposition models and small-scale flow analysis.
\newblock \emph{Journal of Fluid Mechanics}, 826:\penalty0 888--917,
  2017{\natexlab{d}}.

\bibitem[Resseguier et~al.(2021)Resseguier, Picard, M\'emin, and
  Chapron]{resseguier2021ROM}
V.~Resseguier, A.~M. Picard, E.~M\'emin, and B.~Chapron.
\newblock Quantifying truncation-related uncertainties in unsteady fluid
  dynamics reduced order models.
\newblock \emph{SIAM/ASA Journal on Uncertainty Quantification}, 9\penalty0
  (3):\penalty0 1152--1183, 2021.

\bibitem[Towne et~al.(2018)Towne, Schmidt, and Colonius]{towne2018}
A.~Towne, O.~T Schmidt, and T.~Colonius.
\newblock Spectral proper orthogonal decomposition and its relationship to
  dynamic mode decomposition and resolvent analysis.
\newblock \emph{Journal of Fluid Mechanics}, 847:\penalty0 821--867, 2018.

\bibitem[Yim et~al.(2019)Yim, Meliga, and Gallaire]{yim2019}
E.~Yim, P.~Meliga, and F.~Gallaire.
\newblock Self-consistent triple decomposition of the turbulent flow over a
  backward-facing step under finite amplitude harmonic forcing.
\newblock \emph{Proceedings of the Royal Society A}, 475\penalty0
  (2225):\penalty0 20190018, 2019.

\bibitem[Plotter(2005)]{plotterbook}
P.~E. Plotter.
\newblock Stochastic integration and differential equation.
\newblock \emph{Stochastic Modeling and Applied Probability}, 21, 2005.

\bibitem[Jim{\'e}nez(2013)]{jimenez2013nearwallturbulence}
J.~Jim{\'e}nez.
\newblock Near-wall turbulence.
\newblock \emph{Physics of Fluids (1994-present)}, 25\penalty0 (10):\penalty0
  101302, 2013.
\newblock \doi{http://dx.doi.org/10.1063/1.4824988}.

\bibitem[Kadri~Harouna and M{\'e}min(2017)]{kadriharouna2017}
S.~Kadri~Harouna and E.~M{\'e}min.
\newblock {Stochastic representation of the Reynolds transport theorem:
  revisiting large-scale modeling}.
\newblock \emph{{Computers and Fluids}}, 156:\penalty0 456--469, August 2017.
\newblock \doi{10.1016/j.compfluid.2017.08.017}.
\newblock URL \url{https://hal.inria.fr/hal-01394780}.

\bibitem[Gibson et~al.(2019)Gibson, Reetz, Azimi, Ferraro, Kreilos,
  Schrobsdorff, Farano, Yesil, Schütz, Culpo, and Schneider]{channelflow}
J.~F. Gibson, F.~Reetz, S.~Azimi, A.~Ferraro, T.~Kreilos, H.~Schrobsdorff,
  M.~Farano, A.~F. Yesil, S.~S. Schütz, M.~Culpo, and T.~M. Schneider.
\newblock Channelflow 2.0.
\newblock \emph{Manuscript in preparation}, 2019.

\bibitem[Moarref et~al.(2013)Moarref, Sharma, Tropp, and McKeon]{moarref2013}
R.~Moarref, A.~S. Sharma, J.~A. Tropp, and B.~J. McKeon.
\newblock Model-based scaling of the streamwise energy density in
  high-{R}eynolds-number turbulent channels.
\newblock \emph{Journal of Fluid Mechanics}, 734:\penalty0 275–316, 2013.
\newblock \doi{10.1017/jfm.2013.457}.

\bibitem[Cavalieri et~al.(2019)Cavalieri, Jordan, and Lesshafft]{cavalieri2019}
A.~V.~G. Cavalieri, P.~Jordan, and L.~Lesshafft.
\newblock Wave-packet models for jet dynamics and sound radiation.
\newblock \emph{Applied Mechanics Reviews}, 71\penalty0 (2), 2019.

\bibitem[Cavalieri et~al.(2013)Cavalieri, Rodriguez, Jordan, Colonius, and
  Gervais]{cavalieri2013}
A.~V.~G. Cavalieri, D.~Rodriguez, P.~Jordan, T.~Colonius, and Y.~Gervais.
\newblock Wavepackets in the velocity field of turbulent jets.
\newblock \emph{Journal of Fluid Mechanics}, 730:\penalty0 559--592, 9 2013.
\newblock ISSN 1469-7645.
\newblock \doi{10.1017/jfm.2013.346}.

\bibitem[Cossu et~al.(2009)Cossu, Pujals, and Depardon]{cossu2009}
C.~Cossu, G.~Pujals, and S.~Depardon.
\newblock Optimal transient growth and very large–scale structures in
  turbulent boundary layers.
\newblock \emph{Journal of Fluid Mechanics}, 619:\penalty0 79–94, 2009.
\newblock \doi{10.1017/S0022112008004370}.

\bibitem[Pickering et~al.(2021)Pickering, Rigas, Schmidt, Sipp, and
  Colonius]{pickering2021}
E.~Pickering, G.~Rigas, O.~T. Schmidt, D.~Sipp, and T.~Colonius.
\newblock Optimal eddy viscosity for resolvent-based models of coherent
  structures in turbulent jets.
\newblock \emph{Journal of Fluid Mechanics}, 917:\penalty0 A29, 2021.
\newblock \doi{10.1017/jfm.2021.232}.

\bibitem[Moarref et~al.(2014)Moarref, Jovanovi\'c, Tropp, Sharma, and
  McKeon]{moarref2014}
R.~Moarref, M.~R. Jovanovi\'c, J.~A. Tropp, A.~S. Sharma, and B.~J. McKeon.
\newblock A low-order decomposition of turbulent channel flow via resolvent
  analysis and convex optimization.
\newblock \emph{Physics of Fluids (1994-present)}, 26\penalty0 (5):\penalty0
  051701, 2014.
\newblock \doi{http://dx.doi.org/10.1063/1.4876195}.
\newblock URL
  \url{http://scitation.aip.org/content/aip/journal/pof2/26/5/10.1063/1.4876195}.

\bibitem[Lesshafft et~al.(2019)Lesshafft, Semeraro, Jaunet, Cavalieri, and
  Jordan]{lesshafft2019}
L.~Lesshafft, O.~Semeraro, V.~Jaunet, A.~V.~G. Cavalieri, and P.~Jordan.
\newblock Resolvent-based modeling of coherent wave packets in a turbulent jet.
\newblock \emph{Physical Review Fluids}, 4:\penalty0 063901, Jun 2019.

\bibitem[Abreu et~al.(2021)Abreu, Tanarro, Cavalieri, Schlatter, Vinuesa,
  Hanifi, and Henningson]{abreu2021}
L.~I. Abreu, A.~Tanarro, A.~V.~G. Cavalieri, P.~Schlatter, R.~Vinuesa,
  A.~Hanifi, and D.~S. Henningson.
\newblock Spanwise-coherent hydrodynamic waves around flat plates and airfoils.
\newblock \emph{Journal of Fluid Mechanics}, 927:\penalty0 A1, 2021.
\newblock \doi{10.1017/jfm.2021.718}.

\bibitem[Pujals et~al.(2009)Pujals, Garc\'ia-Villalba, Cossu, and
  Depardon]{pujals2009}
G.~Pujals, M~Garc\'ia-Villalba, C.~Cossu, and S.~Depardon.
\newblock A note on optimal transient growth in turbulent channel flows.
\newblock \emph{Physics of Fluids}, 21\penalty0 (1):\penalty0 015109, 2009.
\newblock \doi{10.1063/1.3068760}.
\newblock URL \url{https://doi.org/10.1063/1.3068760}.

\bibitem[Ribeiro et~al.(2020)Ribeiro, Yeh, and Taira]{ribeiro2020}
J.~H.~M. Ribeiro, C.-A. Yeh, and K.~Taira.
\newblock Randomized resolvent analysis.
\newblock \emph{Physical Review Fluids}, 5:\penalty0 033902, Mar 2020.
\newblock \doi{10.1103/PhysRevFluids.5.033902}.
\newblock URL \url{https://link.aps.org/doi/10.1103/PhysRevFluids.5.033902}.

\bibitem[Martini et~al.(2021)Martini, Rodríguez, Towne, and
  Cavalieri]{martini2021}
E.~Martini, D.~Rodríguez, A.~Towne, and A.V.G. Cavalieri.
\newblock Efficient computation of global resolvent modes.
\newblock \emph{Journal of Fluid Mechanics}, 919:\penalty0 A3, 2021.
\newblock \doi{10.1017/jfm.2021.364}.

\end{thebibliography}

\end{document}